\documentclass[mathpazo]{cicp_arxiv}

\usepackage[a4paper,width=160mm,top=35mm,bottom=35mm,bindingoffset=10mm]{geometry}

\usepackage{hyperref}

\usepackage{pgfplots}
\usetikzlibrary{fit,positioning,arrows,automata,calc}
\usepackage{verbatim}
\usepackage{placeins}
\usepackage{titlesec}

\newlength\figureheight
\newlength\figurewidth

\titleformat{\paragraph}
{\normalfont\normalsize\bfseries}{\theparagraph}{1em}{}
\titlespacing*{\paragraph}
{0pt}{3.25ex plus 1ex minus .2ex}{1.5ex plus .2ex}

\usepackage{graphicx}
\usepackage{subcaption}
\usepackage{caption}
\usepackage{amsfonts}
\usepackage{amsmath}
\usepackage{tikz}

\usepackage[ruled]{algorithm2e}

\makeatletter
\renewcommand{\@algocf@capt@plain}{above}
\makeatother

\usepgfplotslibrary{external}
\usepgfplotslibrary{groupplots}

\tikzset{
  main/.style={circle, minimum size = 8mm, thick, draw =black!80, node distance 
= 10mm},
  connect/.style={-latex, thick},
  box/.style={rectangle, draw=black!100}
}

\newcommand{\ee}{\end{equation}}
\newcommand{\be}{\begin{equation}}
\newcommand{\ec}{\end{center}}
\newcommand{\bc}{\begin{center}}
\newcommand{\eea}{\end{eqnarray}}
\newcommand{\bea}{\begin{eqnarray}}
\newcommand{\bd}{\begin{description}}
\newcommand{\ed}{\end{description}}
\newcommand{\bi}{\begin{itemize}}
\newcommand{\ei}{\end{itemize}}
\newcommand{\pa}{\partial}
\newcommand{\bs}{\boldsymbol}
\def\RR{ \mathbb R}
\newcommand{\refeq}[1]{Eq. (\ref{#1})}

\newcommand{\bxx}{\bs{X}}
\newcommand{\bx}{\bs{x}}
\newcommand\bt{\boldsymbol{\theta}}

\usepackage[perpage]{footmisc}

\begin{document}
\title{Physics-constrained, data-driven discovery of coarse-grained dynamics}

\author[L.~Felsberger]{L.~Felsberger\affil{1}, 
P.S.~Koutsourelakis\affil{1}\comma\corrauth}
\address{\affilnum{1}\ Continuum Mechanics Group, Department of Mechanical 
Engineering, 
Technical University of Munich, Boltzmannstrasse 15 85748, Garching, Germany}
\emails{{\tt p.s.koutsourelakis} (P.S.~Koutsourelakis)}


\begin{abstract}
The combination of high-dimensionality and disparity of time scales  encountered in many problems in computational physics has motivated the development of coarse-grained (CG) models.
 In this paper, we advocate the paradigm of data-driven discovery for extracting  governing equations by employing fine-scale simulation data. In particular,  
 we cast the coarse-graining  process under a probabilistic state-space model where the transition law dictates the evolution of the CG state variables and the emission law the coarse-to-fine map. The directed probabilistic graphical model implied, suggests that given values for the fine-grained (FG) variables, probabilistic inference tools must be employed to identify the corresponding values for the CG states and to that end, we employ Stochastic Variational Inference. 
We advocate  a sparse Bayesian learning  perspective which    avoids overfitting and  reveals the most salient features in the CG evolution law.
The formulation adopted enables the quantification of a crucial, and often neglected, component in the CG process, i.e. the predictive uncertainty due to information loss.
Furthermore, it is capable of  reconstructing  the evolution of the full, fine-scale system.
 We  demonstrate the efficacy of the proposed framework  in high-dimensional systems of random walkers.

\end{abstract}


\maketitle

\section{Introduction}
\label{sec:intro}


The present paper is concerned with the discovery of data-driven, dynamic,  stochastic coarse-grained models from 
fine-scale simulations with a view of advancing multiscale modeling.
Many problems in science and engineering are modeled by high-dimensional systems of deterministic or stochastic, 
(non)linear, microscopic evolution laws (e.g  ODEs). Several such  examples are encountered in quantum mechanical models 
\cite{harlim_parametric_2015}, in the    
atomistic simulation of materials \cite{li_coarse-grained_2010}, in  complex flows 
 \cite{majda_normal_2009,majda_physics_2013,li_comparative_2016}, and  in  agent-based models  
\cite{thomas_equation-free_2016}.
Their solution is generally dominated by the smaller time scales involved even though the outputs of interest might 
pertain  to time scales that are greater by several orders of magnitude \cite{givon_extracting_2004}. The combination 
of high-dimensionality and 
disparity of time scales has motivated the development of  coarse-grained formulations. These aim at constructing a 
much lower-dimensional model that is practical to integrate in 
time and can adequately predict the outputs of interest over the time scales of interest.
The literature on this topic is vast and many categorizations are possible. This paper focuses on {\em data-driven} 
strategies 
\cite{horenko_data-based_2007,espanol_obtaining_2011,wan_probabilistic_2013,kondrashov_data-driven_2015,
harmandaris_path-space_2016,lei_data-driven_2016,
schoeberl_predictive_2017,
crosskey_atlas:_2017,yang_learning_2017} where  (short) simulations of the 
original, fine-grained (FG) or full-order system of equations are used in order to  learn (or 
infer) the right coarse-grained 
(CG) model.  This is consistent with the emergence of  data-driven discovery, commonly referred to
as the
fourth paradigm in science \cite{hey_fourth_2009}.
Extracting governing equations from data is a central challenge in a wide variety of physical and
engineering sciences as in climate science, neuroscience, ecology, finance, and epidemiology, where
models (or closures) often remain elusive
\cite{li_deciding_2007,schmidt_distilling_2009,brunton_discovering_2016}.

The challenge in multiscale physical systems such as those encountered in non-equilibrium statistical mechanics  
\cite{grabert_projection_1982}, is even greater as apart from the identification of an effective model, it is crucial 
to 
discover a good set of CG state variables \cite{rohrdanz_discovering_2013}. The latter problem  has been traditionally addressed 
separately from the construction of the CG model, using dimensionality reduction techniques, (e.g. 
\cite{nadler_diffusion_2006}) but in this work we offer a holistic treatment.

The starting point of all (to the best of our knowledge) CG schemes is the specification of a set of coarse (or 
reduced) state variables with respect to which a CG model is prescribed/found. These variables are selected on the 
basis of 
the analysis objectives, or because they are known to be ``slow'' either through physical insight or using rigorous 
statistical learning tools (e.g. dimensionality reduction). The specification of these coarse variables is 
generally done through a projection (or restriction), many-to-one  operator from the fine-scale variables to the 
coarse. For problems exhibiting time-scale separation and when the CG state variables identified correspond to 
slowly-evolving 
features, it has been established that Markovian CG models  are suitable and several numerical strategies have proven 
successful
\cite{weinan_heterogeneous_2007,kevrekidis_equation-free_2003}.
In the absence of scale separation however, memory effects (dissipation) and thermal noise  play an important role. The 
 Mori-Zwanzig (MZ) formalism which was  originally developed in the context of irreversible
statistical mechanics \cite{mori_transport_1965,zwanzig_nonlinear_1973} but has been extended to general systems of  
ODEs (e.g. \cite{chorin_optimal_2000,chorin_problem_2005,stinis_renormalized_2015,legoll_effective_2010}) provides a rigorous mathematical 
foundation. Setting aside the largely unsolved computational 
challenges associated with finding the terms in the Generalized Langevin Equation (GLE) \cite{darve_computing_2009,
hijon_mori-zwanzig_2010} prescribed by MZ, it is in principle a 
perfect scheme, i.e. it is capable of capturing exactly the dynamics of the state variables that are resolved and 
with respect to which a closed system of equations is written. The solution of  non-Markovian system of equations (due 
to the memory term) poses  itself several difficulties as it necessitates storing the history of the CG variables.  
 Interestingly, in many instances, this non-Markovian system can
be mapped onto an, albeit higher-dimensional, Markovian,  system   with additional degrees of freedom 
 corresponding to auxiliary or extended state variables. Such a strategy has been employed in other models exhibiting 
memory, e.g. in plasticity  of solids where the auxiliary variables are physically motivated (internal state 
variables, \cite{coleman_thermodynamics_1967}).  Another difficulty stems from sampling   of the random noise term 
in the GLE which is essential in achieving the correct
equilibrium statistics according to the second fluctuation-dissipation theorem (FDT) \cite{chorin_stochastic_2013}.


Even if a 
good CG model can be identified such that it correctly predicts the time evolution of these coarse variables, it 
is difficult 
(without additional assumptions such as the prescription of a lifting operator, \cite{erban_variable-free_2007}) to 
reconstruct or infer the 
evolution of the full, fine-scale system. Hence if the observables one is interested in predicting, do not exclusively 
depend on the coarse variables selected, the CG model constructed is not useful. In this paper, we advocate a different 
strategy, i.e. we  aim at  constructing a reduced or coarsened description of the FG system that is predictive 
of the whole fine-scale picture. The starting point is the prescription of a (parametrized) {\em probabilistic}, coarse-to-fine map \cite{schoeberl_predictive_2017}.
As a result, the  CG state variables are defined indirectly and are latent (hidden) given the FG simulation data. The formulation is complemented by a (parametrized) {\em probabilistic} evolution law for the CG  variables. 
The goal is not to write a consistent evolution law with respect to these state variables (as one would do with MZ when possible), but rather to identify the right evolution law for the CG states that is most predictive of the FG futures. 
In this manner, CG variables and CG model are learned simultaneously.

The formulation advocated enables the quantification of a crucial, and often neglected, component   in the CG  process \cite{katsoulakis_information_2006}, i.e. the predictive uncertainty due to information 
loss 
that unavoidably takes place. That is, unless there are known redundancies in the full-order, FG  description, 
the coarse variables will provide a lossy compression. This in turn will  
manifest itself in 
uncertainty in the predictions that we can generate given the coarse variables (i.e. there are many microscopic states 
compatible with each value of the CG variables). We emphasize that this has nothing to do with the quality of the 
coarse model nor with its potentially stochastic nature. Even if the exact equations were available (e.g. through MZ) i.e. one was able to perfectly 
predict the time evolution of the coarse state variables, it is not necessary that one would be able to predict 
perfectly the full, fine-scale picture. Naturally, if the CG model was stochastic, the randomness in the evolution of the CG states would  compound the aforementioned predictive uncertainty.

Another potentially significant source of uncertainty in the context of data-driven techniques, pertains to the data 
itself. That is, the CG model (or the parameters thereof) apart from being data-dependent, is learned from finite amounts of fine-scale simulation data. 
As these are generally expensive to procure and only relatively  small time spans are accessible by microscopic simulators (e.g. molecular dynamics), one would like to minimize the number necessary or  even better, to be able to quantify the effect 
of a finite-sized dataset to the uncertainty in the predictive estimates.  In this manner the analyst 
can decide (potentially also from an optimal experimental design perspective \cite{lindley_measure_1956}) whether the predictive accuracy of the CG model is adequate or additional FG simulation data is needed.

In this paper,  we treat the CG model as a probabilistic state-space model 
\cite{cappe_inference_2005,ghahramani_unsupervised_2004,durstewitz_state_2017} where the 
transition law dictates the evolution of the CG state variables and the emission law the coarse-to-fine map. The 
directed probabilistic graphical model implied, suggests that given values for the FG variables, inference tools must be
employed to identify the corresponding values for the CG states \cite{murphy_machine_2012}.  Naturally, 
one of the most critical questions pertains to the form of the CG evolution law which poses a formidable model 
selection issue. Correct identification of the  right-hand side terms involved can also
reveal qualitative features of the coarse-scale evolution such as the type of constitutive relations 
\cite{li_deciding_2007}. It is crucial therefore to be able to search across model types.  In order to avoid 
overfitting as well as endow the estimates with robustness even in the presence of limited data, we invoke the 
principle of parsimony that is reflected in the {\em sparsity}  of the solutions. Sparsity, as a guiding principle in 
the identification of dynamical systems has been employed before in 
\cite{wang_predicting_2011,schaeffer_sparse_2013,proctor_exploiting_2014}. Naturally the topic has a very long history 
in image/signal processing with far reaching contributions in the context of unsupervised learning  
\cite{olshausen_emergence_1996,olshausen_sparse_1997,lee_efficient_2006,goodfellow_spike-and-slab_2012,
 chalk_relevant_2016} which most closely resembles the framework used here. We adopt a sparse Bayesian learning \cite{wipf_sparse_2004,wipf_new_2008} perspective whereby 
appropriate hyperpriors based on Automatic Relevance Determination (ARD, \cite{mackay_probable_1995}) and without the specification of any 
additional hyperparameters, can achieve this goal i.e. regularize the
solution space using a  data-dependent prior distribution that effectively removes redundant or 
superfluous features. The model advocated is combined with Stochastic Variational Inference procedures for the latent CG states \cite{paisley_variational_2012,hoffman_stochastic_2013}.

We emphasize that the observations/data employed are not direct i.e. they do not explicitly correspond to the CG state 
variables as in \cite{daniels_automated_2015,brunton_discovering_2016,wan_reduced-space_2017}. This would effectively require solving a 
regression problem \cite{schaeffer_learning_2017}. Instead, the CG state variables  are actually inferred, simultaneously with the CG model 
form/parameters. As a result they can subsequently be used to reconstruct FG state variables futures either for the 
purposes of computing various observables or simply for consistently re-initializing the data-generating FG simulator.

The structure of the rest of the paper is as follows. In section \ref{sec:method} we present the general methodological framework and introduce the probabilistic graphical model underlying the CG process. We provide a specification of the advocated for systems of random walkers (agents) and discuss the issues of enforcement of conservation laws as well model-structure discovery. We conclude this section with algorithmic details pertaining to the inference (training) and prediction steps. In section \ref{sec:num} we demonstrate the capabilities of the proposed formulation  in three examples involving high-dimensional systems of random walkers. 
%
%
%
%

\section{Methodology}
\label{sec:method}
This section introduces the notational conventions adopted and presents
the proposed modeling and computational framework.  Subscripts with ``f''  denote quantities pertaining to the FG (or fine-scale or microscopic) description whereas those with ``c'' to the CG (or coarse-scale or macroscopic) model. We also generally employ lowercase letters to denote quantities related to the FG model and uppercase for  CG-model-related ones.

\subsection{General probabilistic state-space model}
\label{sec:gen}
We start with the presentation of the basic components of the proposed state-space model in a rather general setting. One of the primary goals is to demonstrate the flexibility of the data-driven strategy advocated, the Bayesian nature of the learning  process as well as its suitability in producing {\em probabilistic} predictions that quantify the information loss and reflect the finiteness of data. 

We consider a (generally  high-dimensional) FG model with state variables $\bs{x} \in \mathcal{M}_f \subset  \RR^{n_f}$ (where $n_f>>1$), the dynamics of which are described by a system of deterministic or stochastic ODEs, i.e.:
\be
\left\{ \begin{array}{l}
\dot{\bs{x}}_t =\bs{f}(\bx_t) \\
\bx_0=\hat{\bx}_0 \textrm{ or } \bx_0 \textrm{ drawn from } p_{f,0}(\bx_0) 
\end{array}
\right.
\label{eq:fg}
\ee
where the initial condition $\bx_0$ is either deterministic and known $\hat{\bx}_0$ or stochastic and  drawn from a given density $p_{f,0}(\bx_0)$.
We assume that the aforementioned microscopic  evolution law can be  integrated in time, albeit with a very small time-step which we denote by $\delta t$\footnote{We do not account for the integration errors in this study as this is outside the scope of the paper, see e.g. \cite{milstein_numerical_1994}.}. For some microscopic systems (e.g. agent-based or chemical-reaction-species models), the evolution laws might  be directly  prescribed in discretized form.

In order to make the discussion more concrete but also to relate with the systems examined in subsequent  sections, we consider in this work microscopic systems which consist of (non)interacting walkers in one-dimension in which case the state vector $\bx$ represents their coordinates. Hence $n_f$ is also the the number of the walkers. It is known \cite{li_deciding_2007} that depending on the specifics of the random walks and their interactions, the emergent behavior of the overall FG system can exhibit a wide range of features and time scales (e.g. shock formation, see section \ref{sec:num}).
The challenge we undertake is to find an effective, CG model that provides an adequate description of the FG system of walkers.

This is generally approached as a two-step process, where the analyst {\em separately}
\bi
\item[1.] identifies a set of CG variables, say $\bxx \in \mathcal{M}_c \subset \RR^{n_c}, ~~(n_c << n_f)$ through a {\em fine-to-coarse} map (restriction operator),
\item[2.] and constructs a closed CG model i.e. an evolution law for the CG variables e.g. $\dot{\bxx}_t=\bs{F}(\bxx_t)$.
\ei
In order to address the difficulties associated with these  tasks, but also  to construct a CG model that is capable of predicting, albeit with uncertainty, the whole FG picture, we adopt here a different strategy which is based on the {\em simultaneous}  learning of:
\bi
\item a {\em probabilistic evolution law} for the CG states $p_c(\bxx_{t+\Delta t} | \bxx_t)$.
\item a {\em probabilistic  coarse-to-fine} map $p_{cf}(\bx_t | \bxx_t)$.
\ei
where $\bs{X}_t$ denotes the CG state variables at time $t$ and $\Delta t >> \delta t$ the coarse time-step used for the evolution law of the CG states. We employ here directly the {\em discretized} evolution law as this will ultimately be used for predictive purposes. We have also prescribed an evolution law that is Markovian in order to simplify the presentation, although this can be relaxed. Before we discuss in detail the aforementioned densities, their parametrization as well as their calibration, we provide some crucial details about the overall formulation.

We note that the CG states are indirectly defined as latent generators that give rise to the FG states through the probabilistic lifting operator implied by $p_{cf}$. The interpretation of these latent variables can only be done through the prism of this generative mapping. According to this, each sequence of FG configurations $\bx_t$ at times $t=k\Delta t,~k=0,1,\ldots,n $ is generated as follows:
\bi
\item one selects a CG initial state $\bxx_0$, drawn for example from a prior density $p_{c,0}(\bxx_0)$. The initial FG state $\bx_0$ is drawn from $p_{cf}(\bx_0 | \bxx_0)$.
\item For $k=0,1,\ldots,n$:
\bi
\item evolve the CG model in time by sampling successive states from $p_c(\bxx_{(k+1)\Delta t} | \bxx_{k \Delta t})$,
\item sample $\bx_{k \Delta t} $ from $p_{cf}(\bx_{k \Delta t} | \bxx_{k \Delta t})$.
\ei
\ei
The aforementioned procedure implies that $\bx_t$ is effectively drawn from the following (conditional) density (given $\bxx_0$):
\be
\bar{p}_f(\bx_{(0:n)\Delta t} | \bxx_0)=   \int p_{cf}(\bx_{0} | \bxx_{0})  \left( \prod_{k=0}^{n-1} p_{c}(\bxx_{(k+1)\Delta t} | \bxx_{k \Delta t}) p_{cf}(\bx_{ (k+1) \Delta t} | \bxx_{(k+1) \Delta t}) \right) ~d\bxx_{\Delta t} d\bxx_{2\Delta t} \ldots d\bxx_{n\Delta t}
\ee
Naturally if the uncertainty in $\bxx_0$ is incorporated then:
\be
\bar{p}_f(\bx_{(0:n)\Delta t} )= \int p_{c,0}(\bxx_0)~p_{cf}(\bx_{0} | \bxx_{0})  \left( \prod_{k=0}^{n-1} p_{c}(\bxx_{(k+1)\Delta t} | \bxx_{k \Delta t}) p_{cf}(\bx_{ (k+1) \Delta t} | \bxx_{(k+1) \Delta t}) \right)   ~d\bxx_0~d\bxx_{\Delta t}  \ldots d\bxx_{n\Delta t}
\ee

\begin{figure}
\centering
\includegraphics[width=.6\textwidth]{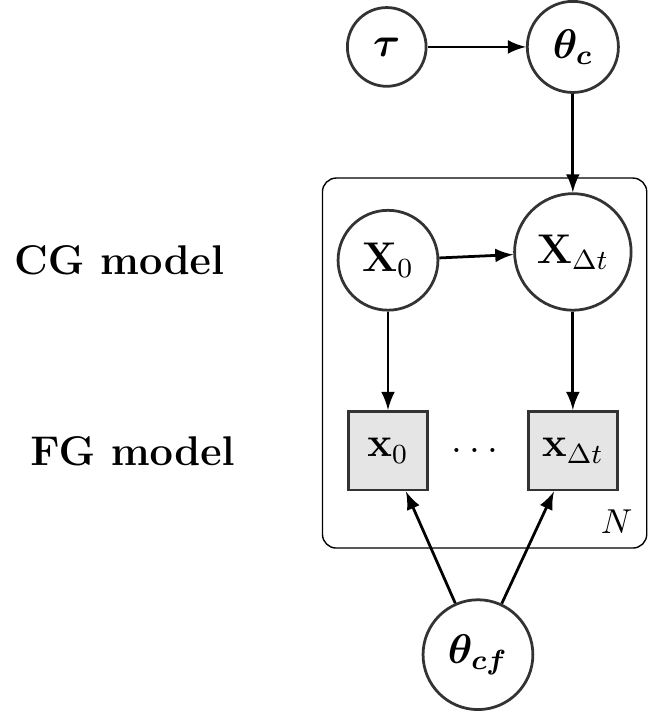} 
 \caption{Proposed probabilistic graphical model (for $n=1$). The dotted lines in the evolution of the FG model indicate that in general, the CG time step $\Delta t$ is a multiple of the FG one  $\delta t$.}
 \label{fig:PGM}
\end{figure}

Suppose the aforementioned densities are parametrized by $\bt=(\bt_c, \bt_{cf})$ i.e. $p_c(\bxx_{t+\Delta t} | \bxx_t, ~\bt_c)$ and $p_{cf}(\bx_t | \bxx_t, ~\bt_{cf})$ and we attempt to minimize the Kullback-Leibler divergence \cite{cover_elements_1991} between the reference density  $p_f(\bx_{(0:n)\Delta t})$ and $\bar{p}_f(\bx_{(0:n)\Delta t} | \bt )$. The former is either a Dirac-delta in the case of deterministic initial condition and deterministic  FG dynamics in \refeq{eq:fg} or a proper density in the case of stochastic initial conditions or stochastic FG evolution law \cite{katsoulakis_information-theoretic_2013} \footnote{As we show in the sequel, it is not necessary to know the density $p_f$ (which, e.g. if Fokker-Planck equations need to be solved, poses an even harder problem) but it suffices to be able to sample from it i.e. simulate the FG  dynamics.}. Then:
\be
\begin{array}{ll}
KL(p_f(\bx_{(0:n)\Delta t}) || \bar{p}_f(\bx_{(0:n)\Delta t} | \bt )   = &  -\int p_f(\bx_{(0:n)\Delta t})  \log  \bar{p}_f(\bx_{(0:n)\Delta t} | \bt  )~ d\bx_{(0:n)\Delta t} \\
& +\int p_f(\bx_{(0:n)\Delta t})  \log  {p}_f(\bx_{(0:n)\Delta t} )~ d\bx_{(0:n)\Delta t} \\
\end{array}
\label{eq:kl}
\ee
Given that the second term, i.e. the negative entropy of $p_f$, is independent of $\bt$, minimizing the KL-divergence is equivalent to maximizing \cite{espanol_obtaining_2011}:
\be
\mathcal{L}(\bt)= \int p_f(\bx_{(0:n)\Delta t})  \log  \bar{p}_f(\bx_{(0:n)\Delta t} | \bt  )~ d\bx_{(0:n)\Delta t} 
\ee
which can be approximated by sampling $N$ times from $p_f(\bx_{(0:n)\Delta t})$ i.e. by simulating the FG dynamics up to $n \Delta t$ in order to obtain the FG sequences $\{\bx_{(0:n)\Delta t}^{(i)} \}_{i=1}^N $. The resulting approximation which we denote by $L_N(\bt)$, effectively represents {\em log-likelihood} of the FG simulated data with respect to the generative model proposed i.e.: 
\be
\begin{array}{ll}
L_N(\bt) & = \sum_{i=1}^N  \log \bar{p}_f(\bx_{(0:n)\Delta t}^{(i)} | \bt  ) \\
 & = \sum_{i=1}^N    \log \int    p_{c,0}(\bxx_0^{(i)} )  p_{cf}(\bx_{0}^{(i)}  | \bxx_{0}^{(i)} )  \left( \prod_{k=0}^{n-1} p_{c}(\bxx_{(k+1)\Delta t}^{(i)}  | \bxx_{k \Delta t}^{(i)} ) p_{cf}(\bx_{ (k+1) \Delta t}^{(i)}  | \bxx_{(k+1) \Delta t}^{(i)} ) \right) ~d\bxx_{(0:n)\Delta t}^{(i)}.
\end{array}
\ee
We note in the expression above that we associate a sequence of latent, CG states  $\bxx_{(0:n)\Delta t}^{(i)}$ to each observed FG sequence $\bx_{(0:n)\Delta t}^{(i)}$. Hence the CG states play the role of pre-images of the FG states.  More importantly, the objective in the aforementioned expression accounts for both
the CG probabilistic evolution law as well as the reconstruction
(lifting) of the FG states from the  the (latent) CG ones.
The $\bt$ that maximize $L_N(\bt)$ would obviously correspond to the Maximum Likelihood estimate $\bt_{MLE}$.
Furthermore the interpretation of the objective as the log-likelihood makes
the passage to Bayesian formulations, straightforward. If for
example we define a prior density $p(\bt)$, then maximizing:
\be
\mathcal{L}_N(\bt)+\log p(\bt)
\ee
would be equivalent to obtaining a Maximum A Posteriori (MAP) estimate $\bt_{MAP}$. The natural progression from point estimates, is a fully Bayesian treatment which involves the {\em posterior density} of $\bt$ that according to Bayes' formula is given by:
\be
p(\bt |~~ \bx_{(0:n)\Delta t}^{(1:N)} ) = \frac{ e^{L_N(\bt)} ~p(\bt) }{\bar{p}_f(\bx_{(0:n)\Delta t}^{(1:N)}) } 
\label{eq:post}
\ee
We note that the posterior is well-defined regardless:
\bi 
\item  of the type of the FG model (since it relies on simulation data),
\item  if one ($N=1$) or more ($N>1$) data sequences, with the same or different lengths, are available,
\item  if these involve one ($n=1$) or more ($n>1$) coarse time-steps $\Delta t$.
\ei
The aforementioned relationship can be concretely represented in the form
of a directed graphical model as depicted in Figure \ref{fig:PGM} (for $n=1$).
%
%
%
%
%
%
%
%
%
%
%
%
%

We discuss in detail  a strategy for approximating this posterior in the next subsections. It is more important to emphasize at this stage that given this posterior, we can produce, not just point estimates of FG state futures by simulating the learned CG model,  but also quantify the  uncertainty in these  predictions. This predictive uncertainty reflects the various sources described previously.
Let $\bx_t^{(I)}$ the FG state of data sequence $I$ ($1\le I \le N$) at a {\em future} time $t=(n+m)\Delta t > n \Delta t$ as compared to the data available. The proposed model provides a {\em predictive posterior} for  $\bx_t^{(I)}$ which can be readily determined by successive applications of the sum and product rules as follows:
\be
\begin{array}{ll}
p(\bx_t^{(I)} |\bx_{(0:n)\Delta t}^{(1:N)} ) & = \int  p(\bx_t^{(I)}, \bxx_t^{(I)}, \bt  |\bx_{(0:n)\Delta t}^{(1:N)} ) ~d\bxx_t^{(I)} ~d\bt \\ 
& =  \int  \underbrace{ p(\bx_t^{(I)} | \bxx_t^{(I)}, \bt  , ~\bx_{(0:n)\Delta t}^{(1:N)} )}_{p_{cf}(\bx_t^{(I)} | \bxx_t^{(I)}, \bt_{cf})}  ~p(\bxx_t^{(I)} | \bt , \bx_{(0:n)\Delta t}^{(1:N)} )~ \underbrace{ p(\bt |\bx_{(0:n)\Delta t}^{(1:N)})}_{posterior ~\textrm{\refeq{eq:post}}}  ~d\bxx_t^{(I)} ~d\bt \\  
& = \int p_{cf}(\bx_t^{(I)} | \bxx_t^{(I)}, \bt_{cf}) ~p(\bxx_t^{(I)} | \bt , \bx_{(0:n)\Delta t}^{(1:N)} ) ~p(\bt |\bx_{(0:n)\Delta t}^{(1:N)})~d\bxx_t^{(I)} ~d\bt 
\label{eq:pred1}
\end{array}
\ee
where $p(\bxx_t^{(I)} | \bt , \bxx_{(0:n)\Delta t}^{(1:N)} )$ is the (conditional, given $\bt$) predictive posterior of the (latent) CG state of the data sequence $I$, at the future time $t$. This can be found using the posterior of $\bxx_{n \Delta t}^{(I)}$ i.e. the posterior of the latent CG state at the time up to which data is available ($n\Delta t$) and by propagating the CG probabilistic evolution law. In particular:
\be
\begin{array}{ll}
 p(\bxx_t^{(I)} | \bt , \bx_{(0:n)\Delta t}^{(1:N)} ) & = \int p(\bxx_{(n+m)\Delta t}^{(I)}, ~\bxx_{(n+m-1)\Delta t}^{(I)}, \ldots, \bxx_{n\Delta t}^{(I)}  | \bt , \bx_{(0:n)\Delta t}^{(1:N)} )~d\bxx_{(n+m-1)\Delta t}^{(I)} \ldots ~ d\bxx_{n\Delta t}^{(I)} \\ 
  & = \int  \prod_{k=0}^{m-1} p_c(\bxx_{(n+k+1)\Delta t}^{(I)} |  \bxx_{(n+k)\Delta t}^{(I)}~, ~\bt_c)~ p(\bxx_{n\Delta t}^{(I)} | \bx_{(0:n)\Delta t}^{(I)}~, \bt)~d\bxx_{(n+m-1)\Delta t}^{(I)} \ldots ~ d\bxx_{n\Delta t}^{(I)} 
\end{array}
\label{eq:pred2}
\ee
The combination of Equations (\ref{eq:pred1}) and (\ref{eq:pred2}) suggests the following procedure in order to sample from the predictive posterior of the future FG state $\bx_t^{(I)} $:
\bi
\item[1.] Sample $\bt=(\bt_c, \bt_{cf})$ from the (marginal) posterior in \refeq{eq:post}
\item[2.] Sample $\bxx_{n \Delta t}^{(I)}$ from the (conditional) posterior $ p(\bxx_{n\Delta t}^{(I)} | \bx_{(0:n)\Delta t}^{(I)}, \bt)$.
\item[3.] Propagate/Sample the CG model until $t=(n+m) \Delta t$ by employing the stochastic evolution law implied by $p_c$ for the $\bt_c$ from step 1.
\item[4.] Sample $\bx_t^{(I)}$ from $p_{cf}(\bx_t^{(I)} | \bxx_t^{(I)}, \bt_{cf})$ for the $\bxx_t^{(I)}$ from step 3 and $\bt_{cf}$ from step 1.
\ei
We present in detail how all these steps can be carried out in the next subsections where $p_c$, $p_{cf}$ and the corresponding posteriors are given specific forms.


\subsection{Specification for microscopic random walkers}

In this subsection, we present the specific form of the components of the state-space model discussed previously for the particular microscopic models considered. Special emphasis is placed on the parametrization adopted that enables the identification of the CG model's form, the enforcement of macroscopic constraints (in this case, conservation laws),  as well as the inference engine employed that makes use of Stochastic Variational Inference methods. We conclude with an algorithm that performs short numerical experiments by appropriately initializing the FG model and ingests  the simulation data generated in order to calibrate the CG model.
We aim also at providing insight that is generalizable to other problem types.

\subsubsection{The coarse-to-fine probabilistic map $p_c$}

The FG models considered in this section consist of identical, random walkers in  a bounded subdomain , say $[y_{min}, y_{max}] \in \RR$  where the FG  variables $\bx_t \in \RR^{n_f}$ represent the coordinate of each of the $n_f$ walkers at time $t$. We will consider different types interactions between the walkers as well as various initial conditions, but in all cases it is assumed that the FG model can be simulated, albeit with a significant computational cost and  with a small time-step $\delta t$. Without loss of generality we employ periodic boundary conditions. For such problems, it is clear that the walker-density represents a reasonable CG state variable \cite{li_deciding_2007} and normally one would attempt to write an appropriate evolution law for it. As discussed earlier, our starting point is the specification of the CG state variables through a coarse-to-fine map as described by $p_{cf}$. In particular, if $\rho(y,t)$ denotes the normalized  density (i.e. $\int_{y_{min}}^{y_{max}}  \rho(y,t) ~dy=1, \forall t$) of the walkers which depends on space $y \in \RR$ and time $t$, let $\bs{\rho}_t=\{ \rho_{t,j} \}_{j=1}^{n_c}$ denote the spatially-discretized vector of the density over $[y_{min}, y_{max}]$  with a step size $\Delta y$ such that $\rho_{t,j}$ expresses the relative number of walkers located in $[y_{min}+(j-1)\Delta y, ~y_{min}+j\Delta y]$ at time $t$, such that at all times:
\be
\sum_{j=1}^{n_c}  \rho_{t,j} =1, \quad ( \rho_{t,j} \ge 0)
\label{eq:cons}
\ee
We note that the equation above enforces a constraint in the evolution of the (discretized) density $\bs{\rho}_t$ i.e. expresses the {\em conservation of mass}. Conservation laws constitute the essential building blocks of macroscopic models in continuum thermodynamics \cite{gurtin_mechanics_2013}. In order to achieve this automatically, i.e. without additional constraints in the CG model, we propose the following softmax transformation:
\be
 \rho_{t,j} = \frac{ e^{ X_{t,j}} }{\sum_{l=1}^{n_c} e^{X_{t,l}}}
 \label{eq:softmax}
 \ee
 where $\bxx_t=\{  X_{t,j} \}_{j=1}^{n_c}$ represent the (real-valued) CG state variables employed.
Given $\bxx_t$, the coarse-to-fine probabilistic map takes the following form of a multinomial density:
\be
p_{cf}(\bx_t | \bxx_t)=  \frac{ n_f !}{m_1!~m_2! \ldots m_{n_c}!} \prod_{j=1}^{n_c} \rho_{t,j}^{m_j} 
\label{eq:pcf}
\ee
where $m_j=m_j(\bx_t)$ are the sufficient statistics of $\bx_t$ that count the number of walkers that at time $t$ are located in the $j$ bin i.e. they are in $[y_{min}+(j-1)\Delta y, ~y_{min}+j\Delta y)$. The underlying assumption is that, given the CG state $\bxx_t$ (and consequently $\bs{\rho}_t$  through \refeq{eq:softmax}), the positions of  the walkers, i.e. the FG states, are {\em conditionally} independent. We emphasize here that this does not imply that the walkers move independently of each other which would be untrue as they can exhibit coherent behavior. This dependence though, is caused indirectly, by the  CG states i.e. the same way masses  located on    separate springs might appear to  vibrate independently even when the springs are attached to the same vibrating  substrate. 

Two remarks are in order at this stage. Firstly, by introducing CG variables that only pertain to the zero-order moment of the particles' distribution (i.e. the density), we have assumed that higher-order moments (e.g. second- (pair) or third-order joint densities or  correlations) are not necessary in predicting the FG states' evolution. The validity of this assumption (in combination with the $p_{c}$ presented in the next subsection) will be assessed given the FG simulation data. If the CG variables adopted are insufficient to predict the future FG configurations, then this should manifest itself in large predictive uncertainty (e.g. \refeq{eq:pred1}). Methods that focus exclusively on finding closures for the CG variables selected (as discussed in the introduction), even if  they are successful in this task, cannot assess {\em the actual predictive value} of these CG variables and associated models in terms of the  original FG system. 

The second observation is that in this particular model, there are no unknown parameters $\bt_{cf}$ that need to be learned from the data. 
Given the {\em exchangeability}  \cite{aldous_exchangeability_1985} of the FG state variables $\bs{x}_t$ arising from the fact that the walkers are identical (i.e. their joint density is invariant to permutations), and according to de Finetti's theorem \cite{jordan_bayesian_2010} $p_{cf}$ can be generally expressed as:
\be
p_{cf}(\bx_t | \bxx_t) = \prod_{j=1}^{n_f} p(x_{t,j} | \bxx_t, \bt_{cf}).
\ee
For non-exchangeable FG variables, any generative, latent variable model \cite{bishop_latent_1999} would be possible. E.g.:
\be
p_{cf}(\bx_t | \bxx_t) = \mathcal{N}(\bx_t ; ~\bs{\mu}+\bs{W} \bxx_t, \bs{S})
\ee
where $\bt_{cf} \equiv\{\bs{\mu}, \bs{W}, \bs{S}\}$ and implies a linear projection on the subspace spanned by the columns of $\bs{W}$ \cite{tipping_probabilistic_1999}.

\subsubsection{The CG probabilistic evolution law $p_c$}

The second component of the proposed model involves the  specification of a CG evolution law. In traditional formulations this would be in the form of (stochastic) PDE for $\rho(y,t)$ such as:
\be
\frac{ \pa \rho}{\pa t}= h(t,\rho, \frac{\pa \rho}{\pa y}, \frac{\pa^2 \rho}{\pa y^2}, \ldots)
\label{eq:rhocont}
\ee
This unavoidably raises several questions with regards to the functional form of the right-hand side $h(.)$, the order of the spatial derivatives of $\rho$ that should appear as well the appropriate values of any parameters  in $h$ \cite{li_deciding_2007}. The difficulties are only amplified if stochastic terms need   to be included. Even if an appropriate CG model was to be identified (as e.g. in simplified cases where closures can be semi-analytically computed directly from the FG model) and was subsequently discretized and integrated forward in time, it would be impossible to retrieve the FG states unless additional assumptions were made. Furthermore, while the discretization errors in the solution of \refeq{eq:rhocont} are well-understood, it is unknown how these errors would affect the quality of the predictions for the FG states.
Since the discretized model is going to be used for prediction purposes, we  also use this in the calibration phase and adopt a probabilistic Markovian model of the hidden CG states $\bs{X}_t$ of the following form:
\be
\begin{array}{ll}
 p_{c}(\bxx_{(k+1)\Delta t} | ~\bxx_{k\Delta t}, ~\bt_c, ~\bs{v})  & = \prod_{j=1}^{n_c} \mathcal{N}(X_{(k+1)\Delta t, ~j}~;~\mu_j(\bxx_{k\Delta t}, \bt_c), ~v_j^{-1})
\end{array}
\label{eq:pc}
\ee
The ``deterministic'' part in the evolution equations is modeled by the functions $\mu_j$ whereas the stochastic fluctuations by the $v_j^{-1}$'s. Naturally correlated Gaussian or non-Gaussian models can be employed. We propose the following form for the $\mu_j$:
\be
\mu_j(\bxx_{t}, \bt_c)= \bs{\bt}_c^T \bs{\phi}^{(j)} (\bxx_t)= \sum_{l=1}^L \theta_{c,l} ~\phi^{(j)}_l (\bxx_t)
\label{eq:pcmean}
\ee
Each of the feature functions account for the dependence with the CG states at the previous time step, but one can readily introduce longer memory. Either way, it is obviously impossible to know a priori which $\bs{\phi}$'s  are relevant in the evolution of the CG states, what types of interactions  are essential  (e.g. first, second-order etc) or how they depend on the coarse-to-fine probabilistic map $p_{cf}$. This underpins an important {\em model selection}
 issue that has been of concern in several coarse-graining studies \cite{rudzinski_coarse-graining_2011,noid_perspective:_2013}.
One strategy for addressing this, is to initiate the search with a small number of features $\bs{\phi}$ and progressively add more. These can be selected from a pool of candidates by employing appropriate criteria. In \cite{bilionis_free_2012} for example, the feature function that causes the largest (expected) decrease (or increase) in the KL-divergence (or the log-likelihood) that we seek to minimize  (or maximize), is added at each step. In this work, we adopt a different  approach whereby we consider simultaneously   a large number of $\bs{\phi}$.   This in turn implies a  vector of unknown  model parameters $\bt_c$  of very large dimension which not only impedes computations but can potentially lead to multiple local maxima and overfitting particularly in the the presence of limited data.
More importantly perhaps, it can obstruct the identification of the most salient features of the CG model which provide valuable physical insight \cite{schoeberl_predictive_2017}.

To address this issue, we propose the use of sparsity-inducing priors that are capable of identifying solutions in which only a (small) subset of $\bt_c$ are non-zero and therefore only the corresponding $\bs{\phi}$'s are active \cite{figueiredo_adaptive_2003}. A lot of
the prior models that have been proposed along these lines can be readily
cast in the context of hierarchical Bayesian models where hyper-parameters
are introduced in the prior. In this work, we adopt the Automatic Relevance
Determination (ARD, \cite{mackay_automatic_1994}) model which consists of the following:
\be
p( \bt_c  | \bs{\tau} ) = \prod_{l=1}^{L} \mathcal{N}(\theta_{c,l} ; ~0, \tau_l^{-1})
\label{eq:ard1}
\ee
This implies that each $\theta_{c,l}$ is a priori independent, and following a zero-mean, Gaussian with a precision hyper-parameter $\tau_l$. The latter are modeled (independently) with a (conjugate) Gamma i.e.:
\be
p(\tau_l) = Gamma(\tau_l ; ~\alpha_0, \beta_0)
\label{eq:ard2}
\ee
 We note that when $\tau_l \rightarrow \infty$, then $\theta_{c,l} \rightarrow 0$. The resulting prior for $\theta_{{c},l}$ arising by marginalizing the hyper-parameter is a heavy-tailed,  Student's $t-$distribution \cite{tipping_relevance_2000}. 
The hyperparameters $\alpha_0, \beta_0$ are effectively the only ones that need to be provided by the analyst. We advocate very small values ($\alpha_0=\beta_0=10^{-10}$ was used in the numerical examples) which correspond to a non-informative prior.
We also a employed independent Gamma priors for the previsions $v_j$ in \refeq{eq:pc} as follows:
\be 
p(v_j)=Gamma(v_j; ~\gamma_0, \zeta_0)
\label{eq:vprior}
\ee
with hyperparameters $\gamma_0, \zeta_0$ which were also set to very small values ($\gamma_0=\zeta_0=10^{-10}$ was used in the numerical examples).

\subsection{Generation of training data}
\label{sec:train}
Given the aforementioned model, we propose the following procedure for generating training data. It consists of  $N$ short, numerical experiments in which the FG model is randomly initialized and propagated for one coarse time-step $\Delta t$ i.e. for $\frac{\Delta t}{\delta t}$ microscopic time-steps. In particular:
\bi
\item For $i=1,\dots,N$:
\bi
\item Sample CG initial state $\bxx_0^{(i)}$ from a density $p_{c,0}(\bxx_0^{(i)})$.
\item Sample FG initial state $\bx_0^{(i)}$ frm $p_{cf}(\bx_0^{(i)} | \bxx_0^{(i)})$.
\item Solve (discretized) FG model in \refeq{eq:fg} for  $m=\frac{\Delta t}{\delta t}$ microscopic time-steps and record final state $\bx_{\Delta t}^{(i)}$
\ei
\ei
The generated data $\{ \bx_{\Delta t}^{(i)} \}_{i=1}^N$ will be used subsequently to draw inferences on the CG model states and parameters (section \ref{sec:inference}).
We note that longer time sequences could readily be generated (albeit at an increased cost). The number of samples $N$ is also something that can be selected adaptively since as we show in the next sections, inferences and predictions can be updated as soon as more data become available. Furthermore, the predictive estimates produced reflect the informational content of the data.  Finally, we note with regards to the density $p_{c,0}(\bxx_0^{(i)})$ from which initial CG states are drawn, that this can be selected  quite flexibly. One can also envisage considering parametrized versions of $p_{c,0}$ where the optimal parameter values can be determined by maximizing information gain criteria in the context of optimal experimental design.

\subsection{Inference}
\label{sec:inference}
In the section we discuss the Bayesian calibration of the proposed model given the synthetic training $\bx_{\Delta t}^{(1:N)}$ data generated previously.
Apart from the parameters $\bt_c$ (and the corresponding hyperparameters $\bs{\tau}$) and $\bs{v}=\{v_j\}$ that control the CG  model form, the latent CG states $\bxx_{\Delta t}^{(1:N)}$ must  be  also inferred. As mentioned in section \ref{sec:train}, the initial CG states $\bxx_0^{(1:N)}$ are known and are omitted  in the ensuing conditional densities  for simplicity.
The joint posterior is given by:
\be
\begin{array}{ll}
 p(\bxx_{\Delta t}^{(1:N)}, \bt_c, \bs{\tau}, \bs{v}| \bx_{\Delta t}^{(1:N)}) & = \frac{ p(\bx_{\Delta t}^{(1:N)} | \bxx_{\Delta t}^{(1:N)}, \bt_c, \bs{\tau}, \bs{v})~p(\bxx_{\Delta t}^{(1:N)}, \bt_c, \bs{\tau}, \bs{v})}{p(\bx_{\Delta t}^{(1:N)})} \\
 & = \frac{ p(\bx_{\Delta t}^{(1:N)} | \bxx_{\Delta t}^{(1:N)})~p(\bxx_{\Delta t}^{(1:N)}, \bt_c, \bs{\tau}, \bs{v})}{p(\bx_{\Delta t}^{(1:N)})}
\end{array}
\label{eq:post}
\ee
With regards to the terms, we note that the likelihood $p(\bx_{\Delta t}^{(1:N)} | \bxx_{\Delta t}^{(1:N)})$ can be expressed as:
\be
\begin{array}{ll}
 p(\bx_{\Delta t}^{(1:N)} | \bxx_{\Delta t}^{(1:N)}) & = \prod_{i=1}^N p(\bx_{\Delta t}^{(i)} | \bxx_{\Delta t}^{(i)}) \\
 & = \prod_{i=1}^N p_{cf}(\bx_{\Delta t}^{(i)} | \bxx_{\Delta t}^{(i)})
\end{array}
\label{eq:like}
\ee
where each of the terms in the product are given by \refeq{eq:pcf}. With regards to the prior $p(\bxx_{\Delta t}^{(1:N)}, \bt_c, \bs{\tau}, \bs{v})$ it decomposes as follows:
\be
p(\bxx_{\Delta t}^{(1:N)}, \bt_c, \bs{\tau},\bs{v})= p(\bxx_{\Delta t}^{(1:N)} | ~\bt_c, \bs{v})~p(\bt_c |~ \bs{\tau})~p(\bs{\tau})~p(\bs{v})
\label{eq:prior1}
\ee
where the last three terms are specified in Equations (\ref{eq:ard1}), \ref{eq:ard2}) and (\ref{eq:vprior}) respectively. Furthermore, the first term depends on the CG evolution law in \refeq{eq:pc} as:
\be
\begin{array}{ll}
 p(\bxx_{\Delta t}^{(1:N)} | ~\bt_c, \bs{v}) & = \prod_{i=1}^N  p(\bxx_{\Delta t}^{(i)} | ~\bt_c, \bs{v}) \\
 & = \prod_{i=1}^N  p_c(\bxx_{\Delta t}^{(i)} | \bxx_0^{(i)},~\bt_c, \bs{v}) \\
\end{array}
\label{eq:prior2}
\ee

As exact inference is impossible and Monte Carlo-based schemes cumbersome, we attempt to find an approximation $q( \bxx_{\Delta t}^{(1:N)}, \bt_c, \bs{\tau}, \bs{v})$  to the posterior $p( \bxx_{\Delta t}^{(1:N)}, \bt_c, \bs{\tau} |~\bx_{\Delta t}^{(1:N)})$. To that end, we adopt a Variational Inference scheme \cite{peierls_minimum_1938,wainwright_graphical_2008} which attempts to find the best approximation within an appropriately selected family of densities such that it minimizes the Kullback-Leibler divergence with the posterior i.e.:
\be
\begin{array}{l}
  KL(q( \bxx_{\Delta t}^{(1:N)}, \bt_c, \bs{\tau}, \bs{v})~~ ||~~ p( \bxx_{\Delta t}^{(1:N)}, \bt_c, \bs{\tau} |~\bx_{\Delta t}^{(1:N)})) \\
  = -\int q( \bxx_{\Delta t}^{(1:N)}, \bt_c, \bs{\tau}, \bs{v})  \log \frac{p( \bxx_{\Delta t}^{(1:N)}, \bt_c, \bs{\tau}, \bs{v} |~\bx_{\Delta t}^{(1:N)})}{q( \bxx_{\Delta t}^{(1:N)}, \bt_c, \bs{\tau}, \bs{v}) } ~d\bxx_{\Delta t}^{(1:N)}~d\bt_c~d\bs{\tau}~d\bs{v} \\
 = -\int q( \bxx_{\Delta t}^{(1:N)}, \bt_c, \bs{\tau}, \bs{v})  \log p(\bx_{\Delta t}^{(1:N)} | \bxx_{\Delta t}^{(1:N)}) ~d\bxx_{\Delta t}^{(1:N)}~d\bt_c~d\bs{\tau}~ d\bs{v} \\
 -\int q( \bxx_{\Delta t}^{(1:N)}, \bt_c, \bs{\tau}, \bs{v})  \log \frac{p( \bxx_{\Delta t}^{(1:N)}, \bt_c, \bs{\tau}, \bs{v}) }{q( \bxx_{\Delta t}^{(1:N)}, \bt_c, \bs{\tau}, \bs{v})} ~d\bxx_{\Delta t}^{(1:N)}~d\bt_c~d\bs{\tau}~d\bs{v} \\
  + \log p(\bx_{\Delta t}^{(1:N)})
\end{array}
\ee
Given that the KL-divergence is always non-negative we obtain that:
\be
\begin{array}{ll}
 \log p(\bx_{\Delta t}^{(1:N)})  \ge &  \int q( \bxx_{\Delta t}^{(1:N)}, \bt_c, \bs{\tau}, \bs{v})  \log p(\bx_{\Delta t}^{(1:N)} | \bxx_{\Delta t}^{(1:N)} ) ~d\bxx_{\Delta t}^{(1:N)}~d\bt_c~d\bs{\tau}~d\bs{v} \\
 & \int q( \bxx_{\Delta t}^{(1:N)}, \bt_c, \bs{\tau}, \bs{v})  \log \frac{p( \bxx_{\Delta t}^{(1:N)}, \bt_c, \bs{\tau}, \bs{v}) }{q( \bxx_{\Delta t}^{(1:N)}, \bt_c, \bs{\tau}, \bs{v})} ~d\bxx_{\Delta t}^{(1:N)}~d\bt_c~d\bs{\tau}~d\bs{v} \\
 & = \mathcal{F}\left( q( \bxx_{\Delta t}^{(1:N)}, \bt_c, \bs{\tau}, \bs{v})\right)  
\end{array}
\label{eq:elbo}
\ee
Hence, finding the $q$ that maximizes the log-evidence lower-bound $\mathcal{F}$   (frequently referred to as ELBO, for Evidence Lower BOund) is equivalent to minimizing the $KL$-divergence, for which the optimal $q$ being the exact posterior.
We adopt a mean-field factorization \cite{parisi_statistical_1988} of the form:
\be
q( \bxx_{\Delta t}^{(1:N)}, \bt_c, \bs{\tau}, \bs{v}) = \prod_{i=1}^N q( \bxx_{\Delta t}^{(i)})~~ q( \bt_c)~q(\bs{\tau})~ q(\bs{v}) 
\label{eq:mf}
\ee
This is turn gives rises to an iterative scheme where each of the densities above are updated, until convergence, while the others are kept fixed.
It can be readily shown  \cite{bishop_pattern_2007} that in this case (based on Equations (\ref{eq:like}), (\ref{eq:prior1}), (\ref{eq:prior2})), the optimal $q^{opt}$ i.e. the ones that maximize the ELBO $\mathcal{F}$ are given by (up to a constant):
\be
\begin{array}{ll}
 \log q^{opt}(\bt_c)  & = < \log  p(\bxx_{\Delta t}^{(1:N)} |  \bt_c, \bs{v}) >_{q(\bxx_{\Delta t}^{(1:N)})~q(\bs{v})}+<\log p(\bt_c | \bs{\tau}>_{q(\bs{\tau})} \\
  & = \sum_{i=1}^N < \log  p_c(\bxx_{\Delta t}^{(i)} | \bxx_{0}^{(i)}, \bt_c, \bs{v}) >_{q(\bxx_{\Delta t}^{(i)})~q(\bs{v})}+<\log p(\bt_c | \bs{\tau})>_{q(\bs{\tau})},
\end{array}
\label{eq:qopttheta}
\ee
\be
\begin{array}{ll}
  \log q^{opt}(\bs{\tau}) = <\log p(\bt_c | \bs{\tau}>_{q(\bt)}+\log p(\bs{\tau}),
\end{array}
\label{eq:qopttau}
\ee
\be
\begin{array}{ll}
  \log q^{opt}(\bs{v}) & = < \log  p(\bxx_{\Delta t}^{(1:N)} |  \bt_c, \bs{v}) >_{q(\bxx_{\Delta t}^{(1:N)})~q(\bt_c)}+\log p(\bs{v})\\
  & = \sum_{i=1}^N < \log  p_c(\bxx_{\Delta t}^{(i)} | \bxx_{0}^{(i)}, \bt_c,\bs{v}) >_{q(\bxx_{\Delta t}^{(i)})~q(\bt_c)}+\log p(\bs{v})
\end{array}
\label{eq:qoptv}
\ee
and:
\be
\begin{array}{ll}
 \log q^{opt}(\bxx_{\Delta t}^{(i)}) & =\log  p(\bx_{\Delta t}^{(i} | \bxx_{\Delta t}^{(i)}) + < \log  p(\bxx_{\Delta t}^{(i)} | \bt_c) >_{q(\bt_c)}  \\
 & = \log  p_{cf}(\bx_{\Delta t}^{(i)} | \bxx_{\Delta t}^{(i)}) +< \log p_c(\bxx_{\Delta t}^{(i)} | \bxx_{0}^{(i)}, \bt_c) >_{q(\bt_c)} 
\end{array}
\label{eq:qoptX}
\ee
where the notation $<.>_q$ implies an expectation with respect to the $q$ density in the subscript.
The equations above are obviously coupled, but the first two attain a closed-form solution. In particular, one finds that $q^{opt}(\bt_c)=\mathcal{N}(\bt_c; ~\bs{\mu}_{\bt}, \bs{S}_{\bt})$ where:
\be
\begin{array}{ll}
 \bs{S}_{\bt}^{-1} &  =  \sum_{j=1}^{n_c} <v_j>_{q(\bs{v})} \sum_{i=1}^N  \bs{\phi}^{(j)}(\bxx_{0}^{(i)}) (\bs{\phi}^{(j)}(\bxx_{0}^{(i)}))^T +<\bs{T}>_{q(\bs{\tau})} \\
 \bs{S}_{\bt}^{-1} \bs{\mu}_{\bt}  & = \sum_{j=1}^{n_c}<v_j>_{q(\bs{v})}  \sum_{i=1}^N  \bs{\phi}^{(j)}(\bxx_{0}^{(i)}) < X_{\Delta t, ~j} >_{ q(\bxx_{\Delta t}^{(i)} ) }
\end{array}
\label{eq:qopttheta1}
\ee
and $\bs{T}=diag(\tau_l)$. Furthermore for the $\bs{\tau}$'s, one finds that 
 $q^{opt}(\bs{\tau})= \prod_l Gamma(\tau_l; ~\alpha_l, \beta_l)$ where:
 \be
 \begin{array}{ll}
   \alpha_l & =\alpha_0+\frac{1}{2} \\
   \beta_l & =\beta_0+\frac{1}{2} <\theta_{c,j}^2>_{q(\bt_c)}=\beta_0+\frac{1}{2} (\mu_{\bt,l}^2+\bs{S}_{\bt,ll})
 \end{array}
\label{eq:qopttau1}
\ee
Note that in this case $<\tau_l>_{q(\bs{\tau})}=\frac{\alpha_l}{\beta_l}$. Similarly, one notes that  $q^{opt}(\bs{v})= \prod_j q_j(v_j)=\prod_j Gamma(v_j; ~\gamma_j, \zeta_j)$ where:
 \be
 \begin{array}{ll}
   \gamma_j & =\gamma_0+\frac{N}{2} \\
   \zeta_j & =\zeta_0+\frac{1}{2} <\sum_{i=1}^N (X_{\Delta t, ~j} -\bt_c^T \bs{\phi}^{(j)}(\bxx_0^{(i)}) )^2>_{ q(\bxx_{\Delta t}^{(i)})~q(\bt_c)}
\label{eq:qoptv1}
 \end{array}
\ee
Note that in this case $<v_j>_{q(\bs{v})}=\frac{\gamma_j}{\zeta_j}$. 

The most challenging component  of the update equations  pertains to $q^{opt}(\bxx_{\Delta t}^{(i)})$ due to the form of the likelihood (\refeq{eq:pcf}) as well as the softmax transformation in \refeq{eq:softmax} that enforces the conservation of mass. Consequently, it is impossible to derive in closed form 
 $q^{opt}$. To address this we make use of Stochastic Variational Inference methods \cite{paisley_variational_2012,hoffman_stochastic_2013}. The underlying idea is to approximate the expectations involved with Monte Carlo. 
In particular, if one denotes with $\mathcal{F}_i$ the terms in $\mathcal{F}$ pertaining to $q(\bxx_{\Delta t}^{(i)})$ from \refeq{eq:elbo}, then:
\be
\begin{array}{ll}
 \mathcal{F}_i & = <\log p_{cf}(\bx_{\Delta t}^{(i)} | \bxx_{\Delta t}^{(i)}) >_{q(\bxx_{\Delta t}^{(i)})}
 +  <\log p_c(\bxx_{\Delta t}^{(i)} | \bxx_{0}^{(i)},\bt_c,\bs{v}) >_{q(\bxx_{\Delta t}^{(i)})q(\bt_c) q(\bs{v})} - <\log q(\bxx_{\Delta t}^{(i)})>_{q(\bxx_{\Delta t}^{(i)})}
\end{array}
\label{eq:fvari}
\ee
In our formulation, we assume a multivariate Gaussian (with a full-rank covariance matrix) for each $q(\bxx_{\Delta t}^{(i)})$ i.e.:
\be
q(\bxx_{\Delta t}^{(i)})= \mathcal{N}(\bxx_{\Delta t}^{(i)}; \bs{\mu}_i, \bs{S}_i)
\label{eq:qxi}
\ee

 Subsequently, we approximate the derivatives of $\mathcal{F}_i$ with respect to $\bs{\mu}_i$ and $\bs{S}_i$ using Monte Carlo, and update these parameters using stochastic gradient ascent \cite{robbins_stochastic_1951}. In particular we made use of the {\em ADAM} algorithm \cite{kingma_adam:_2015}, which is one of the most  robust, first-order stochastic optimization techniques.  A critical role in the lower-variance  estimation of the derivatives, was played by the reparametrization trick \cite{kingma_auto-encoding_2014}. 
Details of this step are contained in Appendix B.
 
 The steps involved in the proposed Variational Inference scheme are summarized in Algorithm \ref{alg:alg}. 
 We also note that, as one would guess due to the resemblance with Expecation-Maximization schemes \cite{dempster_maximum_1977}, various relaxations are possible  \cite{neal_view_1998}. For example it is not necessary that the optimal  $q(\bxx_{\Delta t}^{(i)})$  (given $q(\bt_c)$ and $q(\bs{\tau})$) is found at each stage and it suffices  to do one or more steps along the (stochastic) gradients of $\mathcal{F}_i$. Furthermore, it is not necessary that one updates the $q(\bxx_{\Delta t}^{(i)})$  for all data points $i$. It suffices to do so for one or a subset of the data. Finally, it is not necessary that all the data are introduced simultaneously, but it suffices to incorporate them in batches, updating $q(\bt_c)$ and $q(\bs{\tau})$ each time. Such a scheme provides also a natural tempering effect for the posterior \cite{del_moral_sequential_2006}.  In the case of very large large data sets, which as shown in section \ref{sec:num} is not necessary here, further stochastic approximations are possible \cite{titsias_doubly_2014}.
 
 As a final remark with regards to the Variational Inference perspective adopted, we note that (given the form of the approximate posteriors $q$) one can approximate (up to Monte Carlo noise) the ELBO $\mathcal{F}$ in \refeq{eq:elbo}. This is useful in terms of monitoring convergence during the iterative updates in Algorithm \ref{alg:alg} but more importantly perhaps,  it allows to approximate the log model evidence $\log p(\bx_{\Delta t}^{(1:N)})$. The latter is an essential quantitative indicator for scoring competing models (e.g. with different parametrizations) and could serve as the basic objective  for model enrichment. We intend to investigate this in future work. We include here the final expression  of the ELBO $\mathcal{F}$ based on \refeq{eq:elbo} (up to a constant) when using the optimal $q's$ presented earlier. A detailed derivation is contained in Appendix A.
 \be
 \begin{array}{ll}
  \mathcal{F} & = <  \sum_{i=1}^N \log p_{cf}(\bx_{\Delta t}^{(i)} | \bxx_{\Delta t}^{(i)} )>_{q(\bxx_{\Delta t}^{(i)})}> +\frac{1}{2} \sum_{i=1}^N \log | \bs{S}_i|      +\frac{1}{2} \log | \bs{S}_{\bt}| \\
  & - \sum_k \alpha_k \log \beta_k -\sum_j \gamma_j \log \zeta_j
\end{array}
\label{eq:elbofinal}
\ee

 
\subsection{Probabilistic Prediction}
\label{sec:probpred}
We  devote a few lines to specialize the general formulas given in section \ref{sec:gen} in terms of the predictive ability of the proposed model. Consider first one of the data points $x_{\Delta t}^{(I)}$  generated in \ref{sec:train}, whose future evolution  we wish to predict (obviously without employing the FG model). Let $q(\bxx^{(I)}_{\Delta t})$ be the (approximate) posterior of the corresponding hidden CG state and $q(\bt_c)$ the (approximate) posterior for $\bt_c$ (that depends on all training data $\bx_{\Delta t}^{(1:N)}$ as seen in \refeq{eq:qopttheta}).
Then our model provides a {\em predictive} posterior density for $x_t^{(I)}$ at any future time $t= n \Delta t$ ($n>1$). In particular:
\be
\begin{array}{ll}
p(\bx_{n \Delta t}^{(I)} | \bx_{\Delta t}^{(1:N)}) & = \int p_{cf} (\bx_{n \Delta t}^{(I)} | \bxx_{n \Delta t}^{(I)}) \left(\prod_{k=1}^{n-1} p_c( \bxx_{(k+1) \Delta t}^{(I)} |\bxx_{k \Delta t}^{(I)}, \bt_c) \right) ~q(\bxx^{(I)}_{\Delta t}) ~q(\bt_c) ~d\bxx_{n \Delta t}^{(I)}\ldots d\bxx_{ \Delta t}^{(I)}~d\bt_c
\end{array}
\ee
where $p_{cf}$ and $p_c$ are provided by Equations (\ref{eq:pcf}) and (\ref{eq:pc}) respectively.

The second case we  finally consider is that if  {\em new} (not included in the training data)  FG state $\bar{\bx}_0$ whose future evolution we would like to predict using  the CG model trained on the data $\bx_{\Delta t}^{(1:N)}$  discussed previously. We do not discuss a longer FG  sequence (e.g. $\bar{\bx}_0$ and $\bar{\bx}_{\Delta t}$) as this would be incorporated in the training data as above\footnote{The only fundamental difference in this case is that the CG initial state $\bar{\bxx}_0$  is unknown and would need to be inferred in addition to $\bar{\bxx}_{\Delta t}$. This joint inference with the goal of finding a joint approximate posterior $q(\bar{\bxx}_0, \bar{\bxx}_{\Delta t})$, can be carried out using the aforementioned framework e.g. by assuming a joint Gaussian and optimizing the joint mean and covariance.}. 
The first step in this case would be to find the posterior $p(\bar{\bxx}_0 | \bar{\bx}_0)$ which is proportional to:
\be
p(\bar{\bxx}_0 | \bar{\bx}_0) \propto p_{cf}(\bar{\bx}_0 | \bar{\bxx}_0)~p_{c,0}(\bar{\bxx}_0)
\ee
where $p_{c,0}$ is a prior that can be taken to be e.g. a vague Gaussian (in the case of $n_f>>n_c$ its effect is minimal). Given this (it can also be approximated using SVI as before), the {\em predictive} posterior for the evolution at any time $t=n \Delta t$ is given by:
\be
\begin{array}{ll}
p(\bar{\bx}_{n \Delta t} | \bar{\bx}_0, ~ \bx_{\Delta t}^{(1:N)}) & = \int p_{cf} (\bar{\bx}_{n \Delta t} | \bar{\bxx}_{n \Delta t}) \left(\prod_{k=0}^{n-1} p_c( \bar{\bxx}_{(k+1) \Delta t} | \bar{\bxx}_{k \Delta t}, \bt_c) \right) ~p(\bar{\bxx}_{0}| \bar{\bx}_0 ) ~q(\bt_c) ~d\bar{\bxx}_{n \Delta t}\ldots d\bar{\bxx}_{0}~d\bt_c
\end{array}
\ee



%
%
%

%

\begin{algorithm}[H]
\SetAlgoLined
\SetKwInOut{Output}{Output}
\KwData{ $\{ \bx_{\Delta t}^{(i)}\}_{i=1}^N$ }
\Output{$\{\bs{\mu}_i, \bs{S}_i\}_{i=1}^N$,  $\bs{\mu}_{\bt}, \bs{S}_{\bt}$,  $\{\alpha_l,\beta_l\}_{l=1}^{L}$, $\{ \gamma_j, \zeta_j\}_{j=1}^{n_c}$} \
 Initialize $\bs{\mu}_{\bt}, \bs{S}_{\bt}$, $\{\alpha_l,\beta_l\}_{l=1}^{L}$,$\{ \gamma_j, \zeta_j\}_{j=1}^{n_c}$\;
 \While{$\{\bs{\mu}_i, \bs{S}_i\}_{i=1}^N$,  $\bs{\mu}_{\bt}, \bs{S}_{\bt}$,  $\{\alpha_l,\beta_l\}_{l=1}^{L}$, $\{ \gamma_j, \zeta_j\}_{j=1}^{n_c}$ not converged}{
 \For{$i\leftarrow 1$ \KwTo $N$}{\label{forins}
 Update $\bs{\mu}_i, \bs{S}_i$ by maximizing $\mathcal{F}_i$ (\refeq{eq:fvari}) using ADAM (Appendix B).
 }
 Update $\bs{\mu}_{\bt}, \bs{S}_{\bt}$ based on \refeq{eq:qopttheta1}\;
 Update $\{\alpha_l,\beta_l\}_{l=1}^{L}$ based on \refeq{eq:qopttau1}\;
Update $\{ \gamma_j, \zeta_j\}_{j=1}^{n_c}$ based on \refeq{eq:qoptv1}\;
Estimate $\mathcal{F}$ based on \refeq{eq:elbofinal} (Appendix A) \;
 }
 \caption{Proposed EM-algorithm for hidden state and model parameter inference 
using Stochastic Variational Bayes \cite{kingma2013auto}, Automatic Relevance 
Determination, and stochastic optimization (ADAM \cite{kingma_adam:_2015}).}
 \label{alg:alg}
\end{algorithm}

\section{Numerical Illustrations}
\label{sec:num}

%
%
%

In this section, we  present three numerical illustrations demonstrating various aspects and capabilities of the proposed framework.
Before embarking in the presentation of the results, we provide details of the feature functions $\bs{\phi}$ (\refeq{eq:pcmean}) employed which are common in all three cases.
These provide the vocabulary with which the CG evolution is expressed. In particular we consider feature functions implying first-order interactions:
\be
\phi^{(j)}_{m} (\bxx)= X_{j+l}, \quad m=-M,\ldots,0,\ldots M
\label{eq:phi1order}
\ee
as  well as second order i.e.:
\be
\phi^{(j)}_{m_1,m_2} (\bxx)= X_{j+m_1} X_{j+m_2}, \quad m_1,m_2=-M,\ldots,0,\ldots M
\label{eq:phi2order}
\ee
Higher-order feature or functions of different form can also be employed but in the examples considered were never activated and are therefore omitted from this discussion.
Given the discretized nature of the CG evolution law, we note that the interaction range implied by $M$ effectively   defines
the highest order of spatial derivatives which can be represented with finite differences.
We also employed the same variances $v_j=v, \forall j=1,\ldots, n_c$ for all CG state variables (\refeq{eq:pc}) and report this as $v$ in the following.

\subsection{Synthetic example}
\label{sec:ddm}

The primary goal of this example is to assess the accuracy of the inference engine employed, the effect of the number of training data, the capability of the ARD prior in identifying sparse solutions and to  provide insight to the proposed algorithm.
To that end, we assumed values of the parameters of the CG evolution law (i.e. $\bt_c, \bs{v}$ in \refeq{eq:pc}) and generated FG data as described in section \ref{sec:train}. The data produced (for $dim(\bxx)=n_c=24$ and number of walkers $n_f=4800$) were subsequently used in order to assess whether the ground truth could be recovered.

In particular, we considered $M=2$ (Equations (\ref{eq:phi1order}), (\ref{eq:phi2order})) and  set all $\bt_c$ coefficients to zero, except for:
\bi
\item  $m=-1$ and $m=+1$ in \refeq{eq:phi1order} for which the value $0.5$ was assigned
\item    $(m_1=1,m_2=1)$ in  \refeq{eq:phi2order} for which the value $-0.23$ was assigned and for $(m_1=-1,m_2=-1)$ the value $0.21$
\ei
Hence, 26 out of the 30 features available were turned off. We also assumed a  deterministic evolution law i.e. set ${v}^{-1}=v_j^{-1}=0$.

\begin{figure}[!t]
\begin{subfigure}[b]{0.3\textwidth}
 \includegraphics[width=\textwidth]{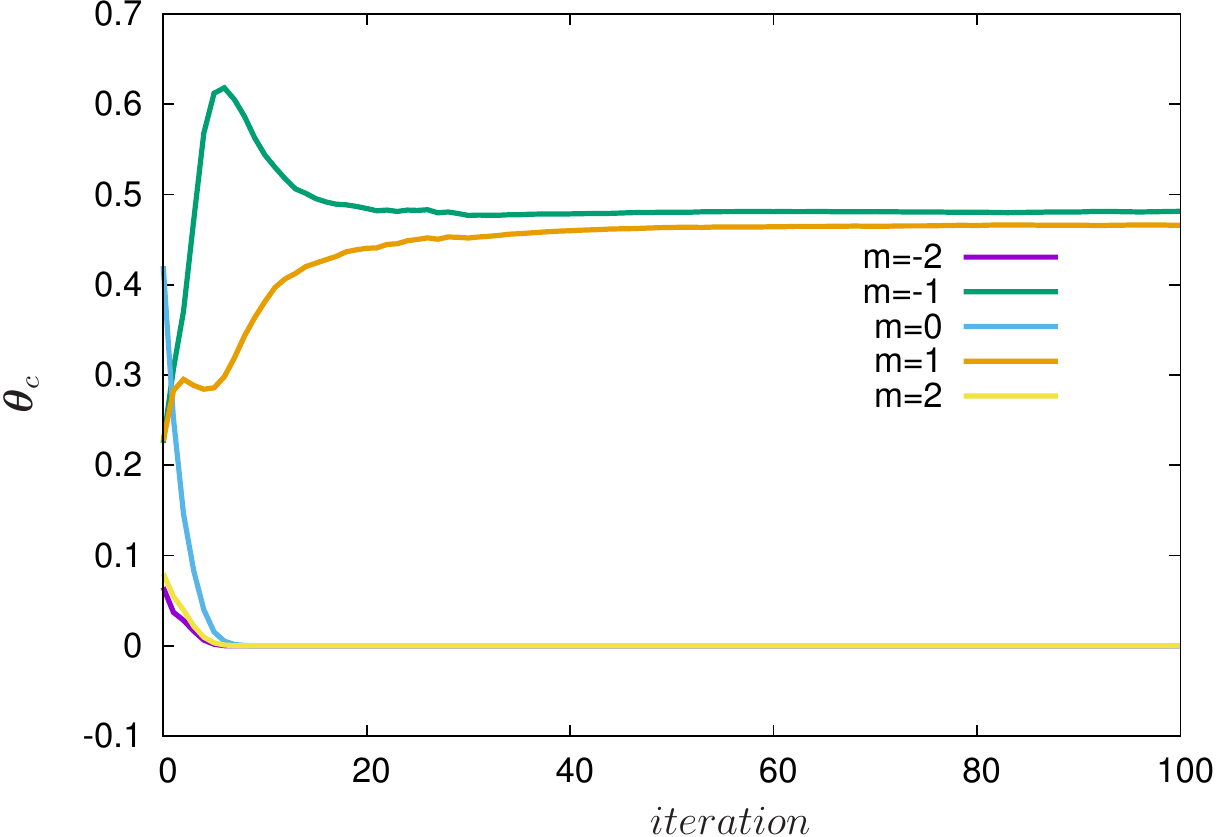}
 \caption{$N=64$}
\end{subfigure}
\begin{subfigure}[b]{0.3\textwidth}
 \includegraphics[width=\textwidth]{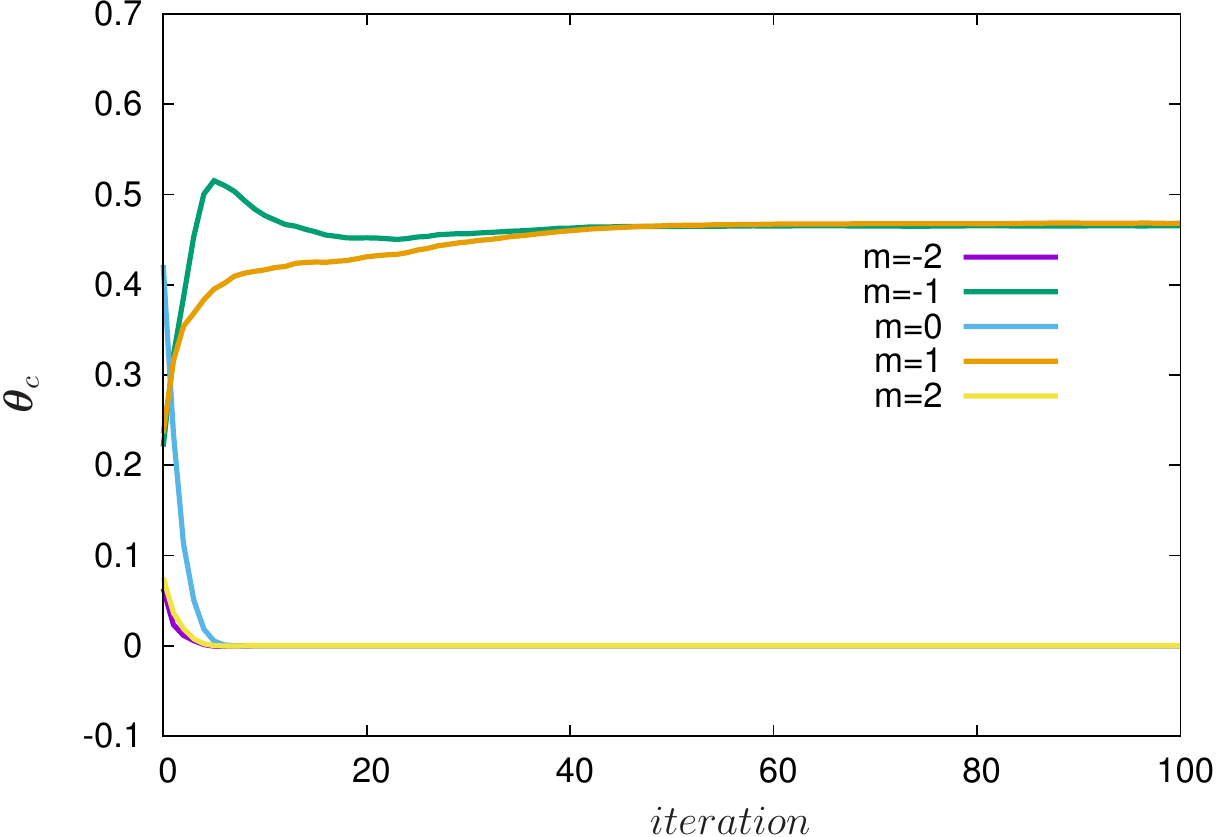}
 \caption{$N=128$}
\end{subfigure}
\begin{subfigure}[b]{0.3\textwidth}
 \includegraphics[width=\textwidth]{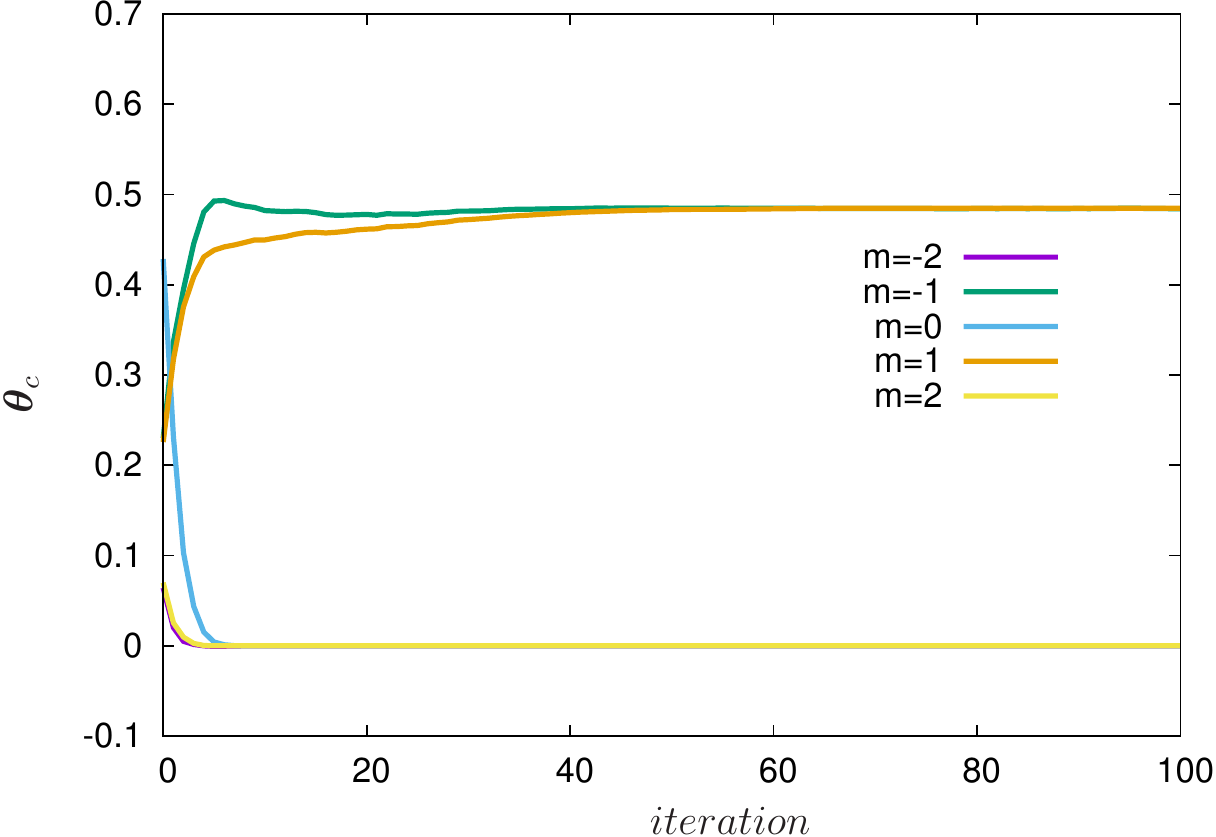}
 \caption{$N=256$}
\end{subfigure}
 \caption{Posterior mean of $\bt_c$ associated with first-order feature functions (\refeq{eq:phi1order}) per iteration of Algorithm \ref{alg:alg}.}
 \label{fig:postacoarse}
\end{figure}

\begin{figure}[!h]
\begin{subfigure}[b]{0.3\textwidth}
 \includegraphics[width=\textwidth]{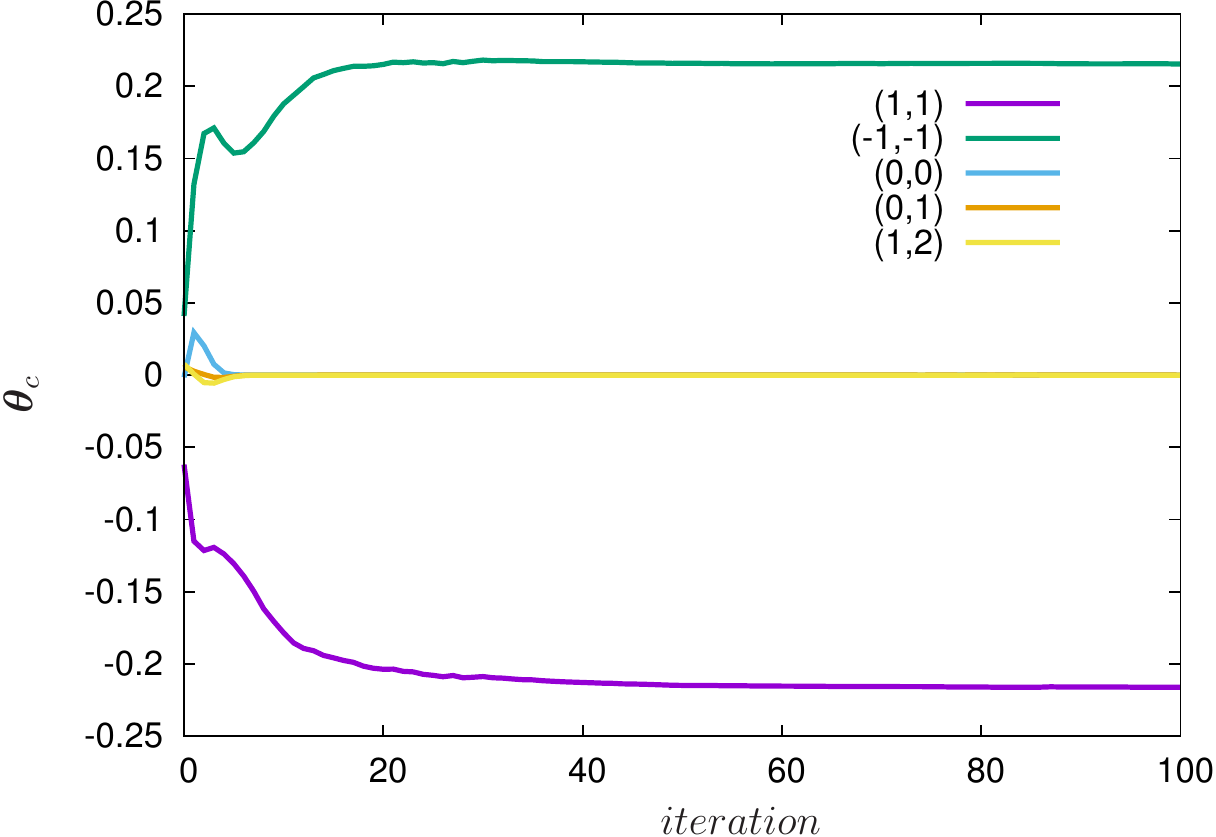}
 \caption{$N=64$}
\end{subfigure}
\begin{subfigure}[b]{0.3\textwidth}
 \includegraphics[width=\textwidth]{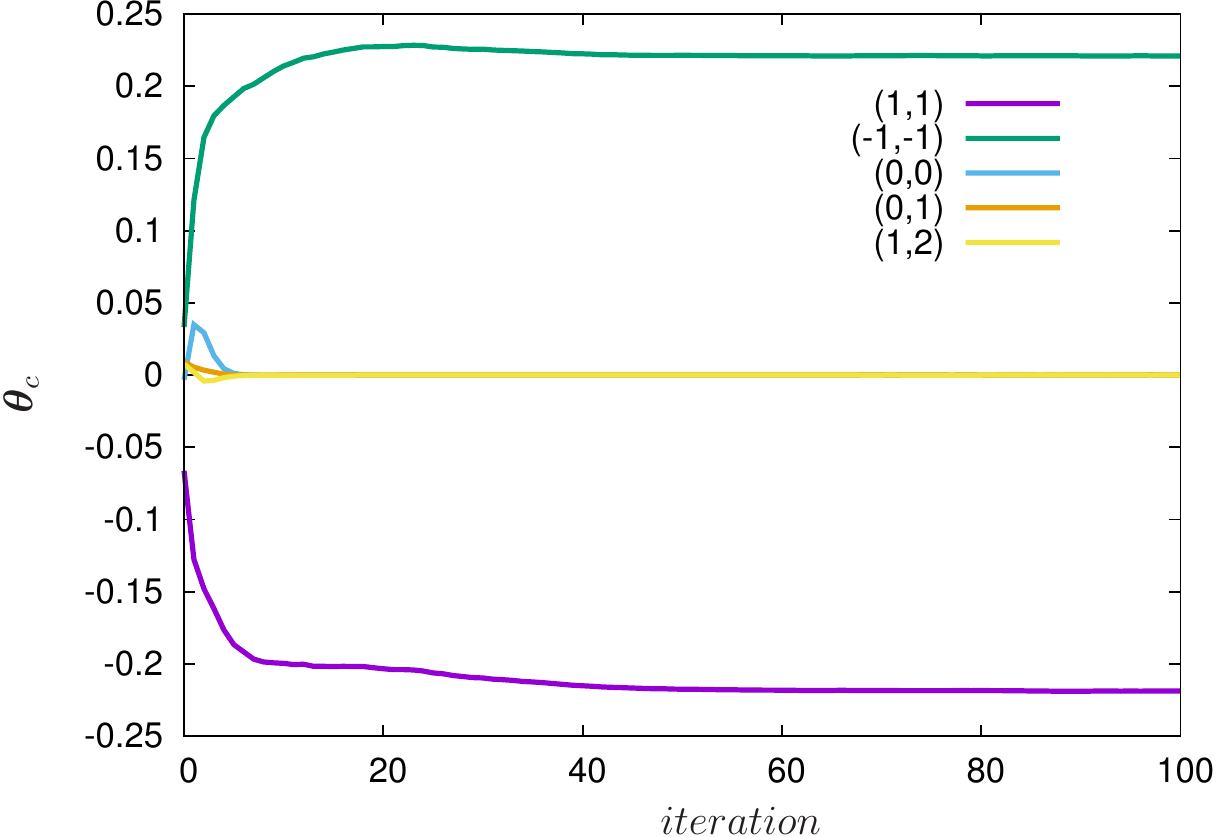}
 \caption{$N=128$}
\end{subfigure}
\begin{subfigure}[b]{0.3\textwidth}
 \includegraphics[width=\textwidth]{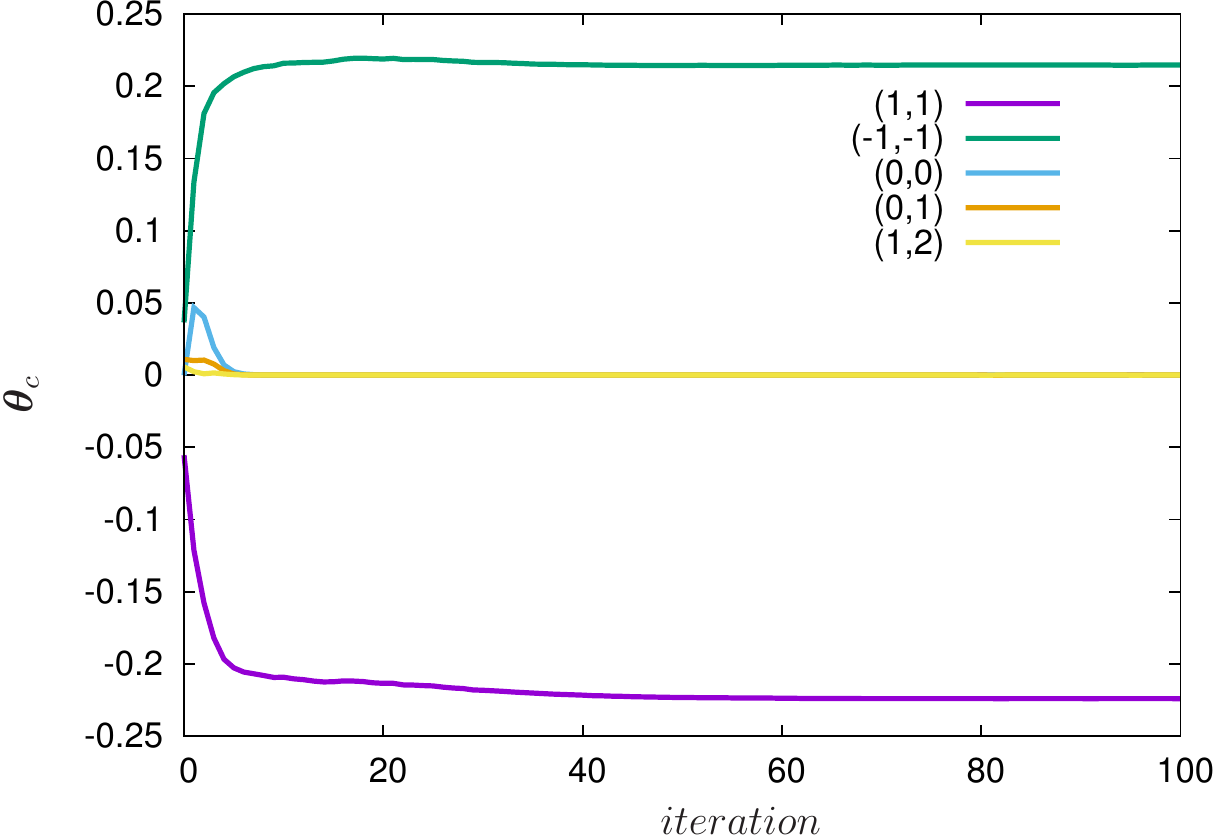}
 \caption{$N=256$}
\end{subfigure}
 \caption{Posterior mean of $\bt_c$ associated with (some) second-order feature functions (\refeq{eq:phi2order}) per iteration of Algorithm \ref{alg:alg}.}
 \label{fig:postbcoarse}
\end{figure}

\begin{figure}
\includegraphics[width=.5\textwidth]{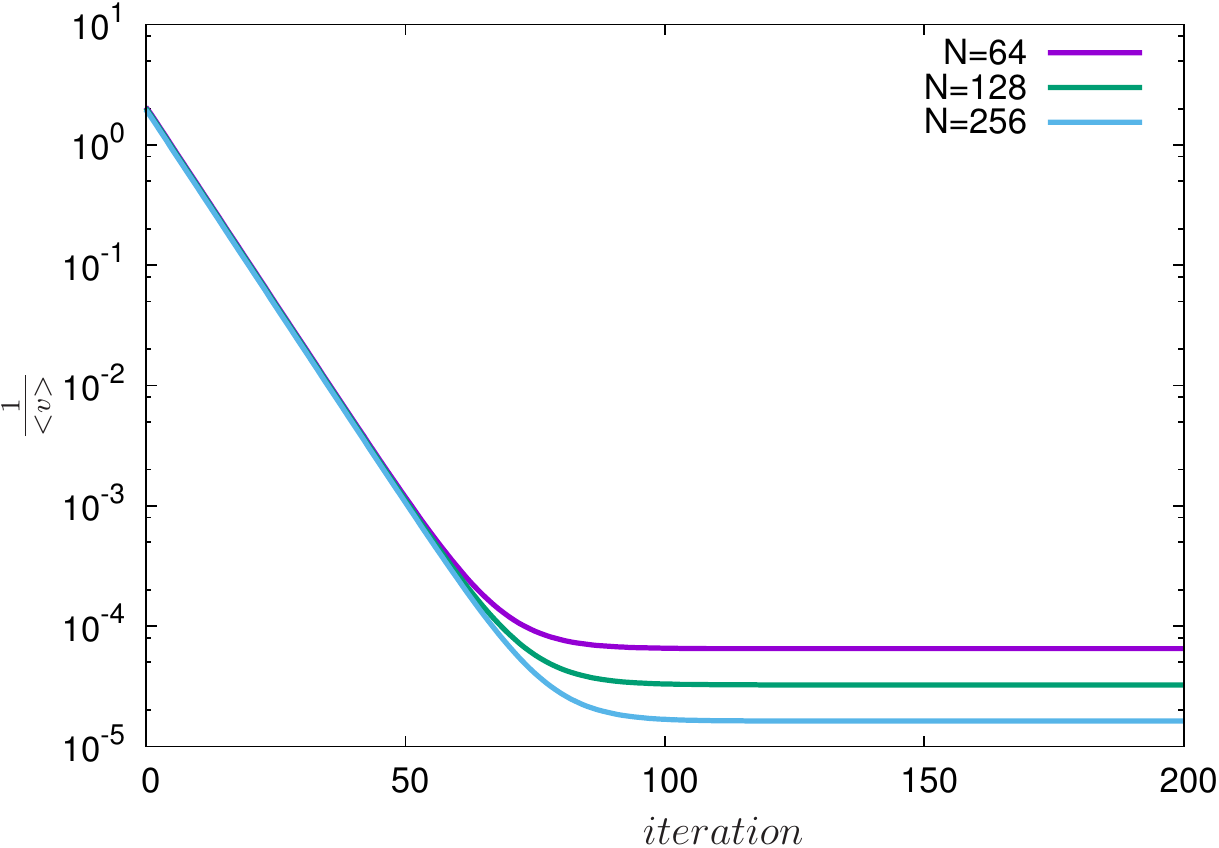}
 \caption{Reciprocal of the posterior mean of $v$ per iteration of Algorithm \ref{alg:alg}.}
 \label{fig:postsigmacoarse}
\end{figure}

\begin{figure}
\begin{minipage}[c]{0.5\textwidth}
\centering
\begin{subfigure}[b]{\textwidth}
\includegraphics[width=\textwidth]{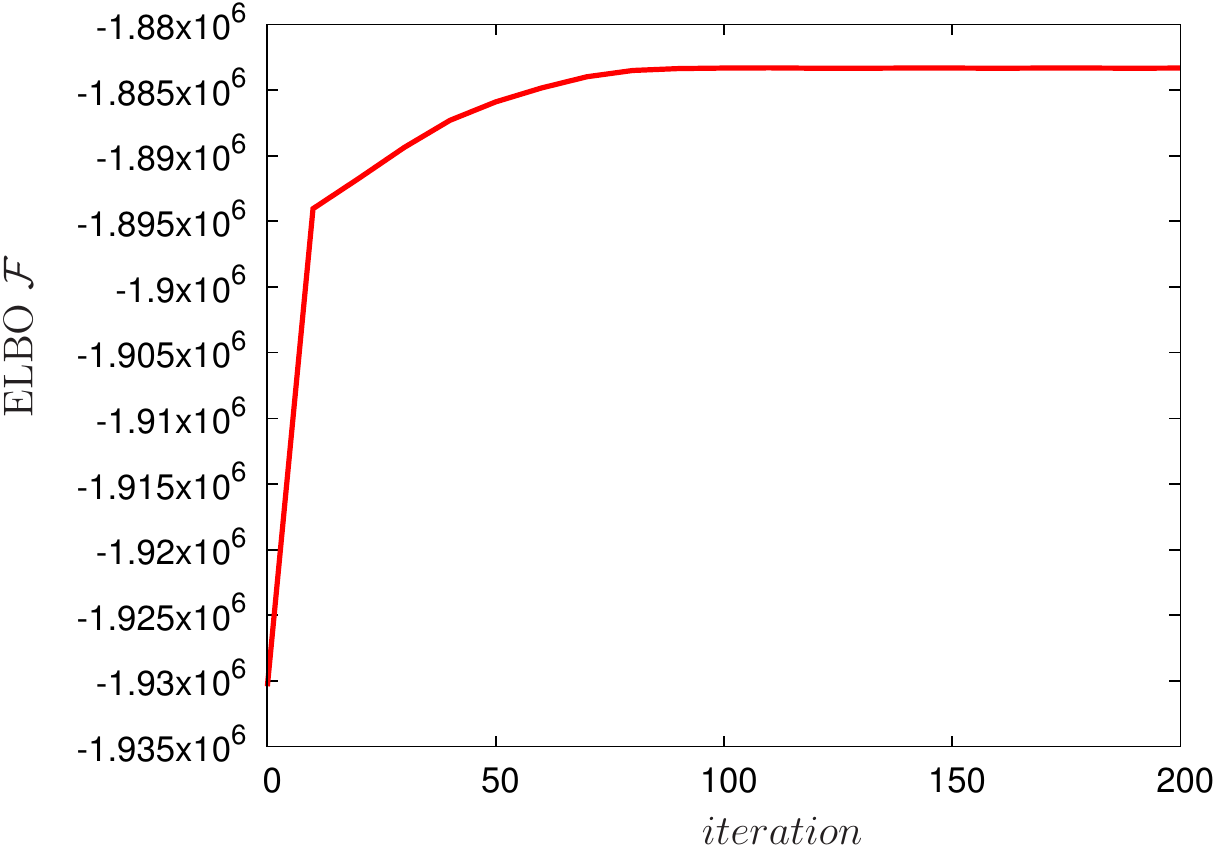}
\caption{$N=128$}
 \end{subfigure}
 \end{minipage}
 \begin{minipage}[c]{0.5\textwidth}
\centering
\begin{subfigure}[b]{\textwidth}
 \includegraphics[width=\textwidth]{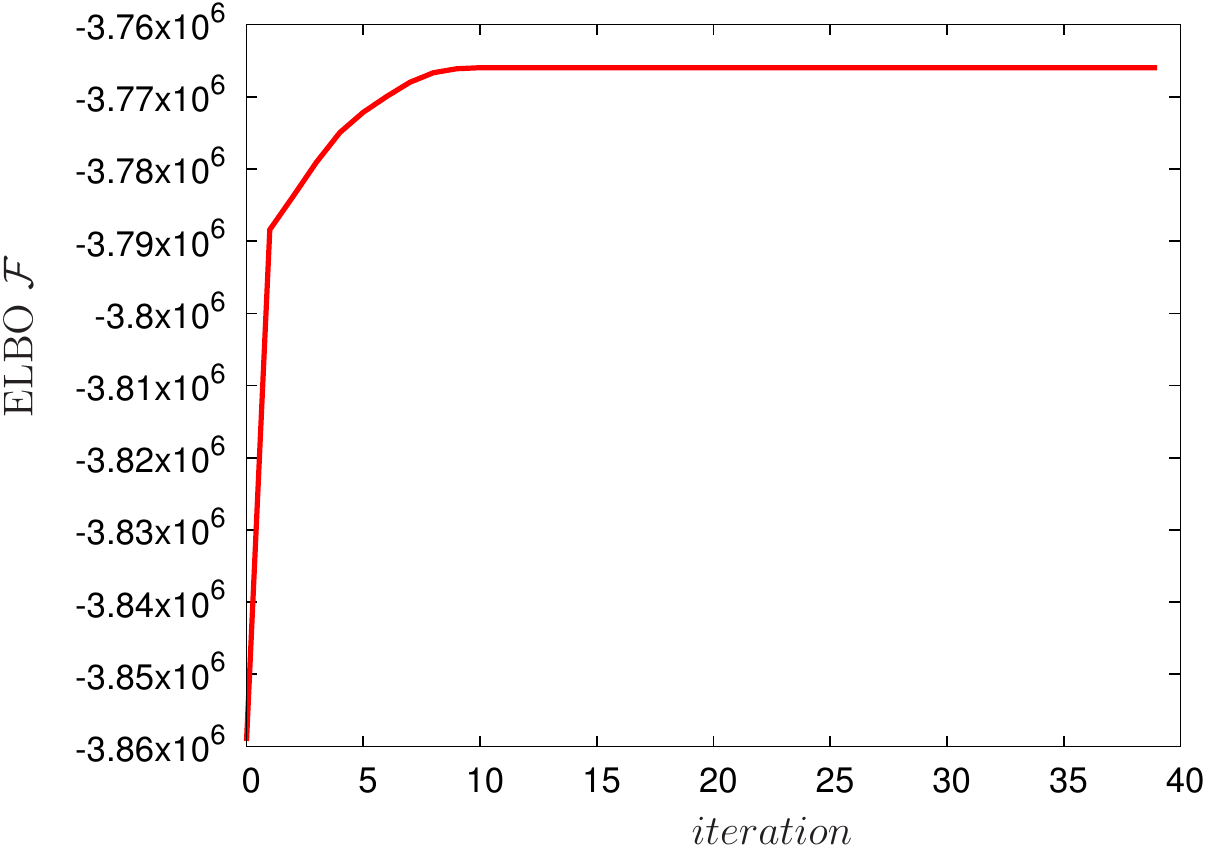}
\caption{$N=256$}
 \end{subfigure}
 \end{minipage}
 \caption{Evolution  of the ELBO  $\mathcal{F}$ as estimated at each iteration based on \refeq{eq:elbofinal} for $N=128$ (left) and $N=256$ (right).}
 \label{fig:elbo}
\end{figure}

\begin{figure}
 \includegraphics[width=.45\textwidth]{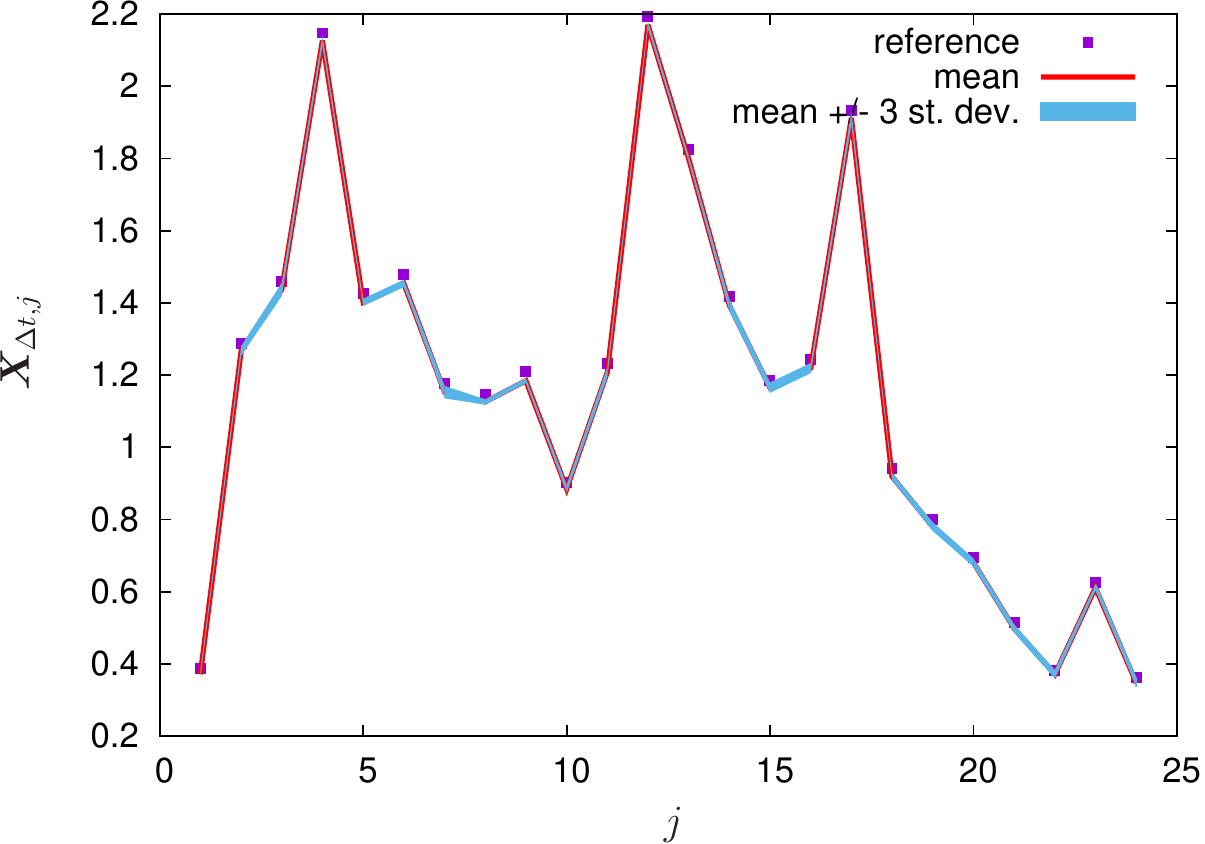}
 \hfill
 \includegraphics[width=.45\textwidth]{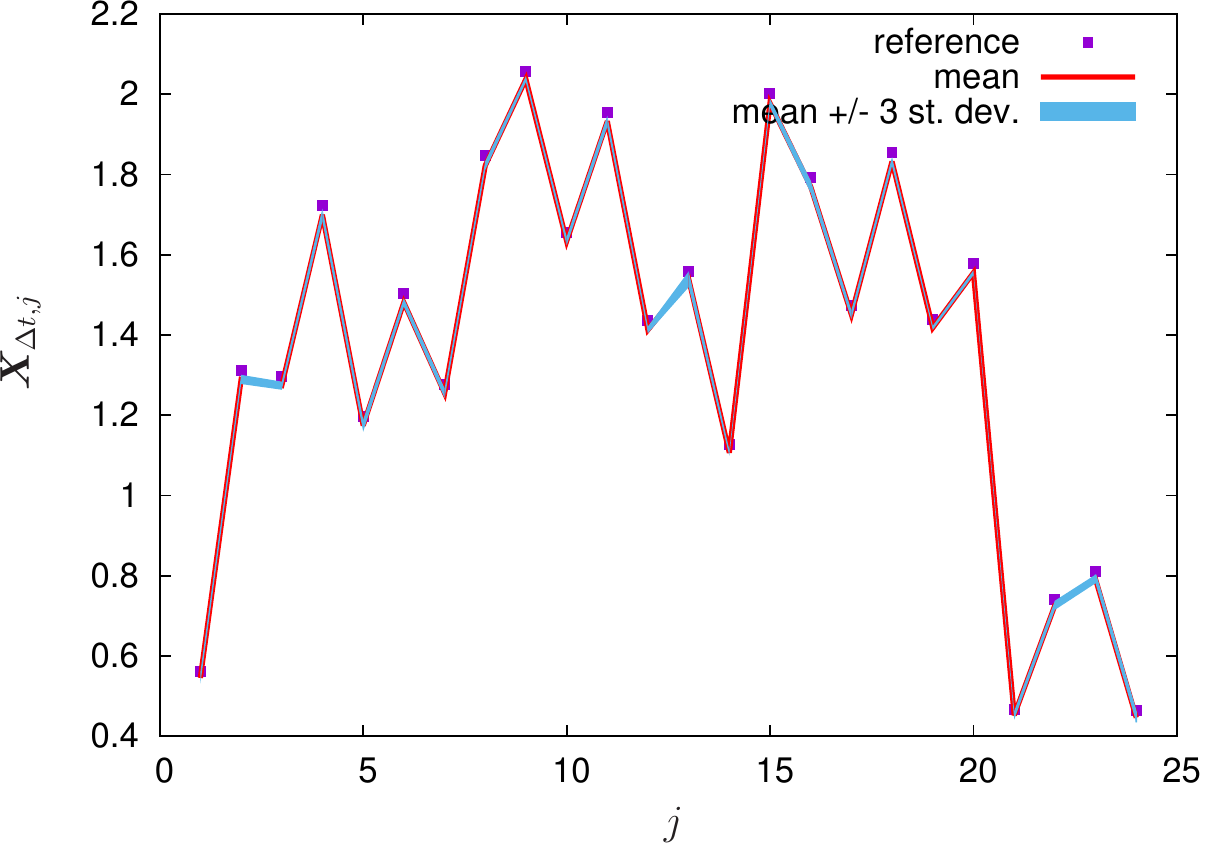}
 \caption{Comparison of reference $X_{\Delta t, j}$ with posterior mean and standard deviation as inferred using FG data using Algorithm \ref{alg:alg}, for two different cases.}
 \label{fig:postXt}
\end{figure}

Figures \ref{fig:postacoarse} and \ref{fig:postbcoarse} depict the evolution of the posterior means  of $\bt_c$ associated with  first-order and (some) second-order feature functions for three data sizes, namely $N=64, 128, 256$. We observe that in all cases the ground-truth values are fairly accurately recovered and the level of accuracy improves with increasing data, as expected. More importantly perhaps, the model correctly identifies the active features and the $\theta_c$ associated with the inactive ones are set to zero (this also holds for the second-order features not depicted   in Figure \ref{fig:postbcoarse} due to space limitations).
Figure \ref{fig:postsigmacoarse} depicts the evolution of the reciprocal of the posterior mean $<v>$ of the CG model's precision (\refeq{eq:pc}) which converges to very small values approaching the reference value $0$.
Figure \ref{fig:elbo} depicts the ascending evolution (as expected)  of the ELBO  $\mathcal{F}$ as estimated at each iteration based on \refeq{eq:elbofinal},  for $N=128$ and $256$ (qualitatively this is the same as for other $N$ although the range of values can differ drastically).
Figure \ref{fig:postXt} compares the true CG state vector $\bxx_{\Delta t}$ with the results of the Variational posterior approximation which is trained purely on FG data. One can clearly observe very good agreement and tight uncertainty  bounds as reflected in the small posterior variances.

\begin{figure}[!h]
\begin{minipage}[c]{0.5\textwidth}
\centering
\begin{subfigure}[b]{.9\textwidth}
\includegraphics[width=\textwidth]{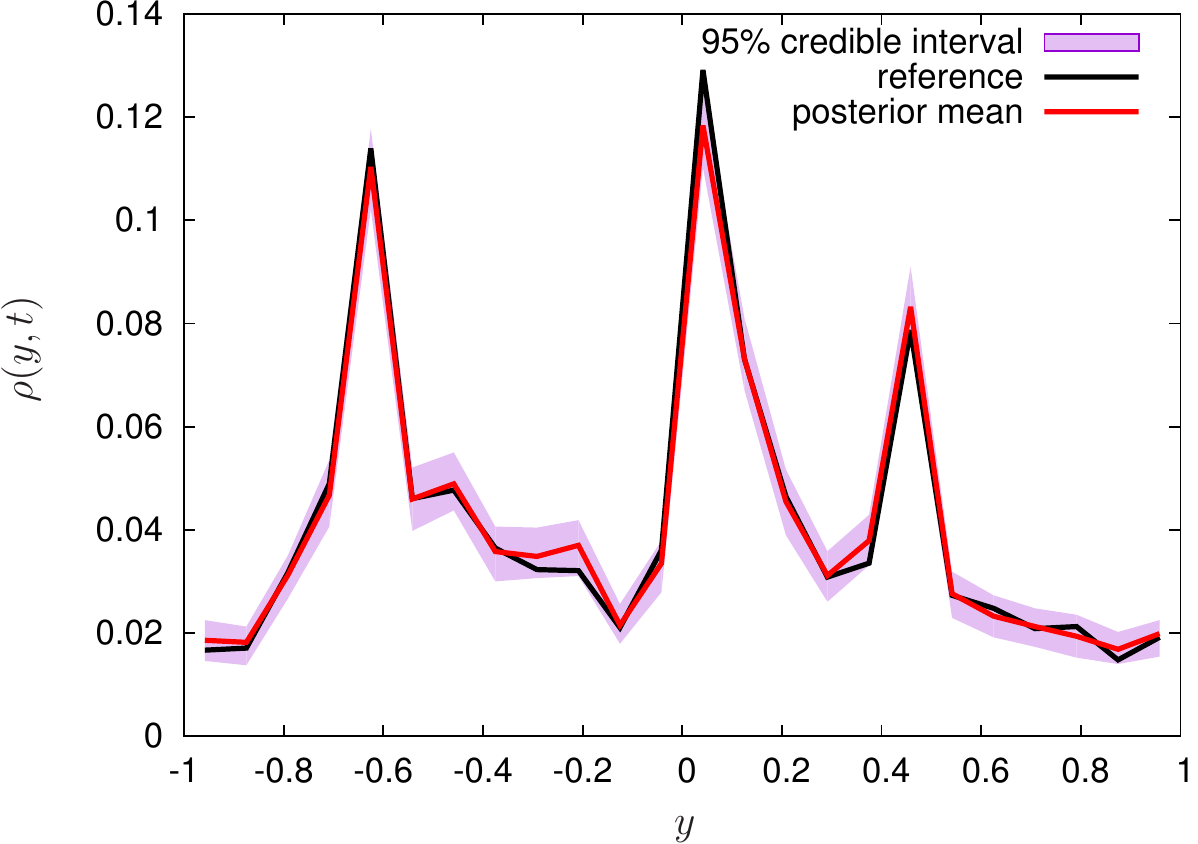}
\caption{$t=2\Delta t$}
 \end{subfigure}
 \begin{subfigure}[b]{.9\textwidth}
\includegraphics[width=\textwidth]{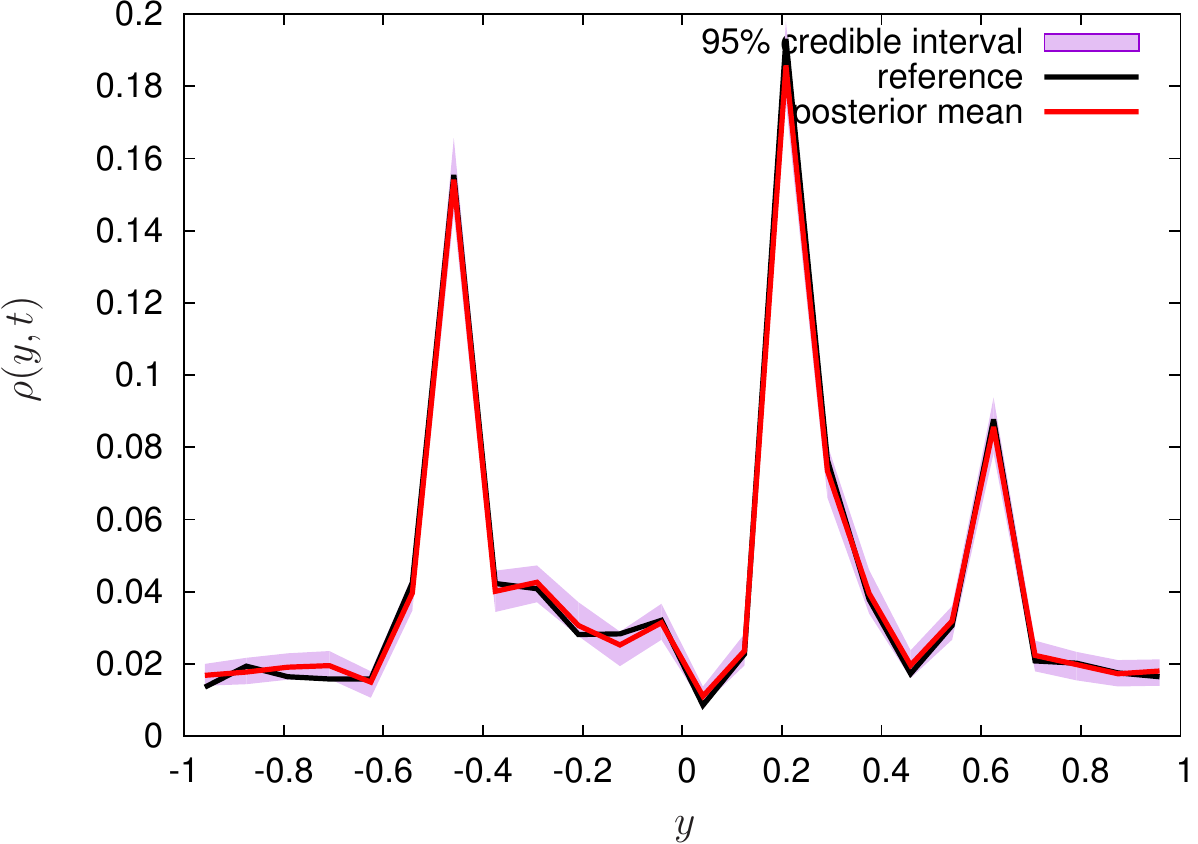}
 \caption{$t=4\Delta t$}
 \end{subfigure}
 \begin{subfigure}[b]{.9\textwidth}
\includegraphics[width=\textwidth]{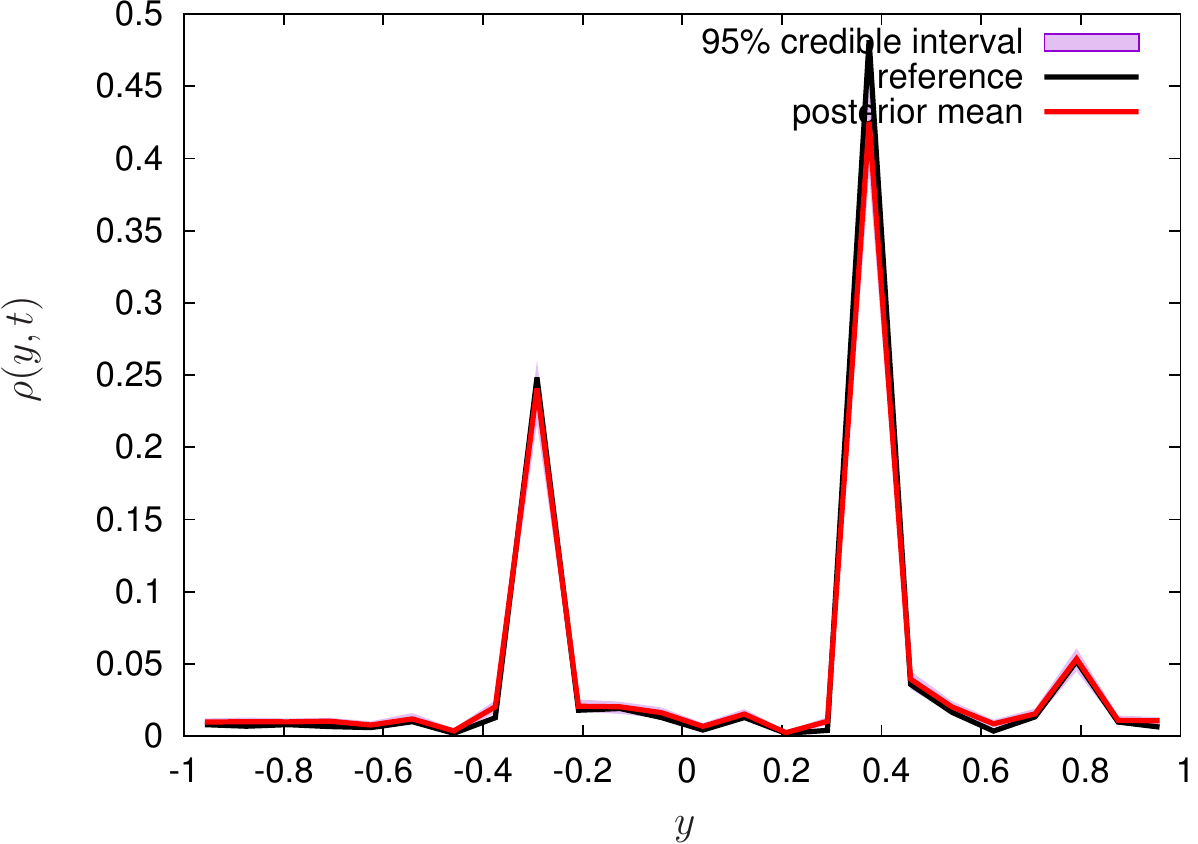}
 \caption{$t=6\Delta t$}
 \end{subfigure}
 \begin{subfigure}[b]{.9\textwidth}
\includegraphics[width=\textwidth]{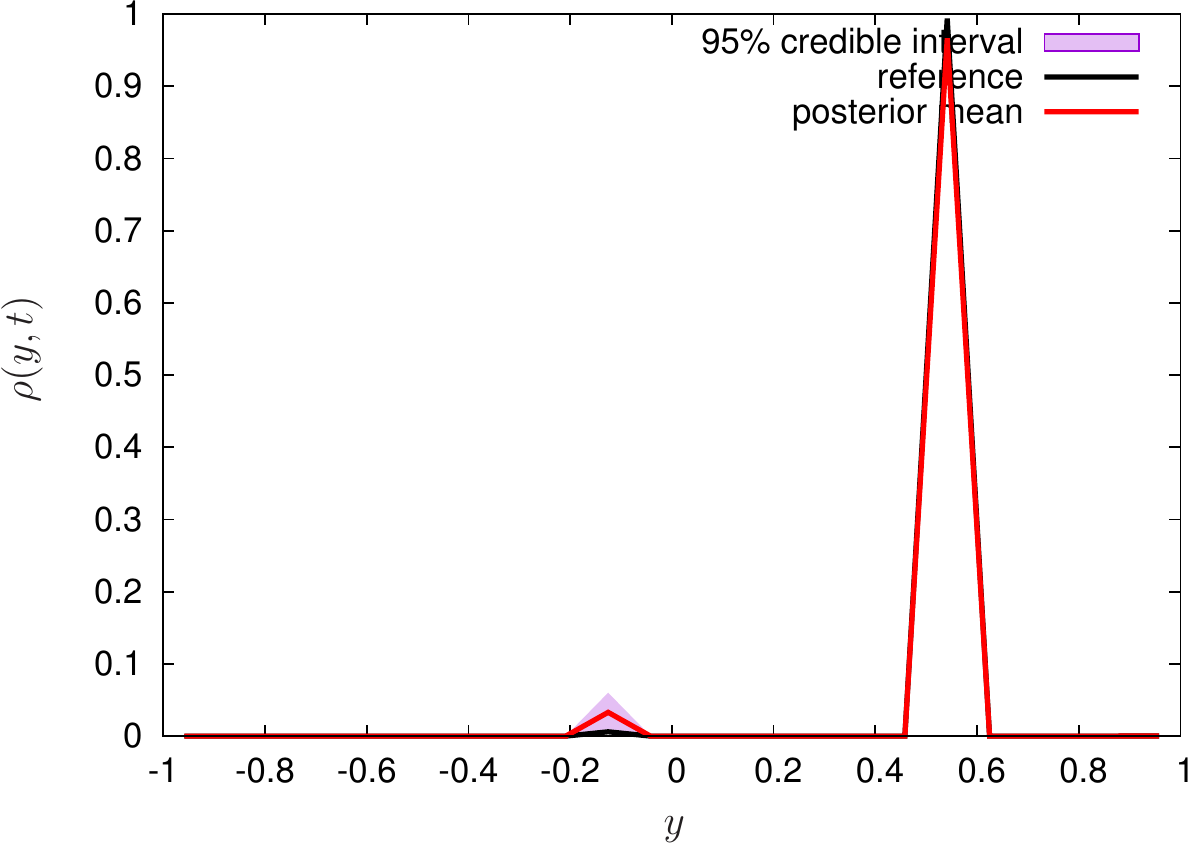}
 \caption{$t=8\Delta t$}
 \end{subfigure}
 \end{minipage}
 \begin{minipage}[c]{0.5\textwidth}
\centering
\begin{subfigure}[b]{.9\textwidth}
\includegraphics[width=\textwidth]{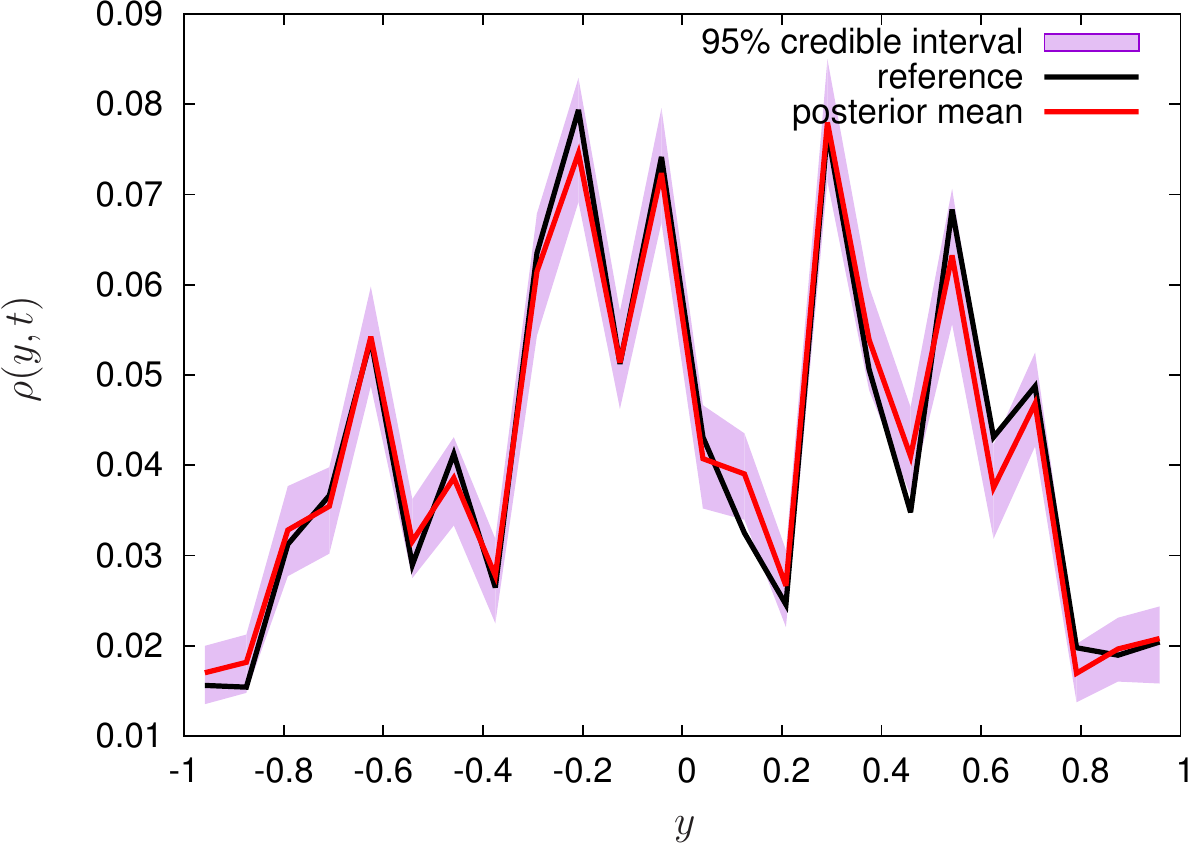}
 \caption{$t=2\Delta t$}
 \end{subfigure}
 \begin{subfigure}[b]{.9\textwidth}
\includegraphics[width=\textwidth]{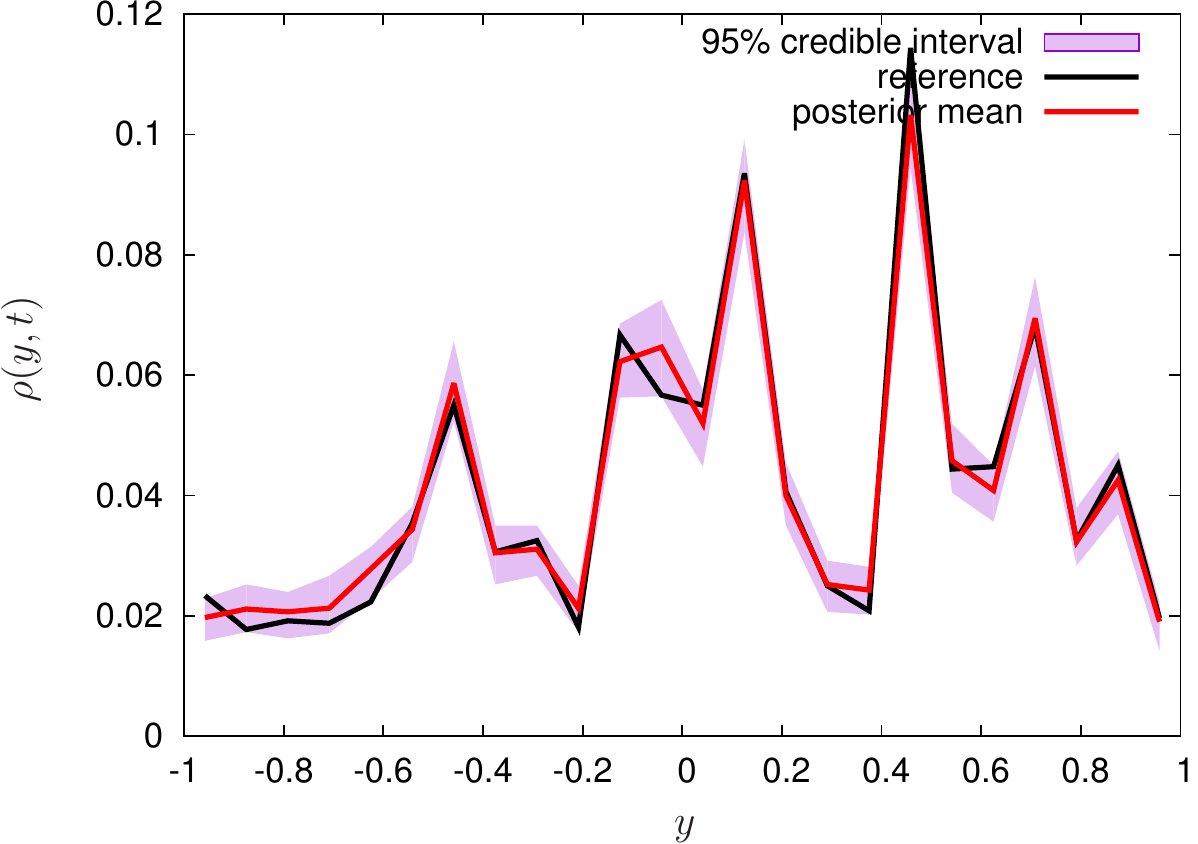}
 \caption{$t=4\Delta t$}
 \end{subfigure}
 \begin{subfigure}[b]{.9\textwidth}
\includegraphics[width=\textwidth]{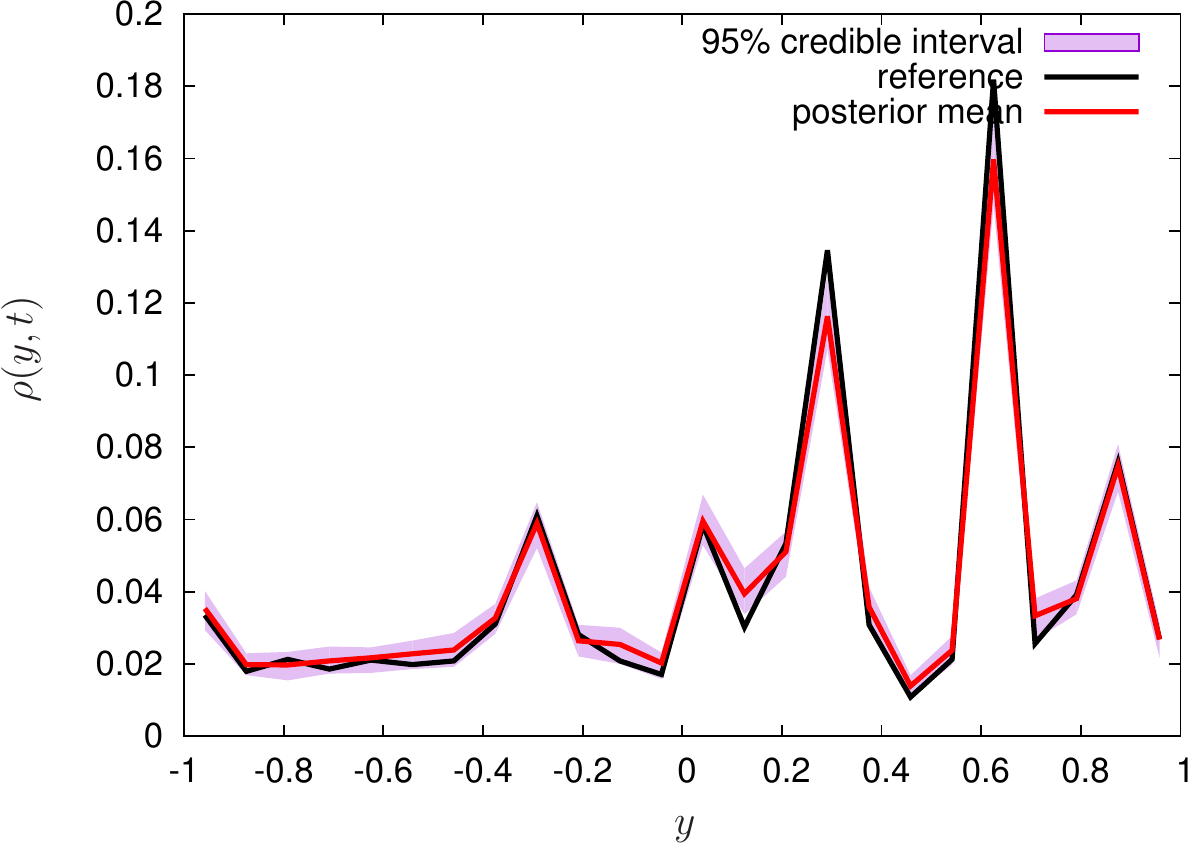}
 \caption{$t=6\Delta t$}
 \end{subfigure}
 \begin{subfigure}[b]{.9\textwidth}
\includegraphics[width=\textwidth]{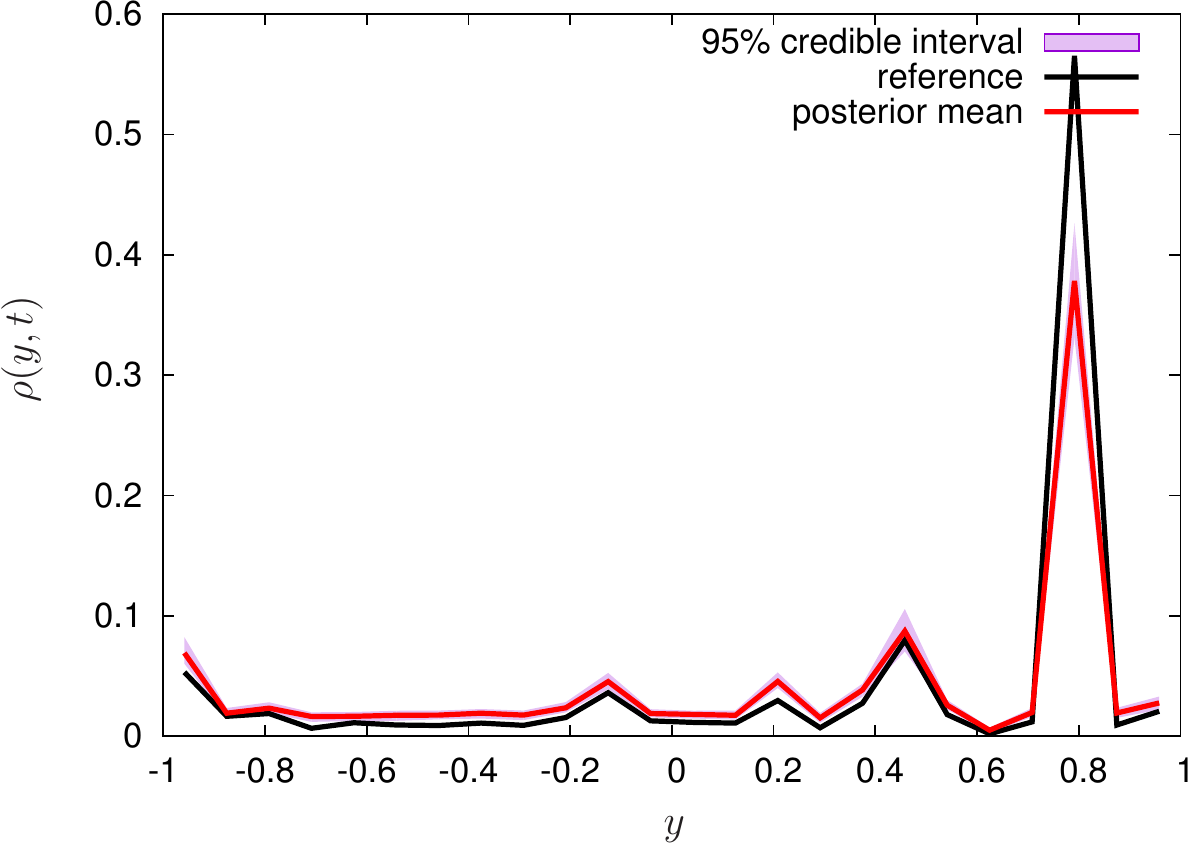}
 \caption{$t=8\Delta t$}
 \end{subfigure}
 \end{minipage}
 \caption{Predictive estimates of walker density for $24$ bins (same as in CG evolution), for various times in the future i.e. $t=2\Delta t, 4 \Delta t, 6\Delta t, 8\Delta t$ for two different initial conditions (columns). The reference density profile was computed by simulating the FG model of walkers using the FG time step $\delta t$.  ($N=256$) }
 \label{fig:predrho}
\end{figure}

\begin{figure}[!h]
\begin{minipage}[c]{0.5\textwidth}
\centering
\begin{subfigure}[b]{.9\textwidth}
\includegraphics[width=\textwidth]{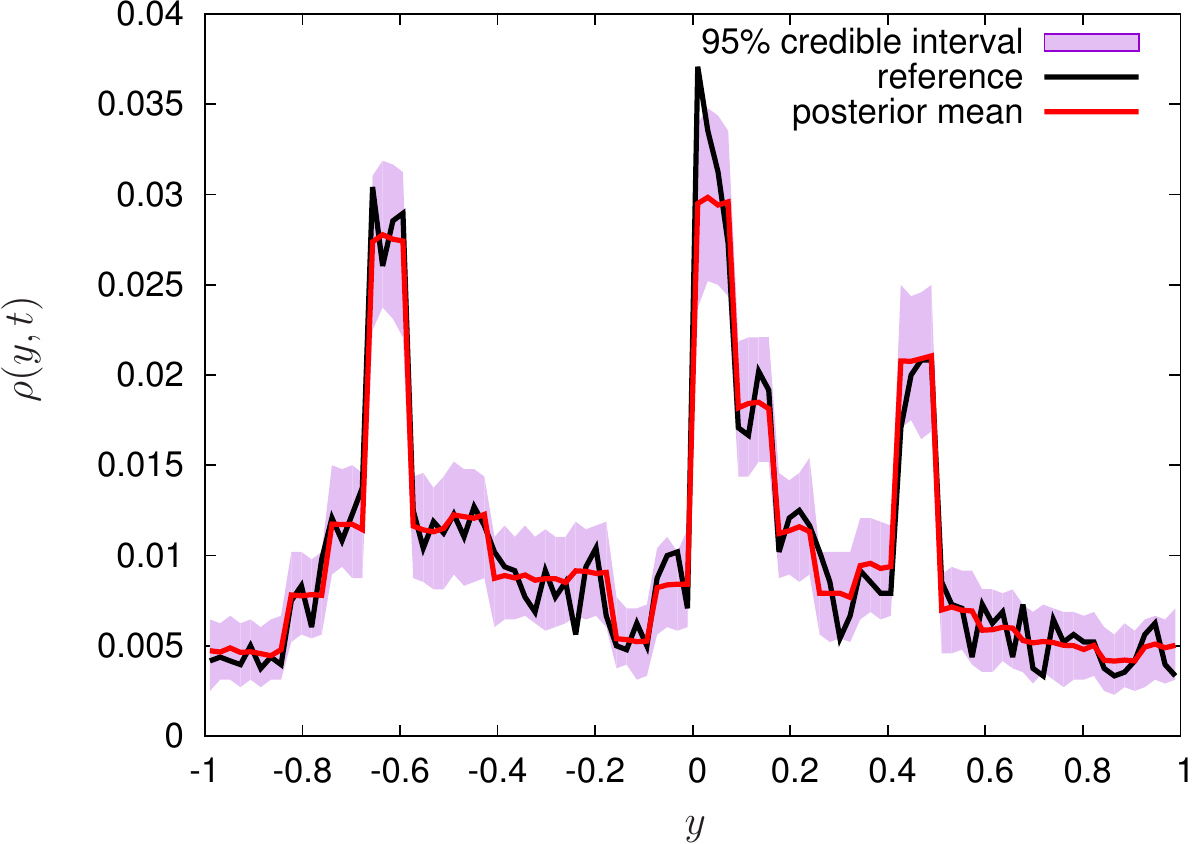}
\caption{$t=2\Delta t$}
 \end{subfigure}
 \begin{subfigure}[b]{.9\textwidth}
\includegraphics[width=\textwidth]{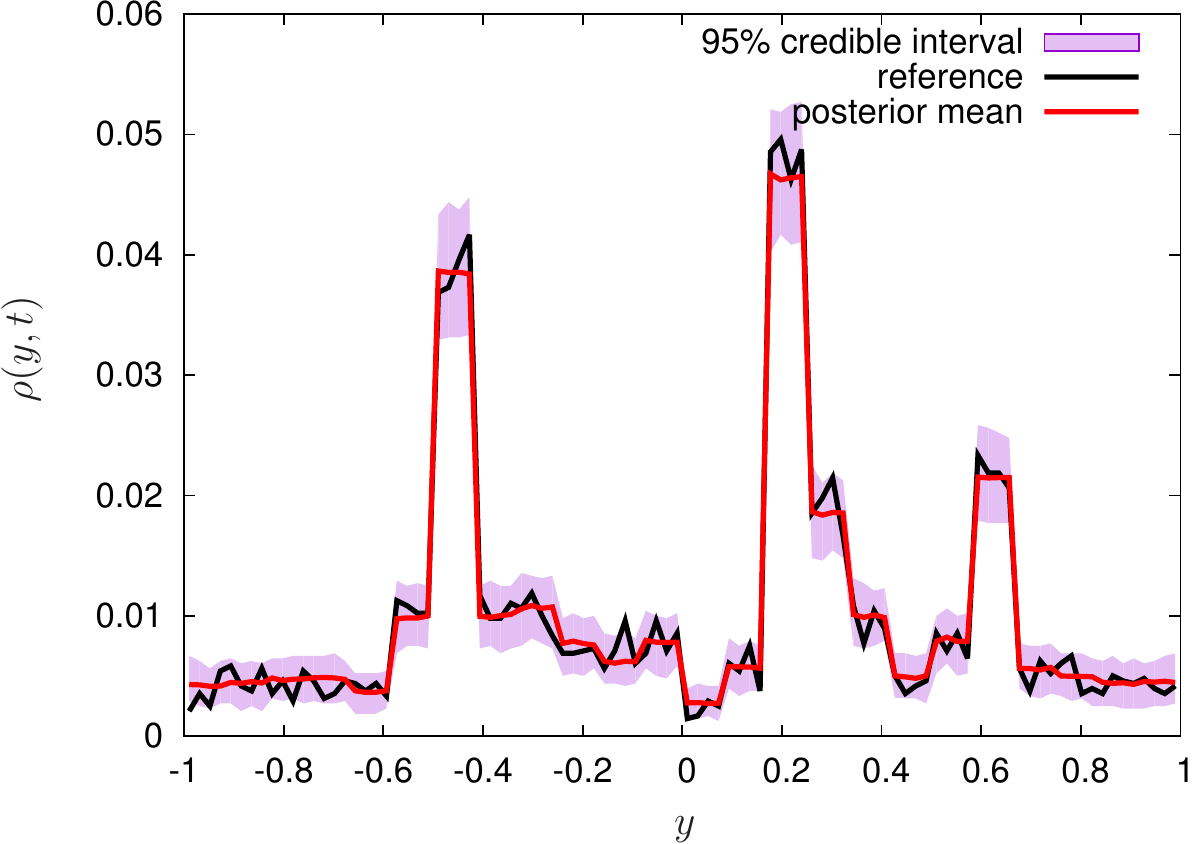}
 \caption{$t=4\Delta t$}
 \end{subfigure}
 \begin{subfigure}[b]{.9\textwidth}
\includegraphics[width=\textwidth]{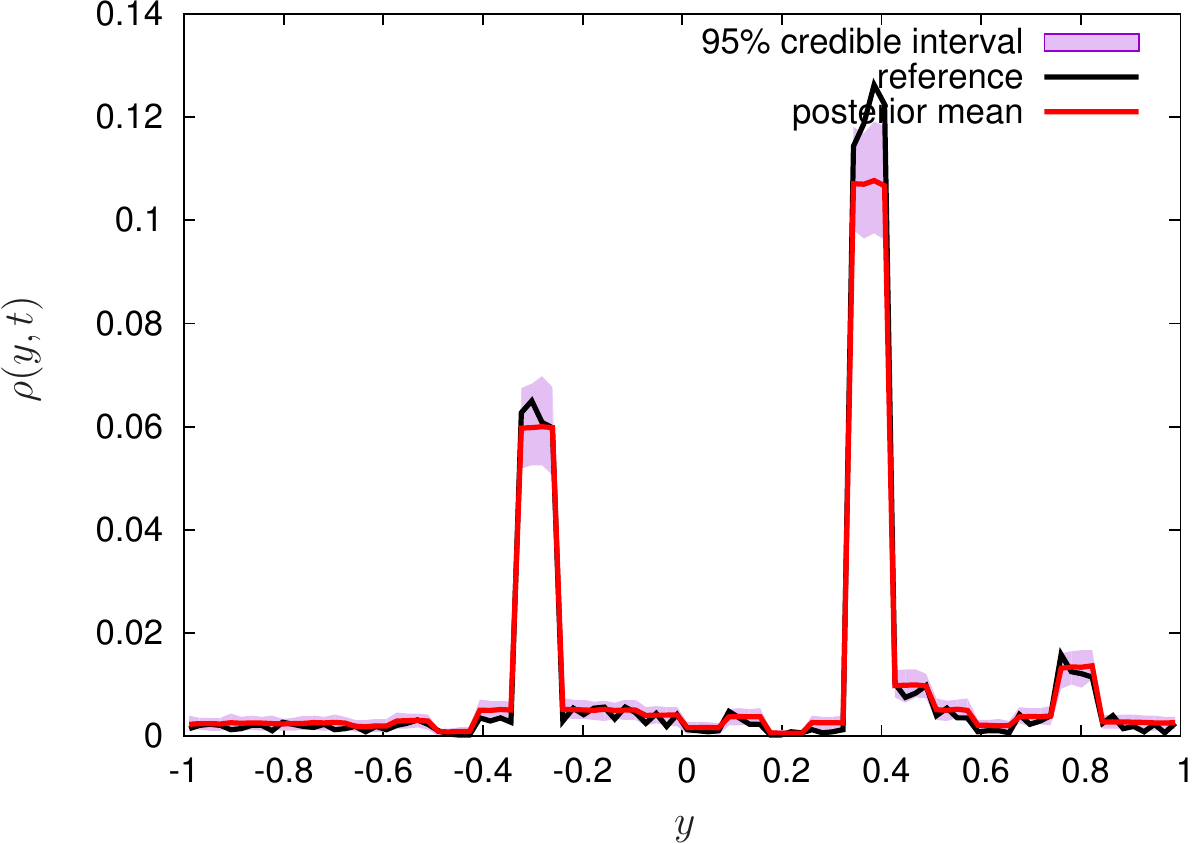}
 \caption{$t=6\Delta t$}
 \end{subfigure}
 \begin{subfigure}[b]{.9\textwidth}
\includegraphics[width=\textwidth]{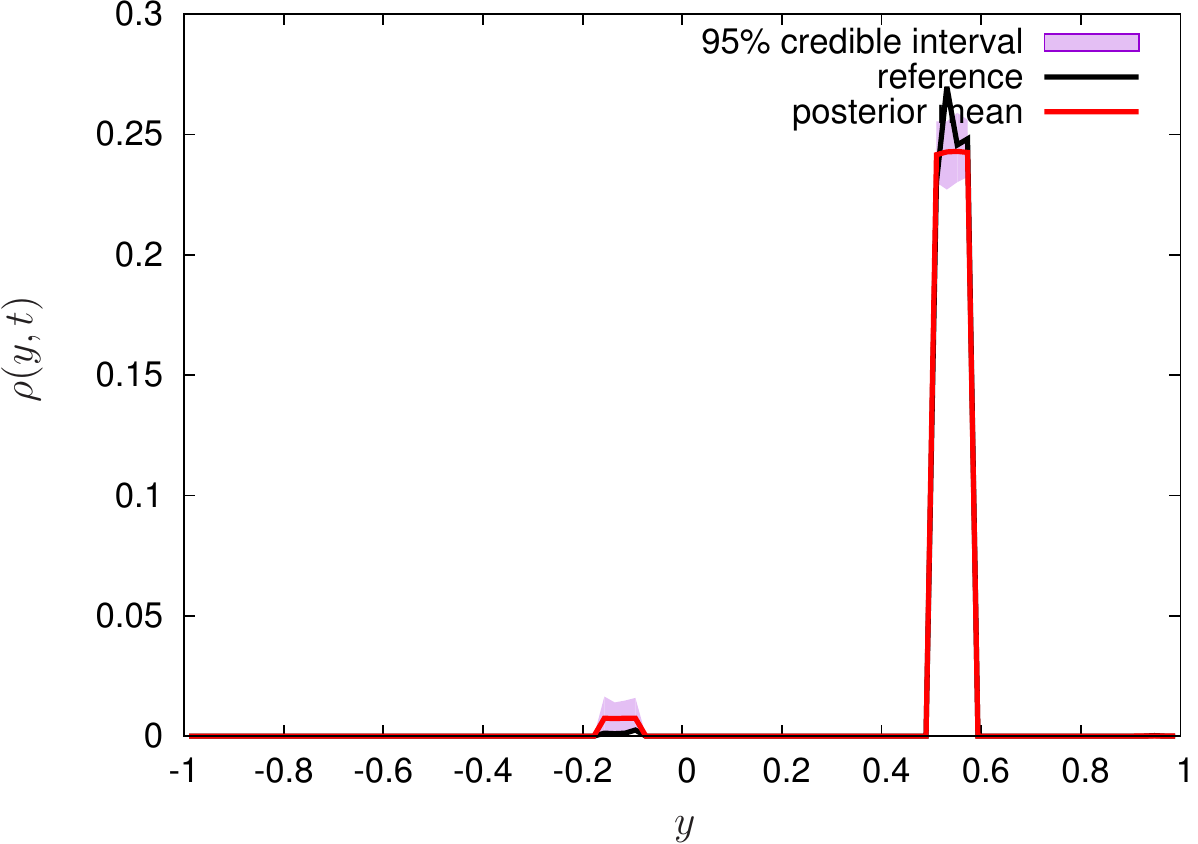}
 \caption{$t=8\Delta t$}
 \end{subfigure}
 \end{minipage}
 \begin{minipage}[c]{0.5\textwidth}
\centering
\begin{subfigure}[b]{.9\textwidth}
\includegraphics[width=\textwidth]{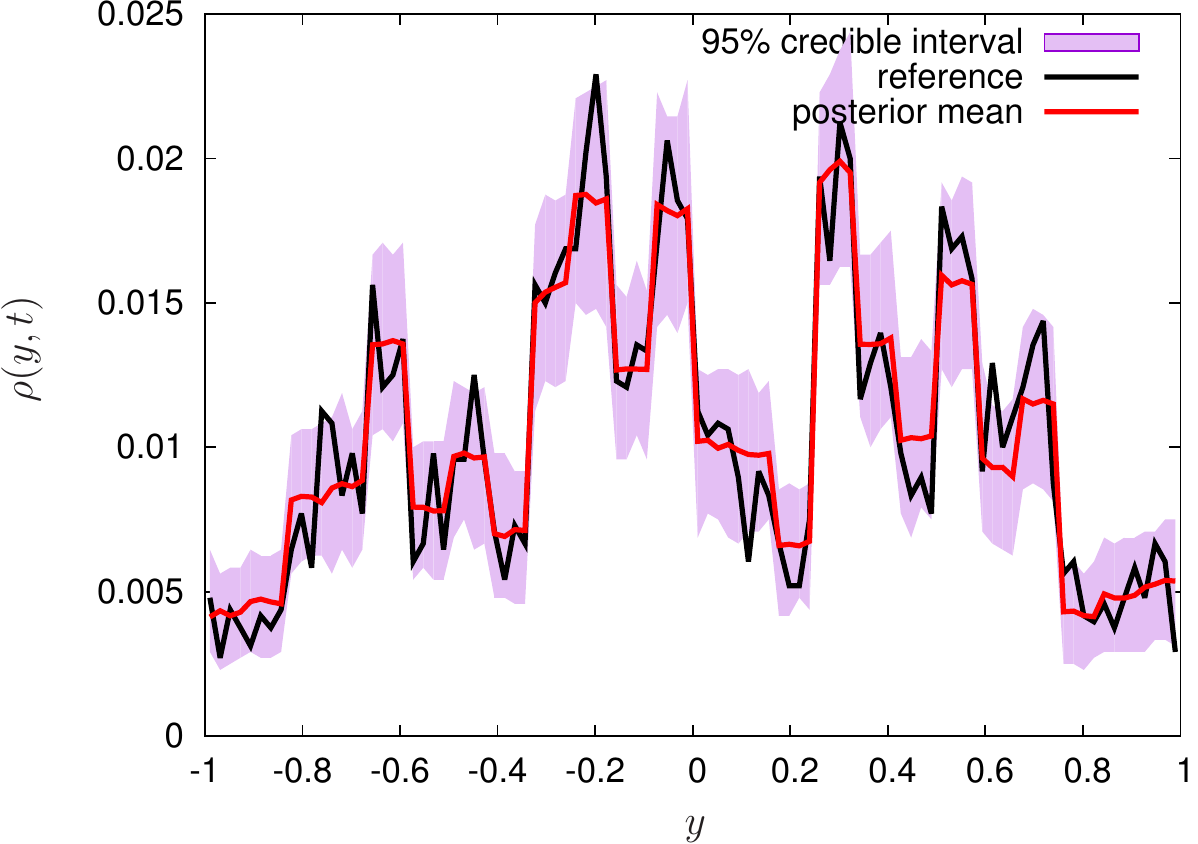}
 \caption{$t=2\Delta t$}
 \end{subfigure}
 \begin{subfigure}[b]{.9\textwidth}
\includegraphics[width=\textwidth]{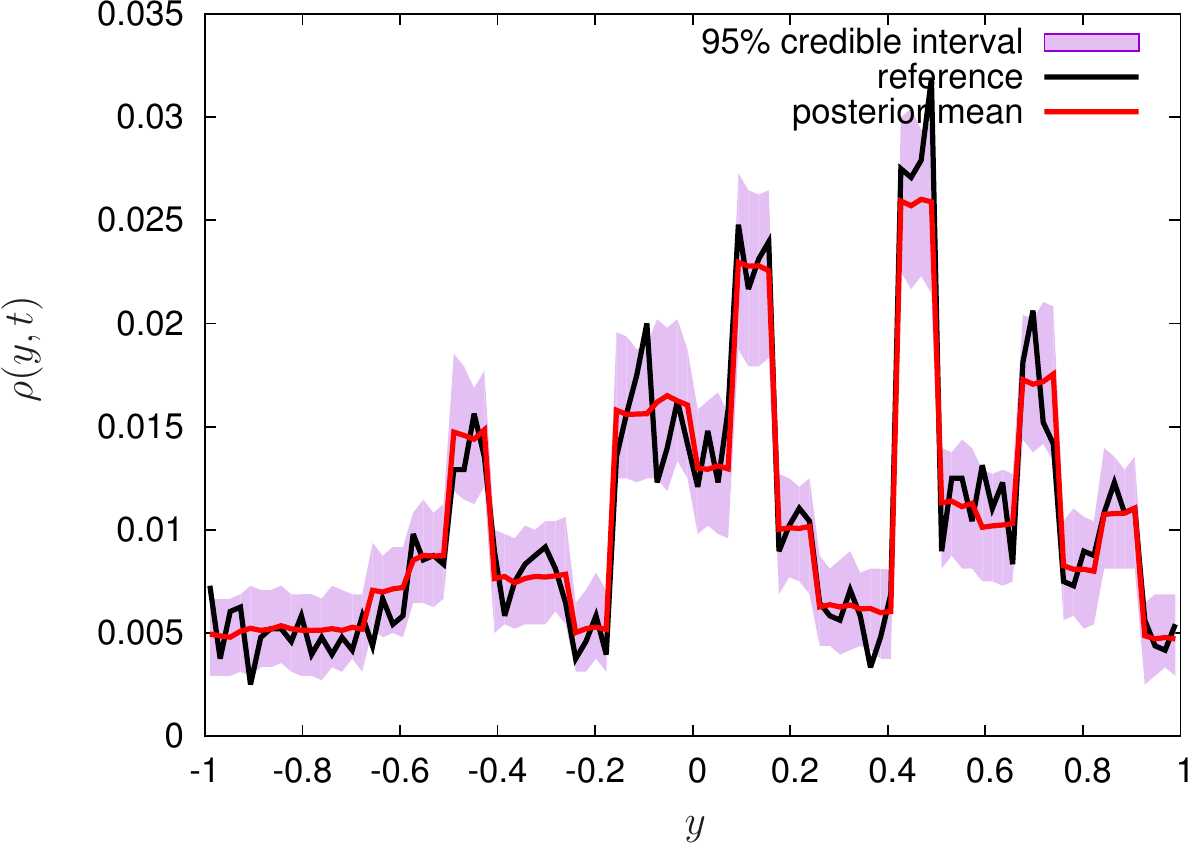}
 \caption{$t=4\Delta t$}
 \end{subfigure}
 \begin{subfigure}[b]{.9\textwidth}
\includegraphics[width=\textwidth]{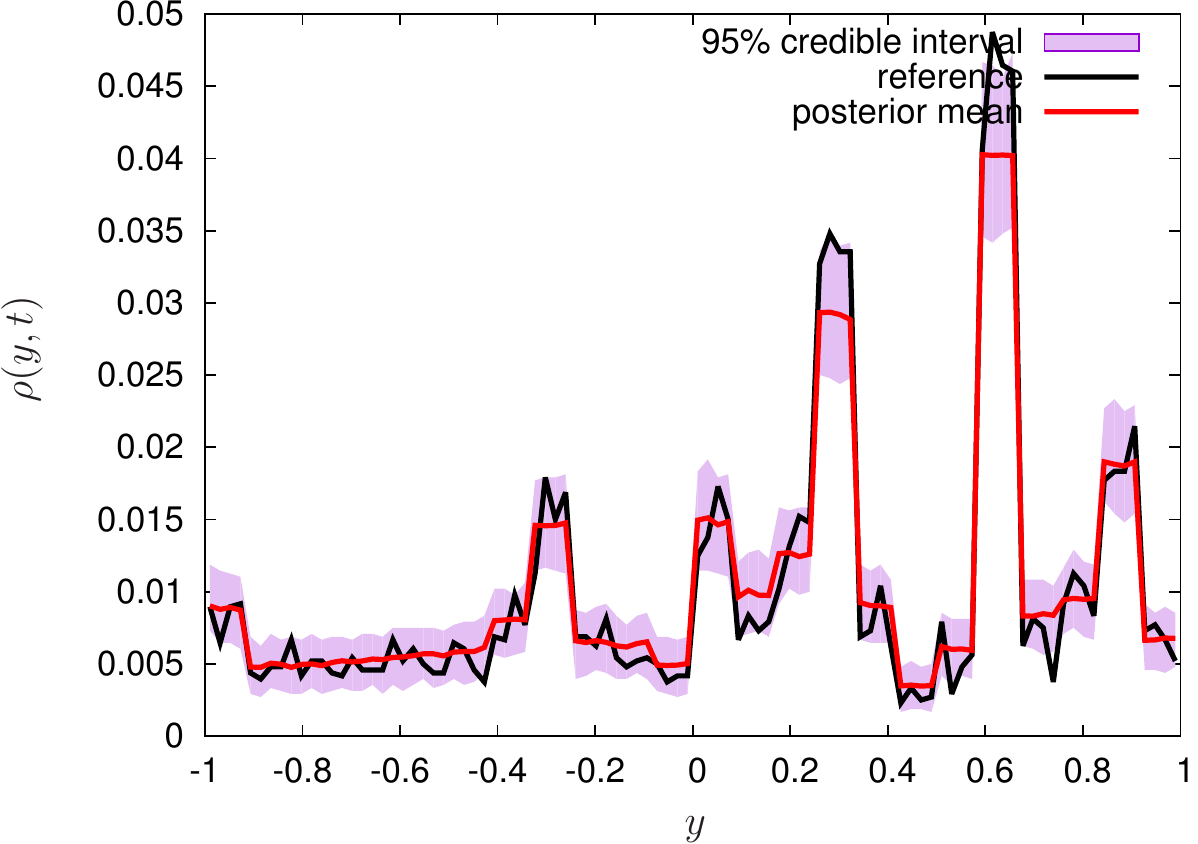}
 \caption{$t=6\Delta t$}
 \end{subfigure}
 \begin{subfigure}[b]{.9\textwidth}
\includegraphics[width=\textwidth]{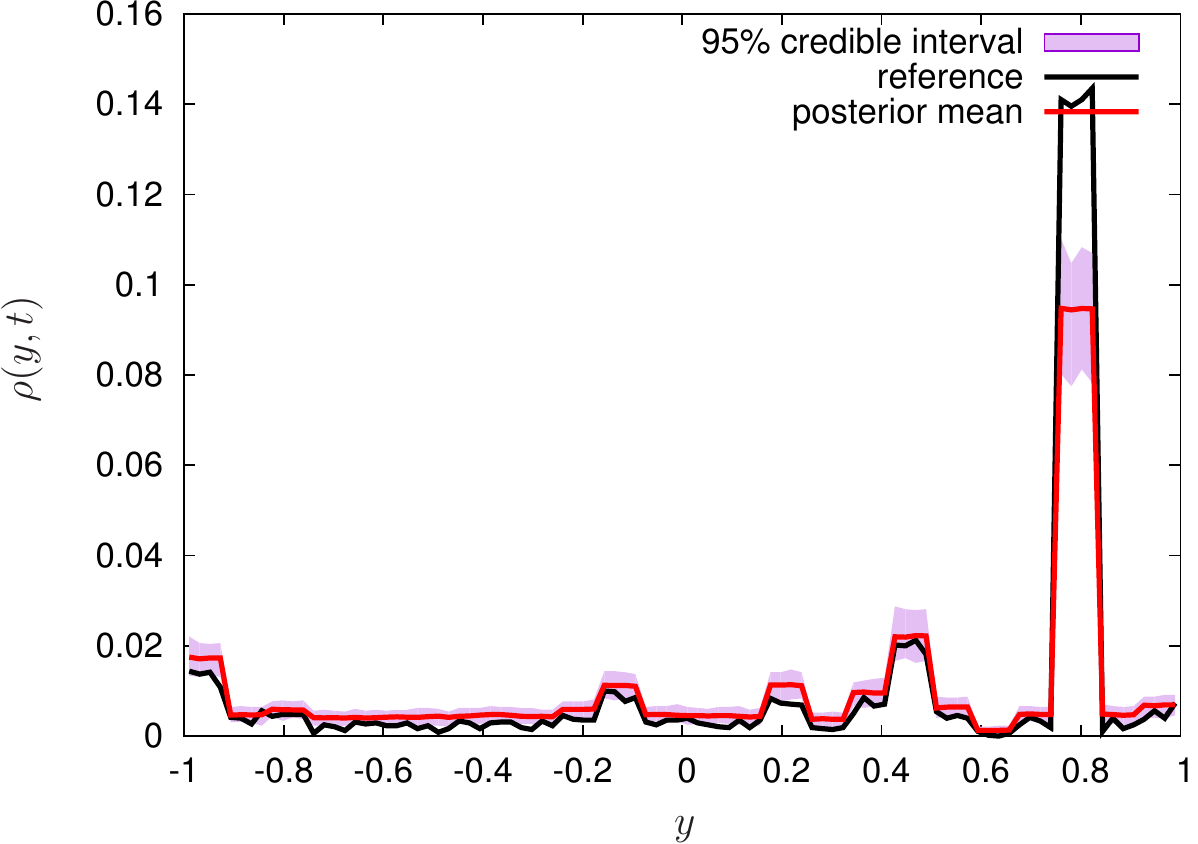}
 \caption{$t=8\Delta t$}
 \end{subfigure}
 \end{minipage}
 \caption{Predictive estimates of walker density {\em for $96$ bins} ($4$ times greater than the $24$ used in CG evolution), for various times in the future i.e. $t=2\Delta t, 4 \Delta t, 6\Delta t, 8\Delta t$ for two different initial conditions (columns). The reference density profile was computed by simulating the FG model of walkers using the FG time step $\delta t$.  ($N=256$) }
 \label{fig:predrhofine4}
\end{figure}

The remaining figures pertain to predictive estimates. In particular,  in Figure \ref{fig:predrho}, we compare the evolution of the walker density at various future times and for two different initial profiles  with the predictive posterior estimates (section \ref{sec:probpred}).  One observes that the 95\% credible interval almost always envelops the reference, even when the reference profile corresponds to all the walkers being concentrated in a single bin. In these illustrations, we employed  the same number of bins ($24$) as those used in the CG description even though the predictive estimates are based on a full reconstruction of the FG  states $\bx_t$ as described in section \ref{sec:probpred} and not the CG states $\bxx_t$. To further demonstrate this aspect of the model, in Figure \ref{fig:predrhofine4}, we depict the same predictive estimates of the walker's density as reconstructed from the FG states $\bx_t$ and using a binning that is $4$ times finer than the CG states (i.e. $96$ bins). Again, highly accurate estimates are observed with one important difference. The credible intervals are wider i.e. reflect increased predictive uncertainty. This can be explained by the fact that when finer level details are sought, it is unavoidable that  greater uncertainty is present. 

\begin{figure}[!h]
\begin{minipage}[c]{0.5\textwidth}
\centering
\begin{subfigure}[b]{.9\textwidth}
\includegraphics[width=\textwidth]{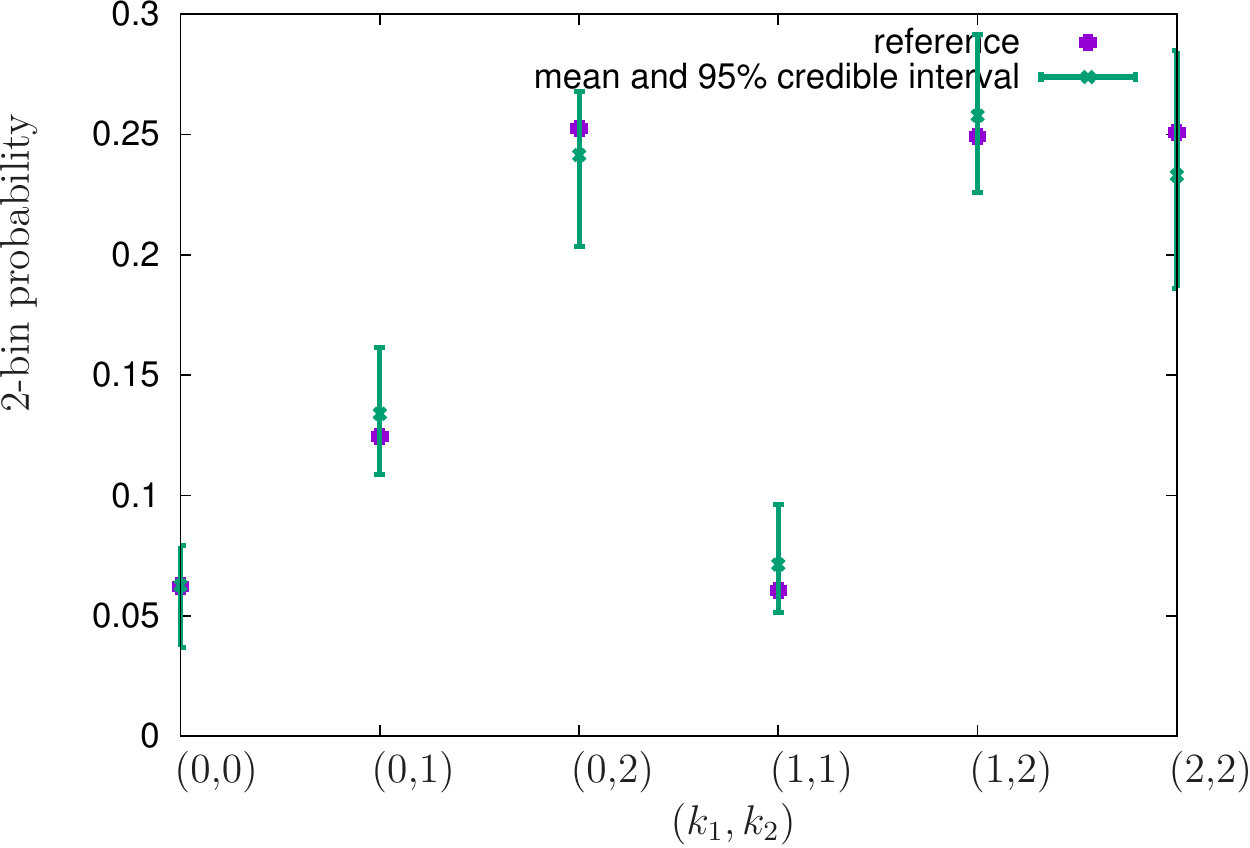}
\caption{$t=2\Delta t$}
 \end{subfigure}
 \begin{subfigure}[b]{.9\textwidth}
\includegraphics[width=\textwidth]{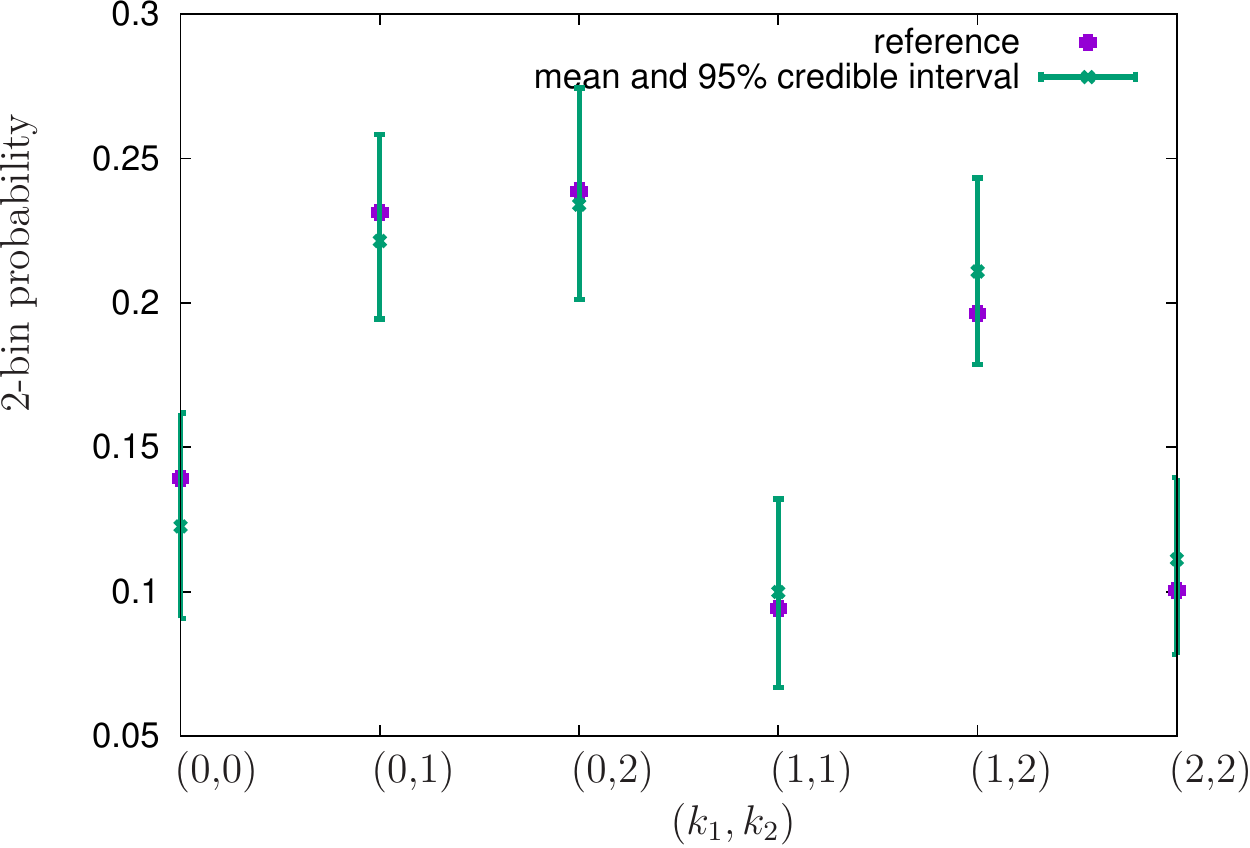}
 \caption{$t=3\Delta t$}
 \end{subfigure}
 \begin{subfigure}[b]{.9\textwidth}
\includegraphics[width=\textwidth]{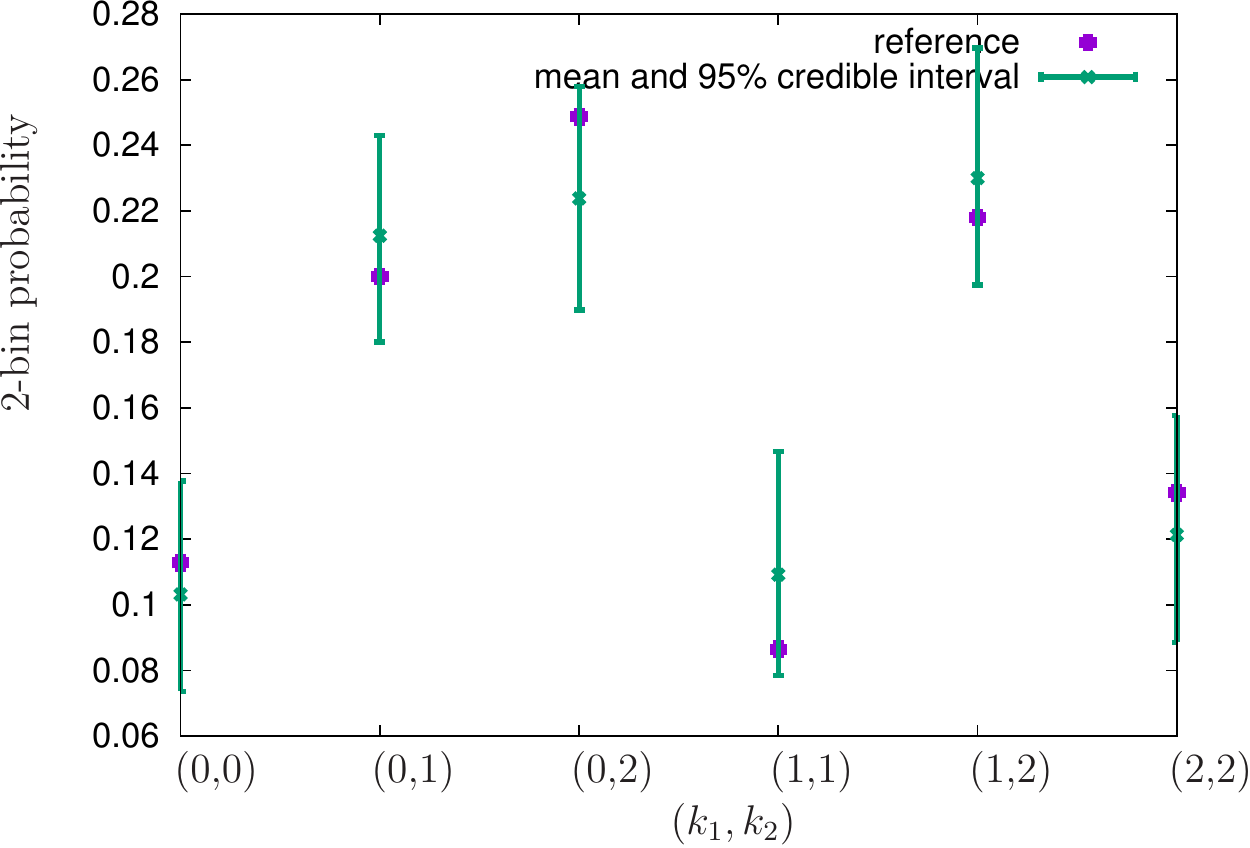}
 \caption{$t=5\Delta t$}
 \end{subfigure}
 \begin{subfigure}[b]{.9\textwidth}
\includegraphics[width=\textwidth]{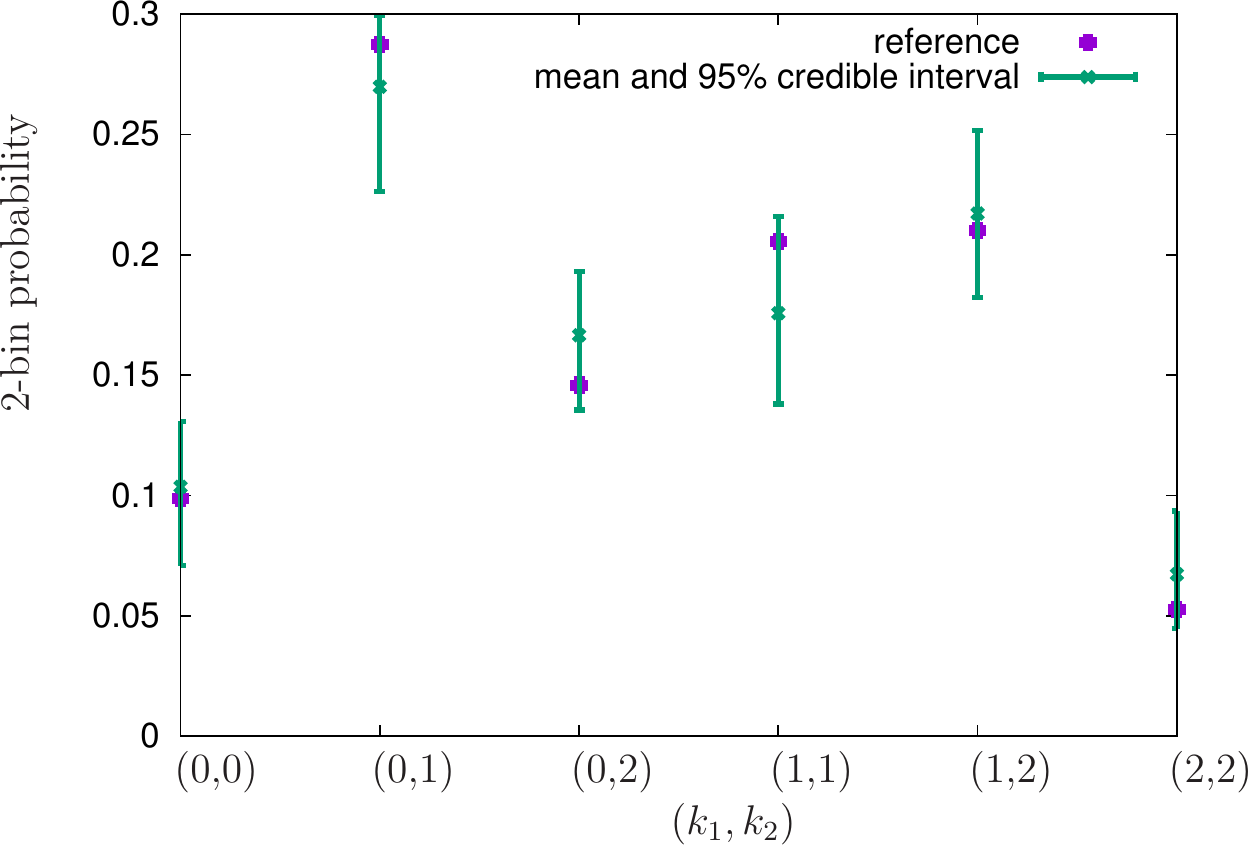}
 \caption{$t=7\Delta t$}
 \end{subfigure}
 \end{minipage}
 \begin{minipage}[c]{0.5\textwidth}
\centering
\begin{subfigure}[b]{.9\textwidth}
\includegraphics[width=\textwidth]{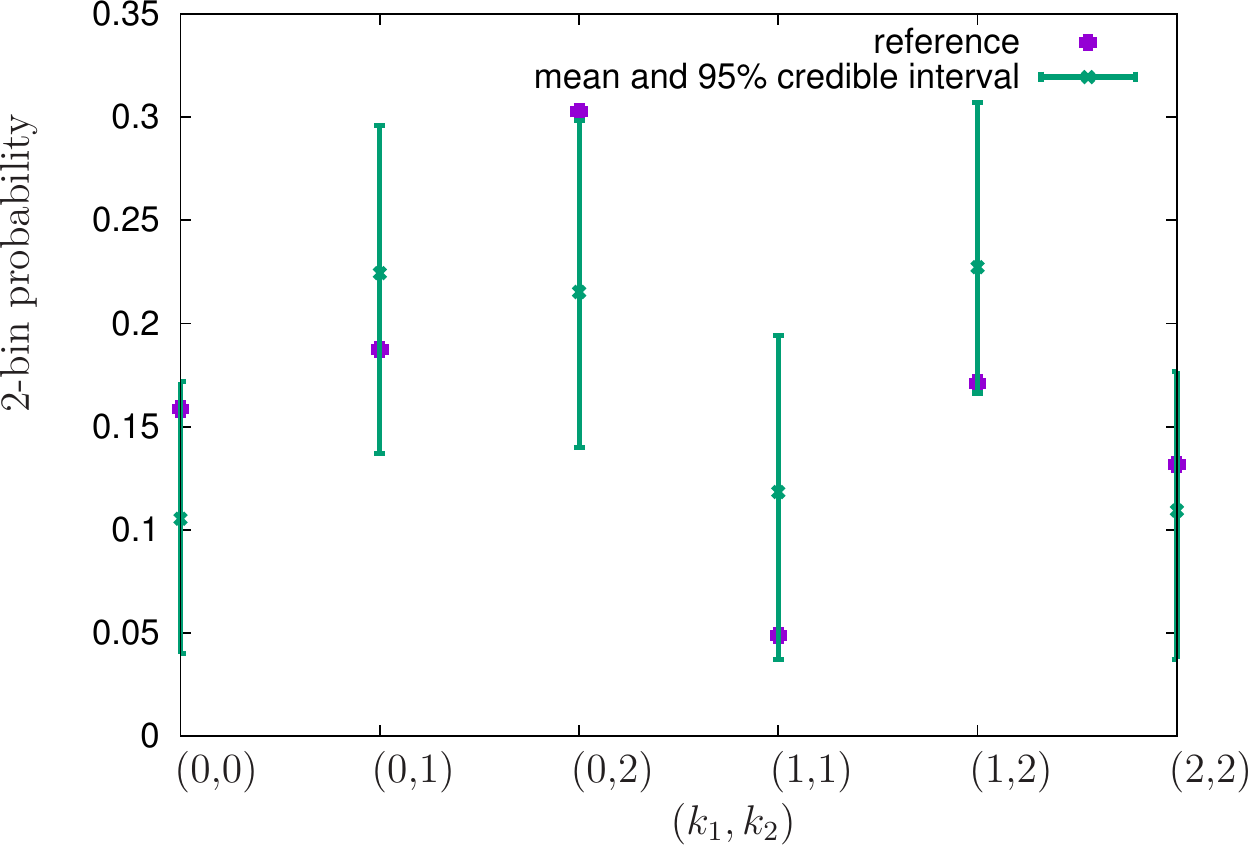}
\caption{$t=2\Delta t$}
 \end{subfigure}
 \begin{subfigure}[b]{.9\textwidth}
\includegraphics[width=\textwidth]{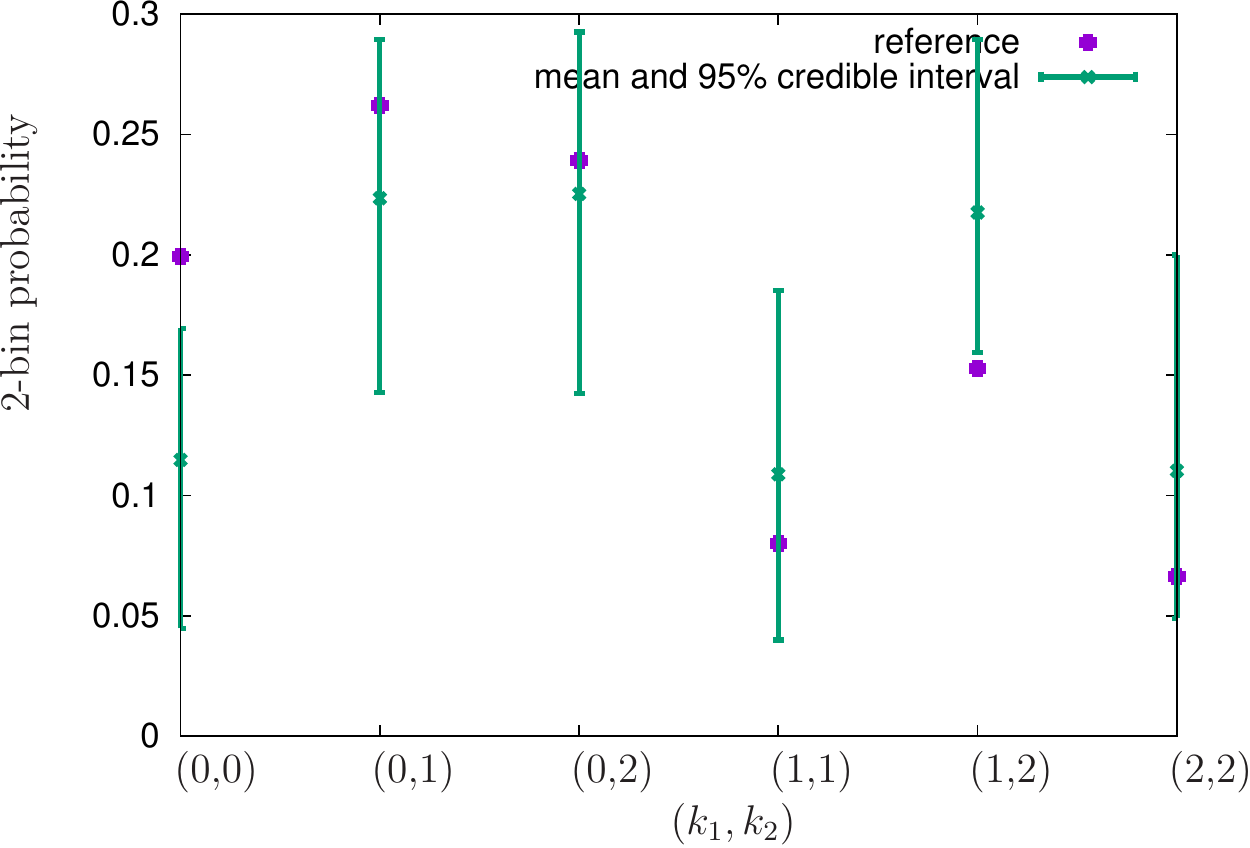}
 \caption{$t=3\Delta t$}
 \end{subfigure}
 \begin{subfigure}[b]{.9\textwidth}
\includegraphics[width=\textwidth]{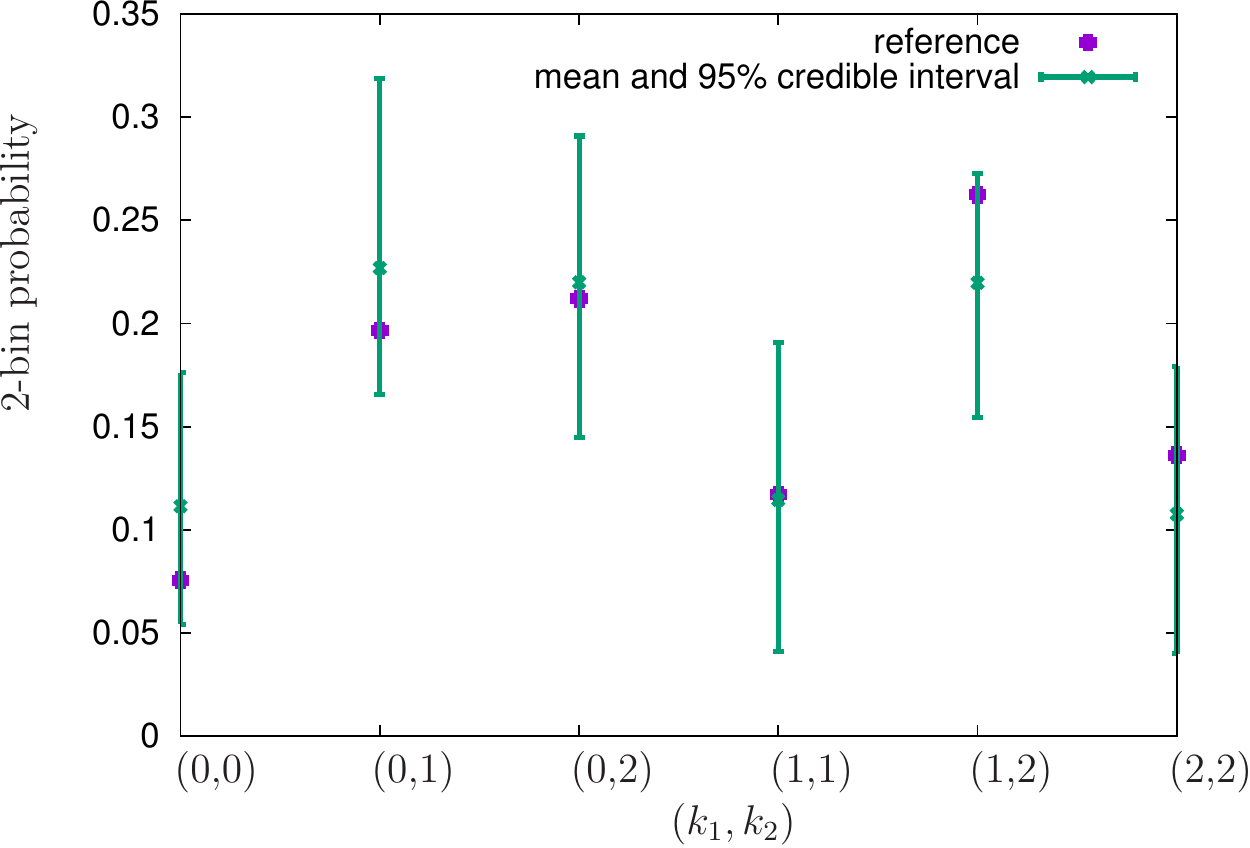}
 \caption{$t=5\Delta t$}
 \end{subfigure}
 \begin{subfigure}[b]{.9\textwidth}
\includegraphics[width=\textwidth]{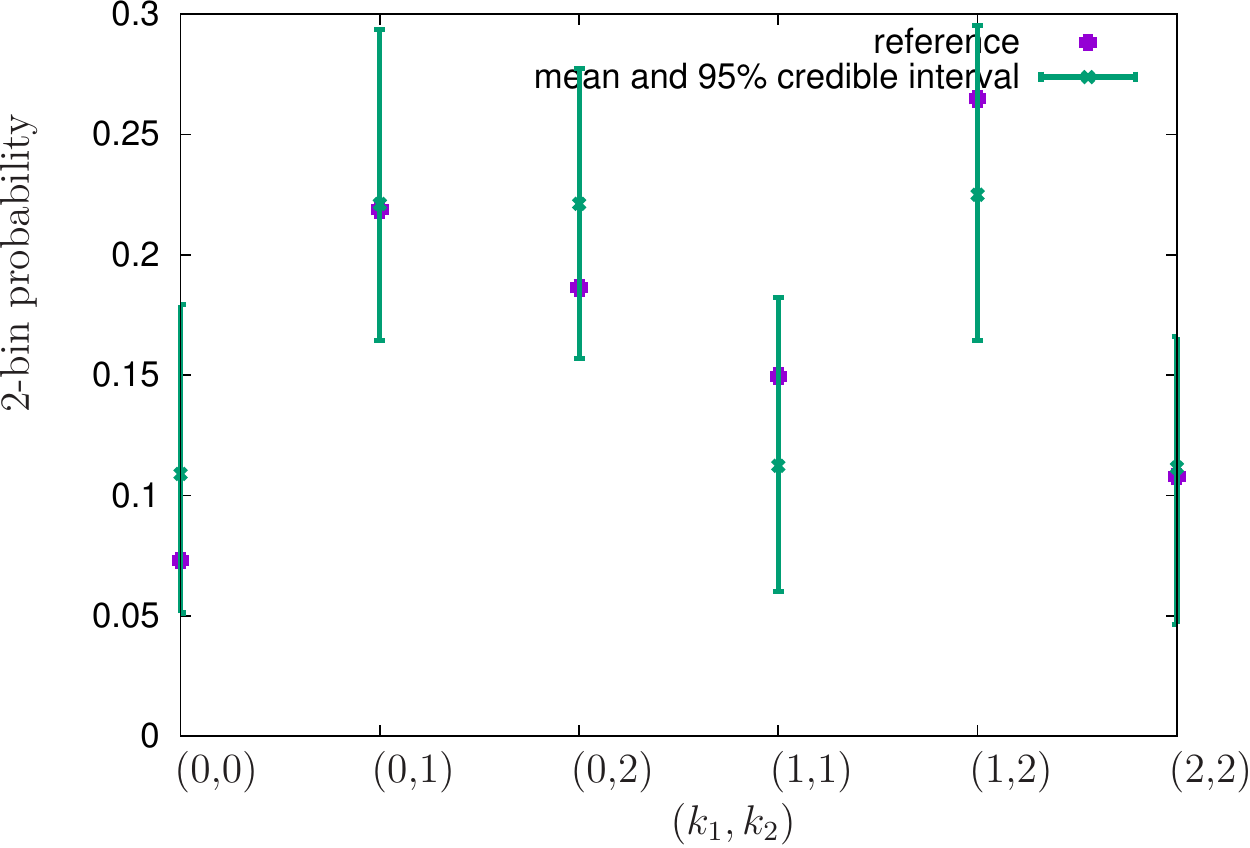}
 \caption{$t=7\Delta t$}
 \end{subfigure}
 \end{minipage}
 \caption{Predictive estimates of 2-bin probability for various times in the future i.e. $t=2\Delta t, 3 \Delta t, 5\Delta t, 7\Delta t$ . The left column corresponds to  the  original binning, whereas the right column to  4 times finer binning.  The reference density profile was computed by simulating the FG model of walkers using the FG time step $\delta t$. ($N=256$) }
 \label{fig:pred2drho}
\end{figure}

Finally, in Figure \ref{fig:pred2drho} we depict predictive estimates pertaining to second-order statistics. In particular, we focus on the probability of finding two walkers, simultaneously, at two bins  $(k_1,k_2)$ (we call this 2-bin probability) and compare reference values (i.e. those obtained by simulating the FG model) with the probabilistic predictive estimates of the proposed model. We do so for some sample values of $(k_1,k_2)$, for different time instants in the future, and for two different numbers of bins i.e. $24$ (left column) and $96$ (right column). Despite the fact that the CG model pertains to the walker density, accurate predictive estimates of the second-order FG statistics can still be obtained. As before, the predictive uncertainty is increased  when a finer binning is employed.

\subsection{Advection-Diffusion example}

The next system  considered consists at the fine-scale of $n_f$ walkers which perform random jumps by $\delta y$ either to left or right with probabilities $k_l \cdot 
\delta t$ and $k_r \cdot \delta  t$ (no interaction). Therefore the probability of staying at the same position is $1-(k_l+k_r) \delta t$. It is well-known (\cite{cottet_vortex_2000}) that, in the limit ($n_f \to \infty, \delta x, \delta t \to 0$), this yields an advection-diffusion process with a diffusion constant $D=(k_l+k_r) \frac{ \delta y^2}{2 \delta t}$ and an advection velocity of $v_a=\frac{k_r-k_l}{\delta t}$ \footnote{In this example, the following values were used $k_l=0.195$, $k_r=0.205$}. Clearly when $ k_r=k_l$  a pure diffusion is obtained.
In the following example, we employed $n_f=2400$ walkers, $\delta y=3.875 \times 10^{-3}$, $\delta t=2.5 \times 10^{-3}$.  The latter two numbers should be compared with the size of the problem domain which is $[-1,1]$ and the CG time-step which was set to $\Delta t =1$ i.e. $400$ times larger than the FG one.

At the CG level we employed $n_c=24$ variables (i.e. $n_f/n_c=100$) and as described in section \ref{sec:train} initialized the walkers using $N$ randomly selected initial densities and propagated them for $400$ FG time-steps $\delta t$ (i.e. one CG time-step $\Delta t$) in order to generate $N$ training data.
Given the availability of closed-form expressions for the evolution   law  of the walker-density $\rho(y,t)$ (in the limit of scale separation), we wish to assess whether the data-driven, identified CG model resembles it. We note that due to the softmax transform (\refeq{eq:softmax}) such a comparison is not straightforward (we write the CG evolution law for $\bxx_t$ and not for the density $\rho(y,t)$). In this investigation we employed the same first- and second-order  feature function (Equations (\ref{eq:phi1order}), (\ref{eq:phi2order})) for $M=6$. This gave rise to $182$ features in total and the same number of coefficients $\bt_c$ which emphasizes the need for potent model selection tools.

\begin{figure}
\includegraphics[width=.5\textwidth]{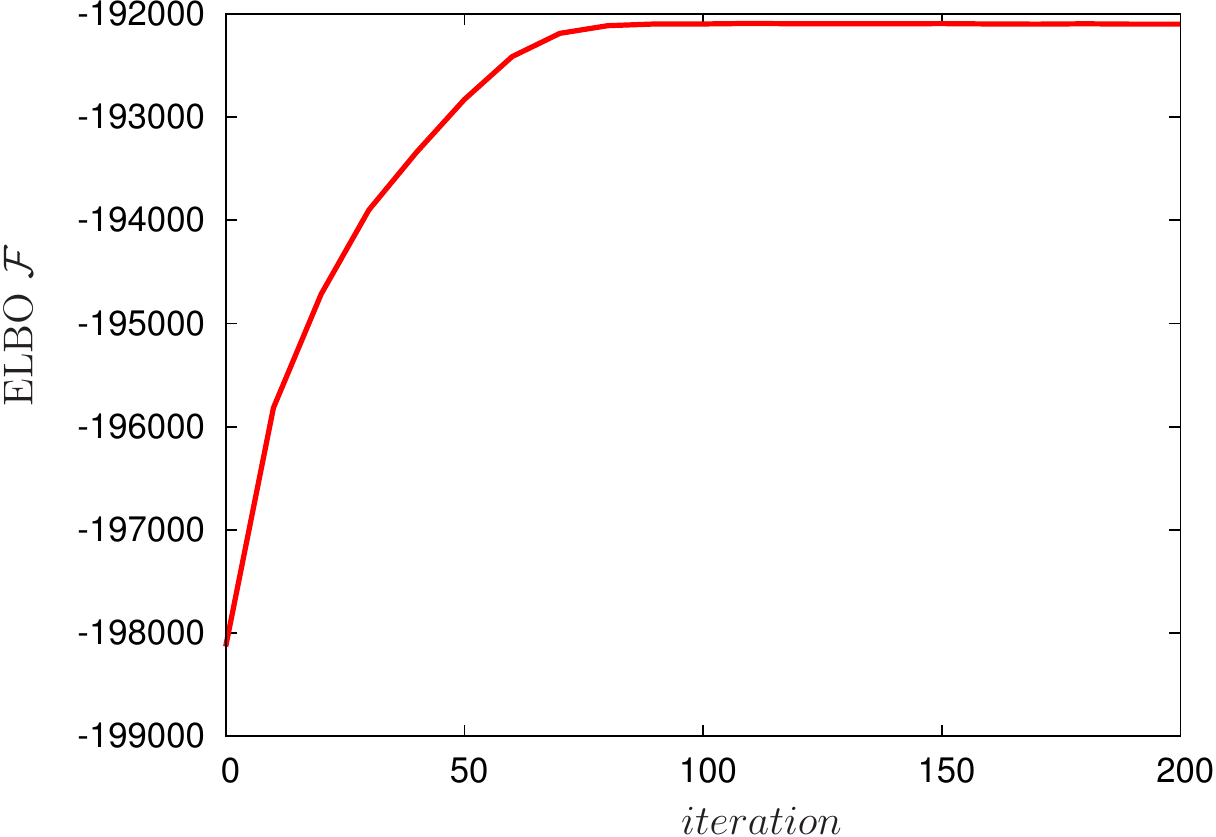}
 \caption{Evolution  of the ELBO  $\mathcal{F}$ as estimated at each iteration based on \refeq{eq:elbofinal} for $N=64$.}
 \label{fig:adelbo}
\end{figure}

\begin{figure}
\begin{minipage}[c]{0.3\textwidth}
\centering
\begin{subfigure}[b]{\textwidth}
\includegraphics[width=\textwidth]{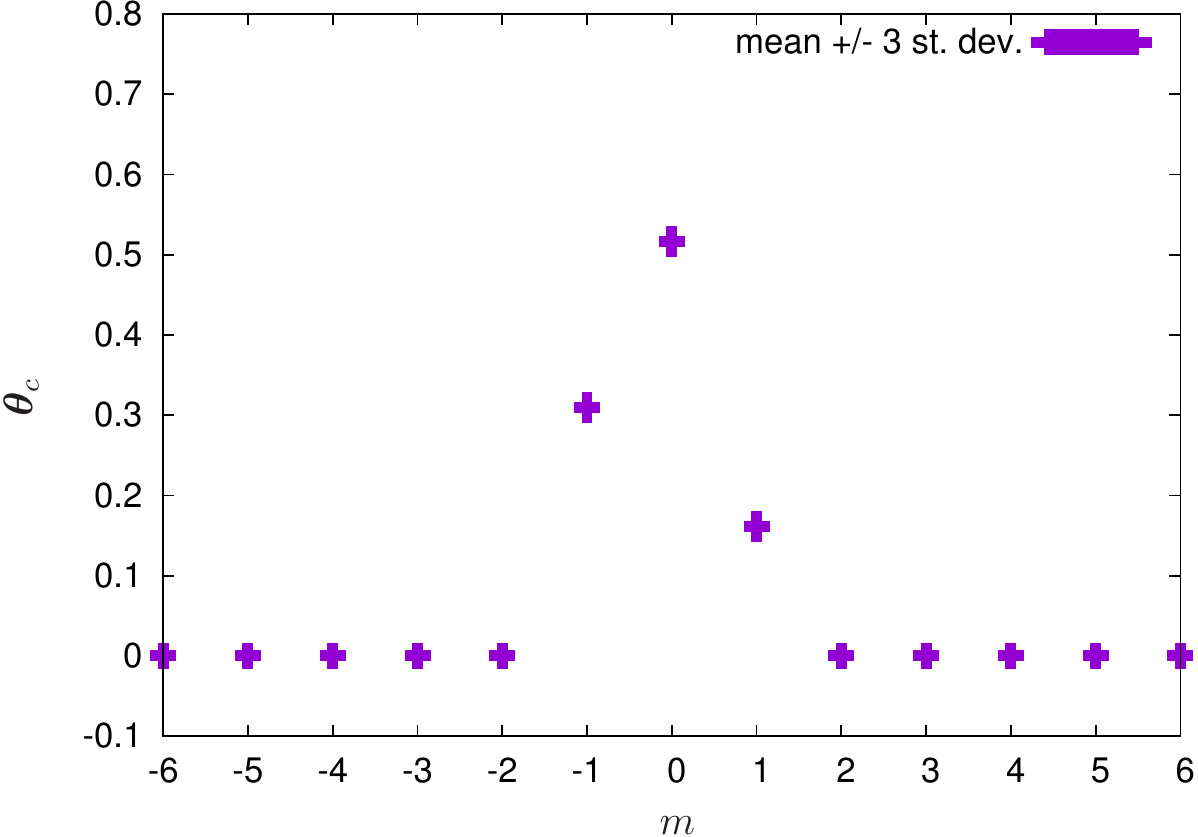}
\caption{$N=32$}
\end{subfigure}
\end{minipage}
\begin{minipage}[c]{0.3\textwidth}
\centering
\begin{subfigure}[b]{\textwidth}
\includegraphics[width=\textwidth]{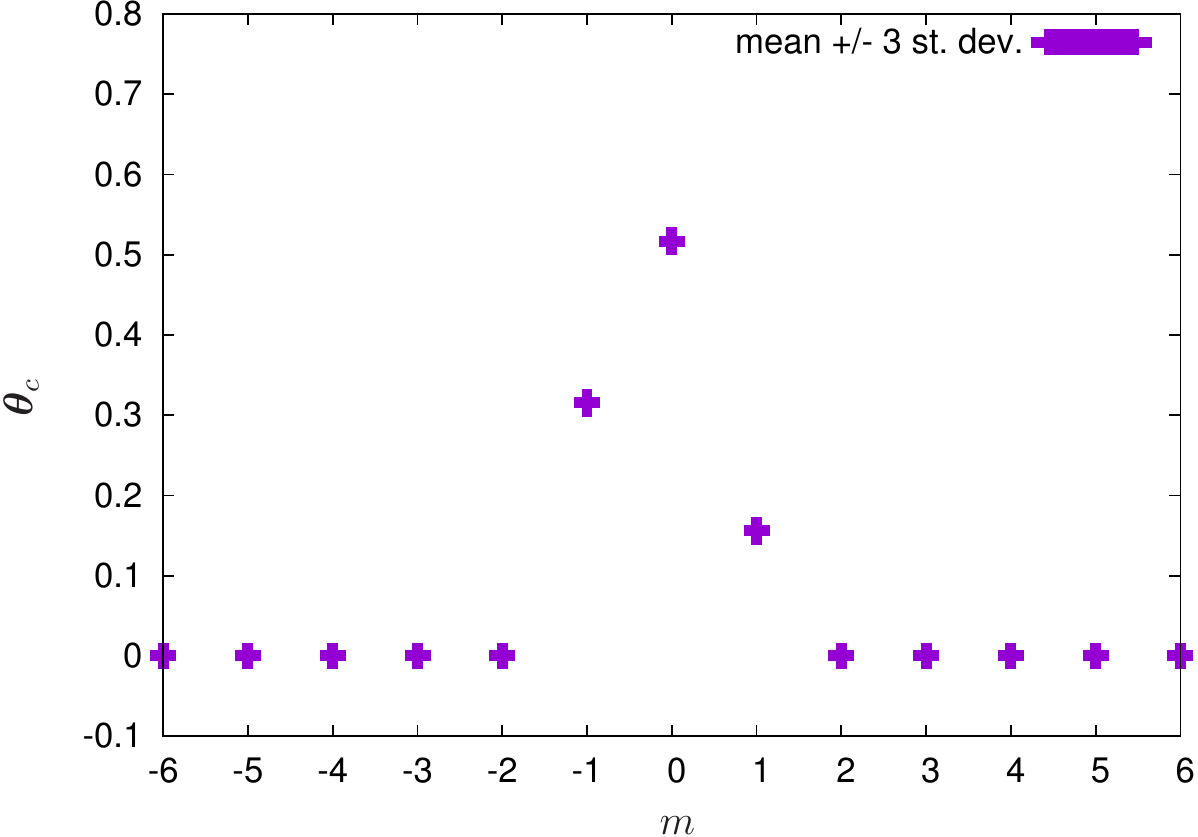}
\caption{$N=64$}
\end{subfigure}
\end{minipage}
\begin{minipage}[c]{0.3\textwidth}
\centering
\begin{subfigure}[b]{\textwidth}
\includegraphics[width=\textwidth]{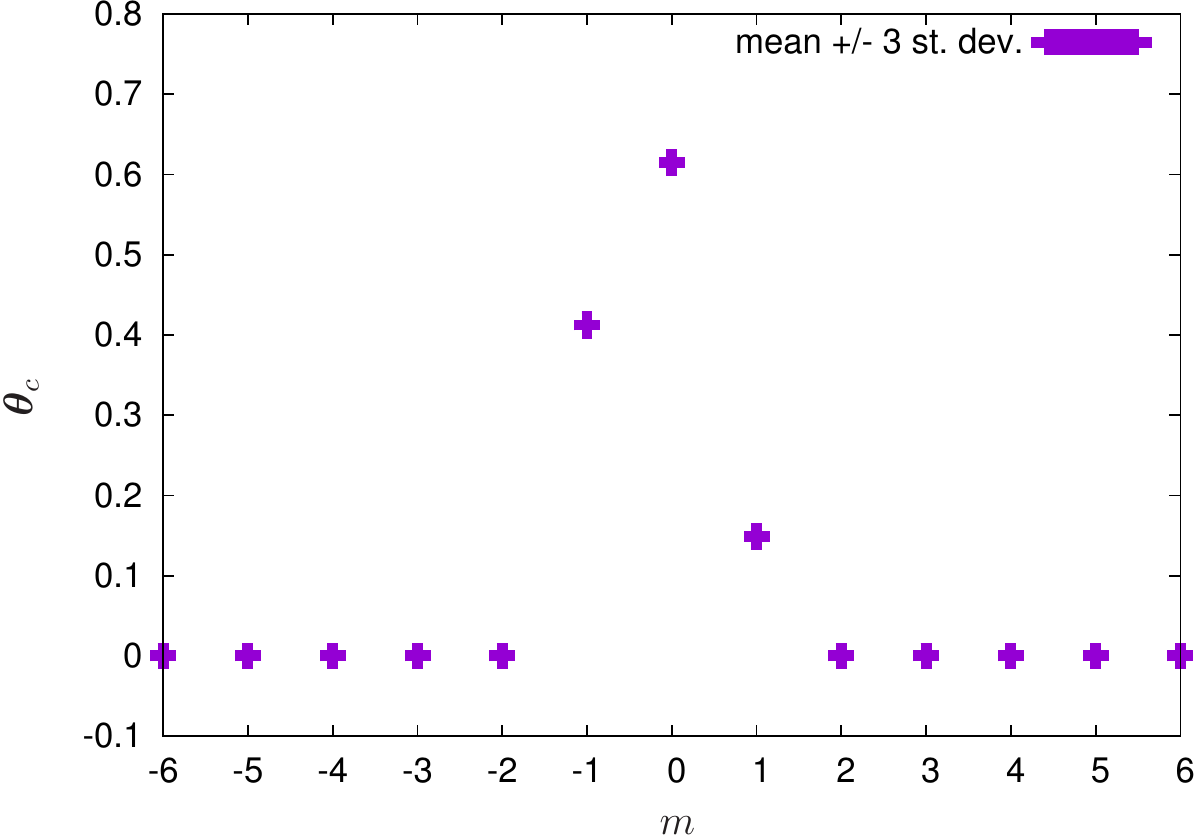}
\caption{$N=128$}
\end{subfigure}
\end{minipage}
 \caption{Posterior estimates for $N=64$ of $\bt_c$ associated with first-order feature functions (\refeq{eq:phi1order}).}
 \label{fig:adpostalpha}
\end{figure}

Figure \ref{fig:adelbo} depicts the evolution   of the ELBO  $\mathcal{F}$ as estimated at each iteration based on \refeq{eq:elbofinal} for $N=64$ training data sequences.
Figure \ref{fig:adpostalpha} depicts the posterior mean and standard deviation of the $\bt_c$ associated with first-order feature functions for various values of $m$ (\refeq{eq:phi1order}) and for three different data-sizes $N=32,~64, ~128$. We note that the only active ones are for   $m=0$ and $m=\pm 1$ which effectively correspond to a finite difference approximation of spatial  derivatives up to order 2. This is consistent with finite difference schemes of the advection-diffusion equation. Furthermore the asymmetry of the values for $m=\pm 1$ is consistent with the advective term of  the equation. The aforementioned characteristics are identified correctly even with $N=32$ and, as expected, the posterior uncertainty reduces as $N$ increases. Parameters $\bt_c$ associated with second-order features were all deactivated by the algorithm and are therefore omitted in the illustrations. 
%
%

\begin{figure}[!h]
\begin{minipage}[c]{0.5\textwidth}
\centering
\begin{subfigure}[b]{.9\textwidth}
\includegraphics[width=\textwidth]{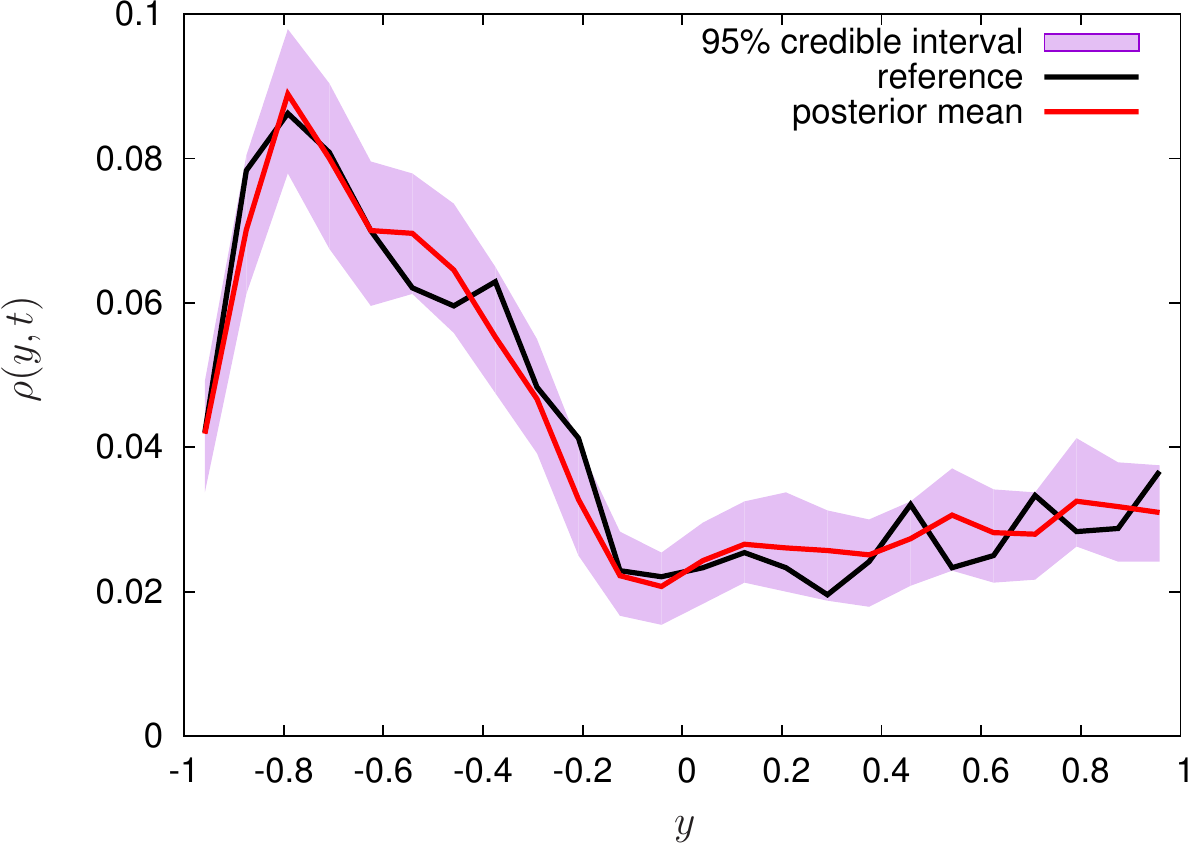}
 \caption{$t=2\Delta t$}
 \end{subfigure}
 \begin{subfigure}[b]{.9\textwidth}
\includegraphics[width=\textwidth]{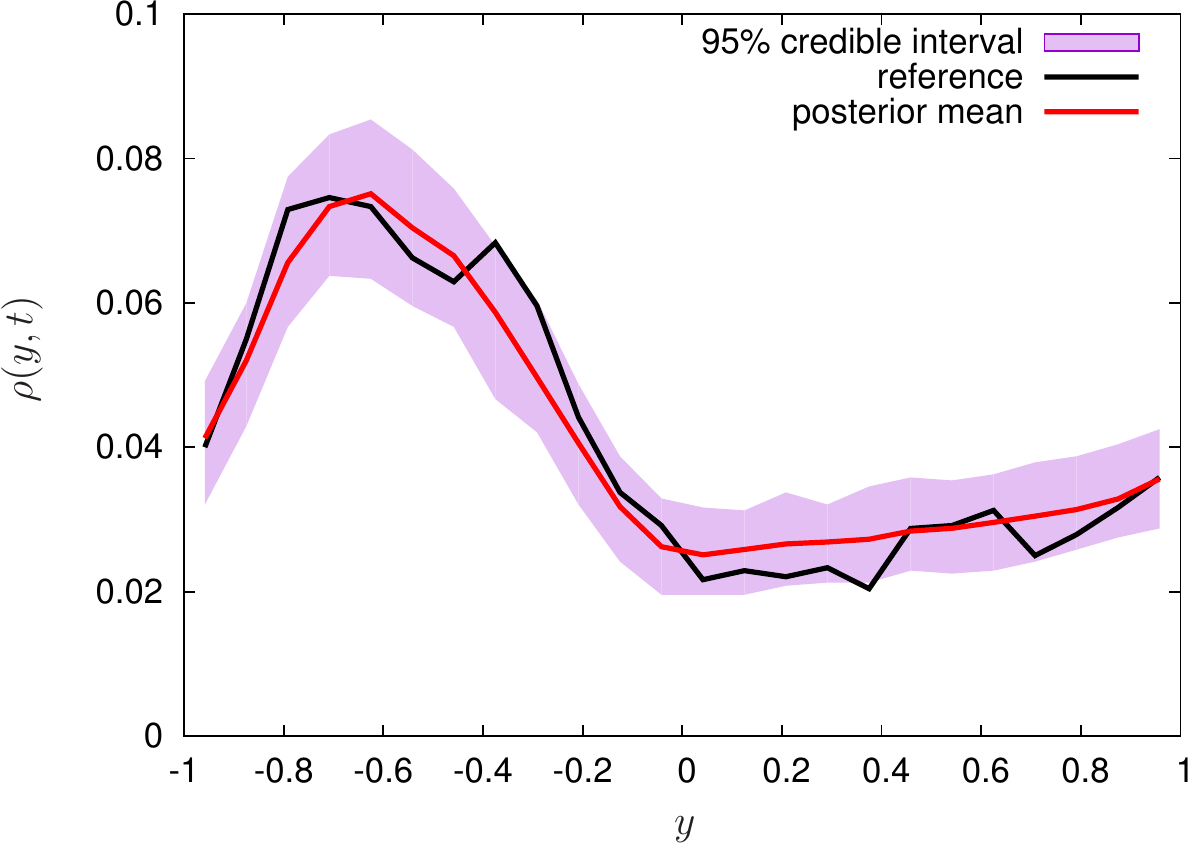}
 \caption{$t=6\Delta t$}
 \end{subfigure}
 \begin{subfigure}[b]{.9\textwidth}
\includegraphics[width=\textwidth]{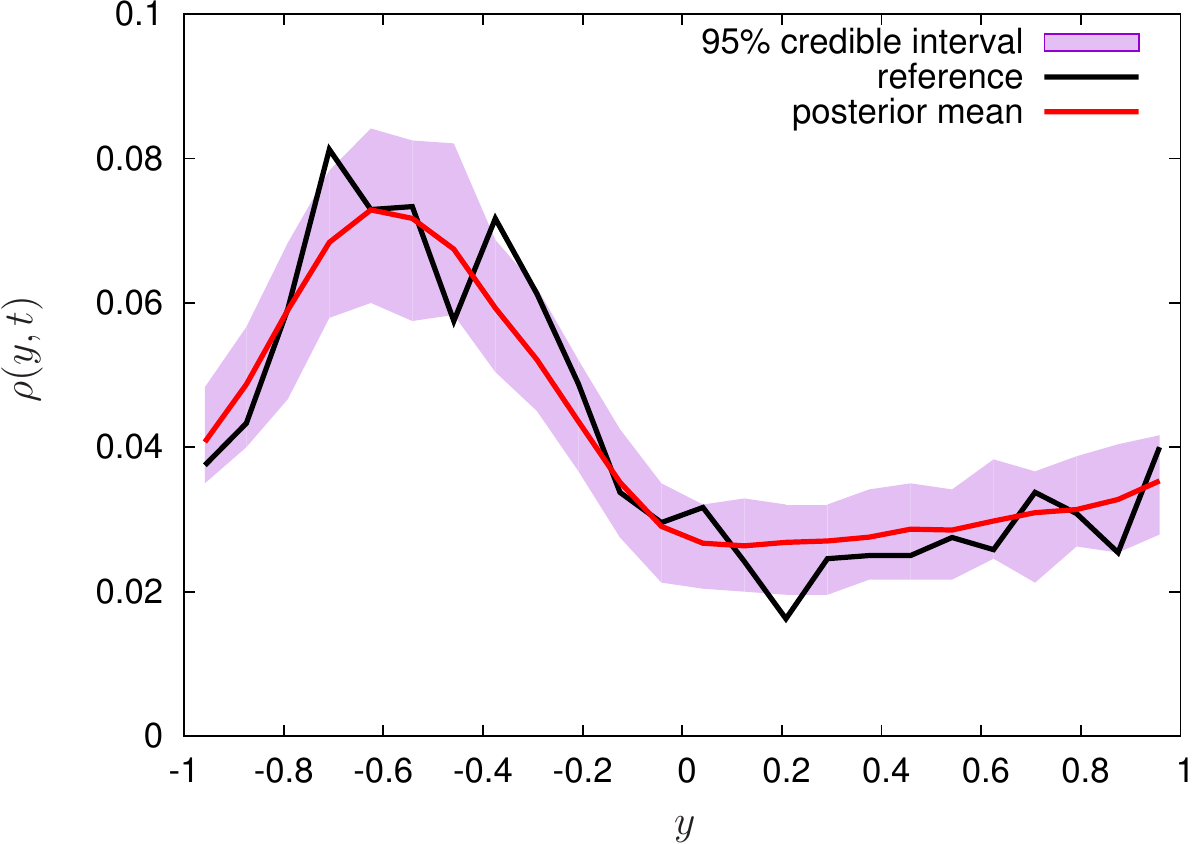}
 \caption{$t=8\Delta t$}
 \end{subfigure}
 \begin{subfigure}[b]{.9\textwidth}
\includegraphics[width=\textwidth]{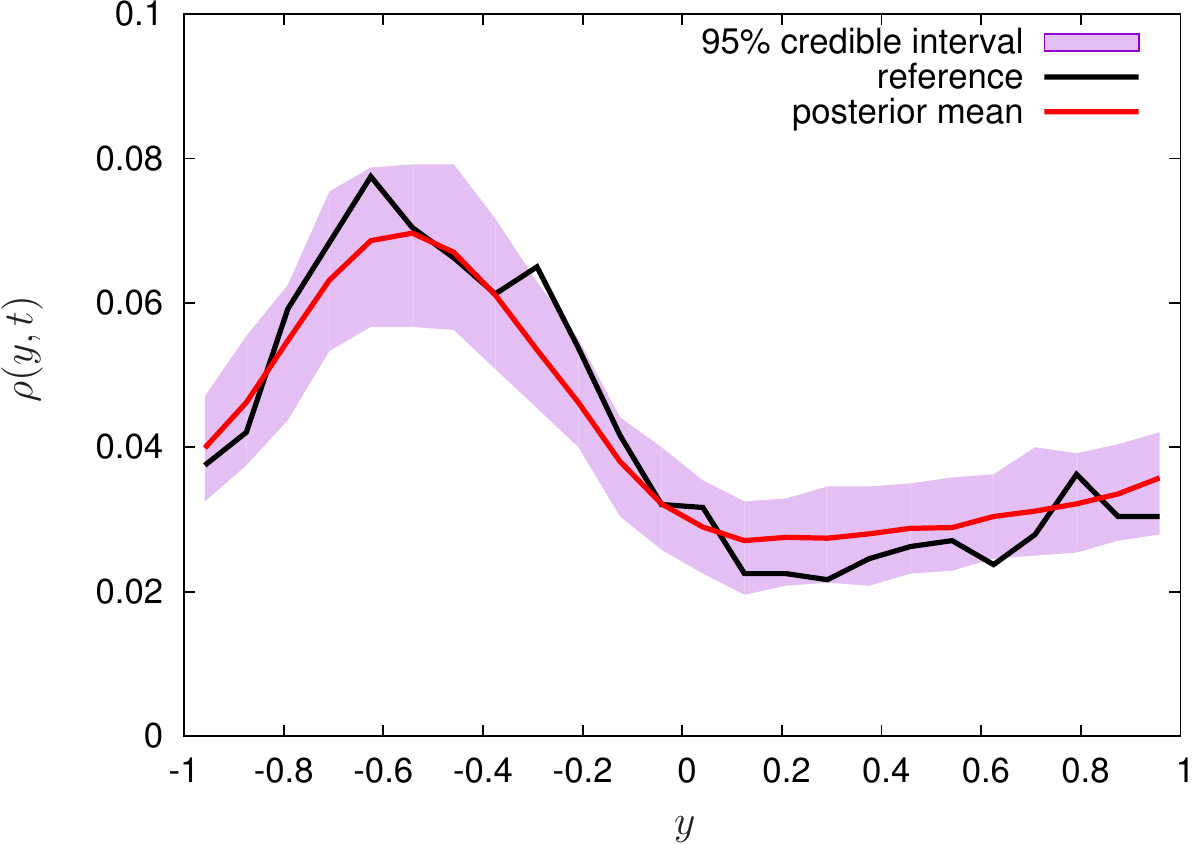}
 \caption{$t=10\Delta t$}
 \end{subfigure}
 \end{minipage}
 \begin{minipage}[c]{0.5\textwidth}
\centering
\begin{subfigure}[b]{.9\textwidth}
\includegraphics[width=\textwidth]{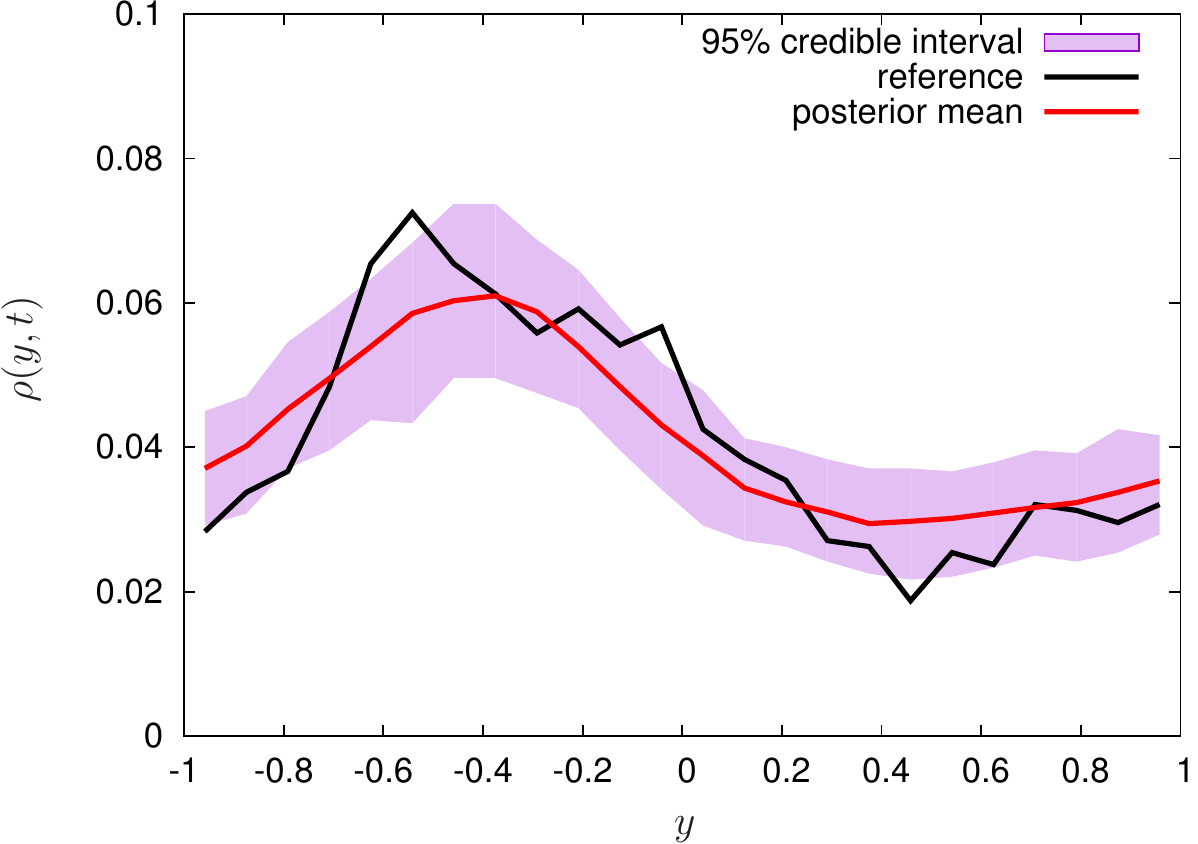}
 \caption{$t=20\Delta t$}
 \end{subfigure}
 \begin{subfigure}[b]{.9\textwidth}
\includegraphics[width=\textwidth]{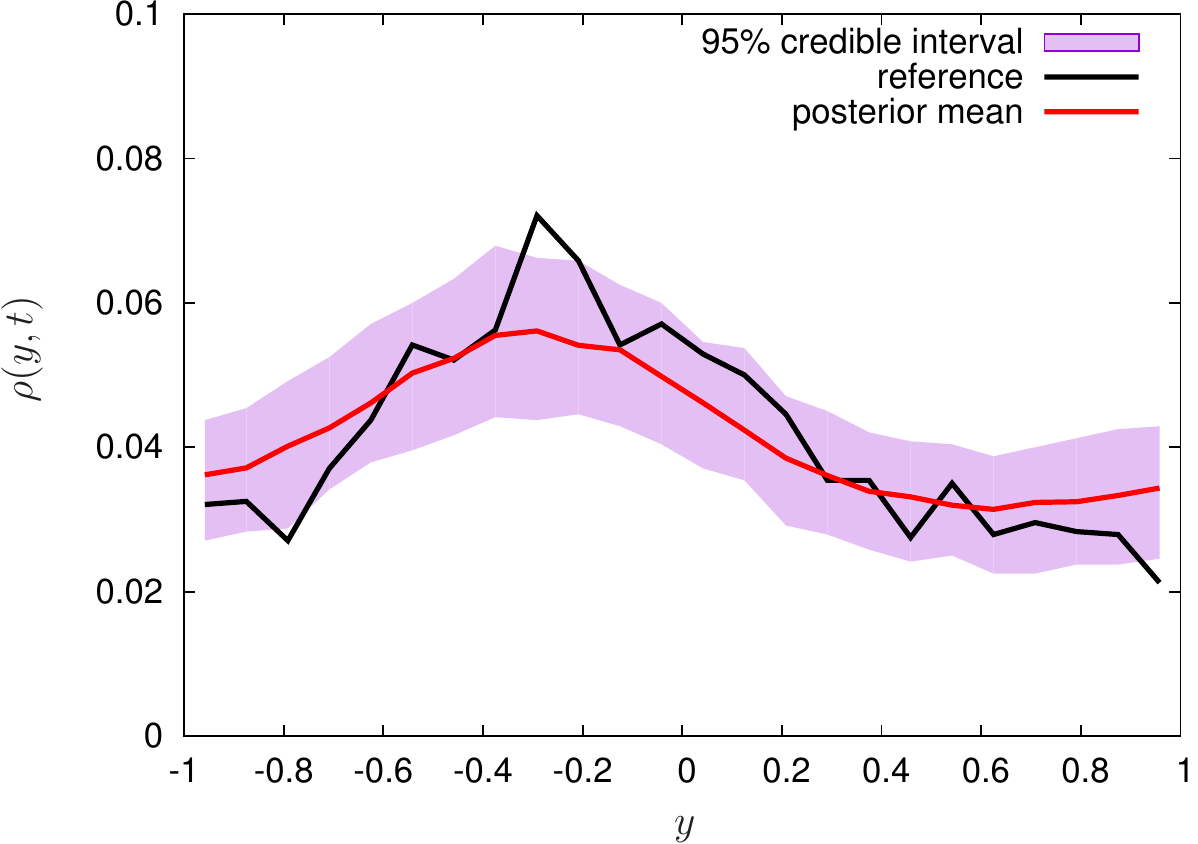}
 \caption{$t=30\Delta t$}
 \end{subfigure}
 \begin{subfigure}[b]{.9\textwidth}
\includegraphics[width=\textwidth]{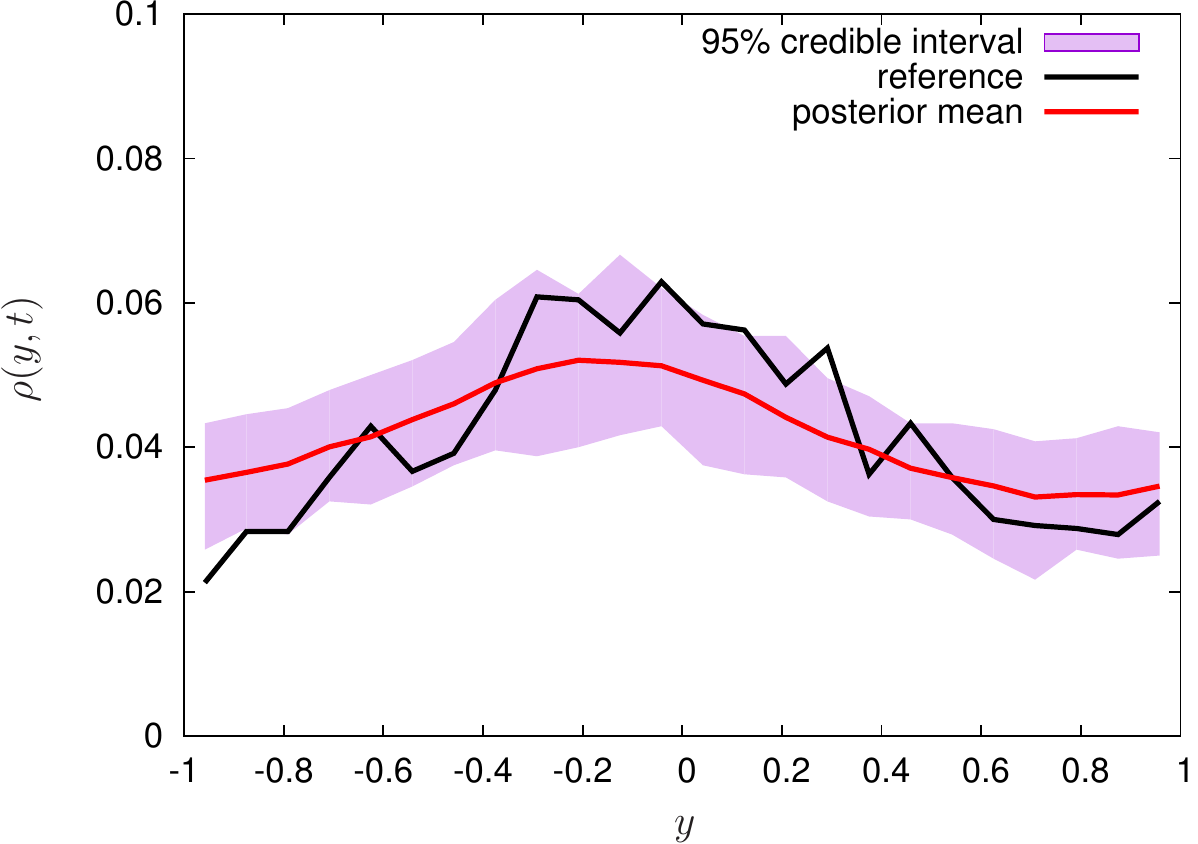}
 \caption{$t=40\Delta t$}
 \end{subfigure}
 \begin{subfigure}[b]{.9\textwidth}
\includegraphics[width=\textwidth]{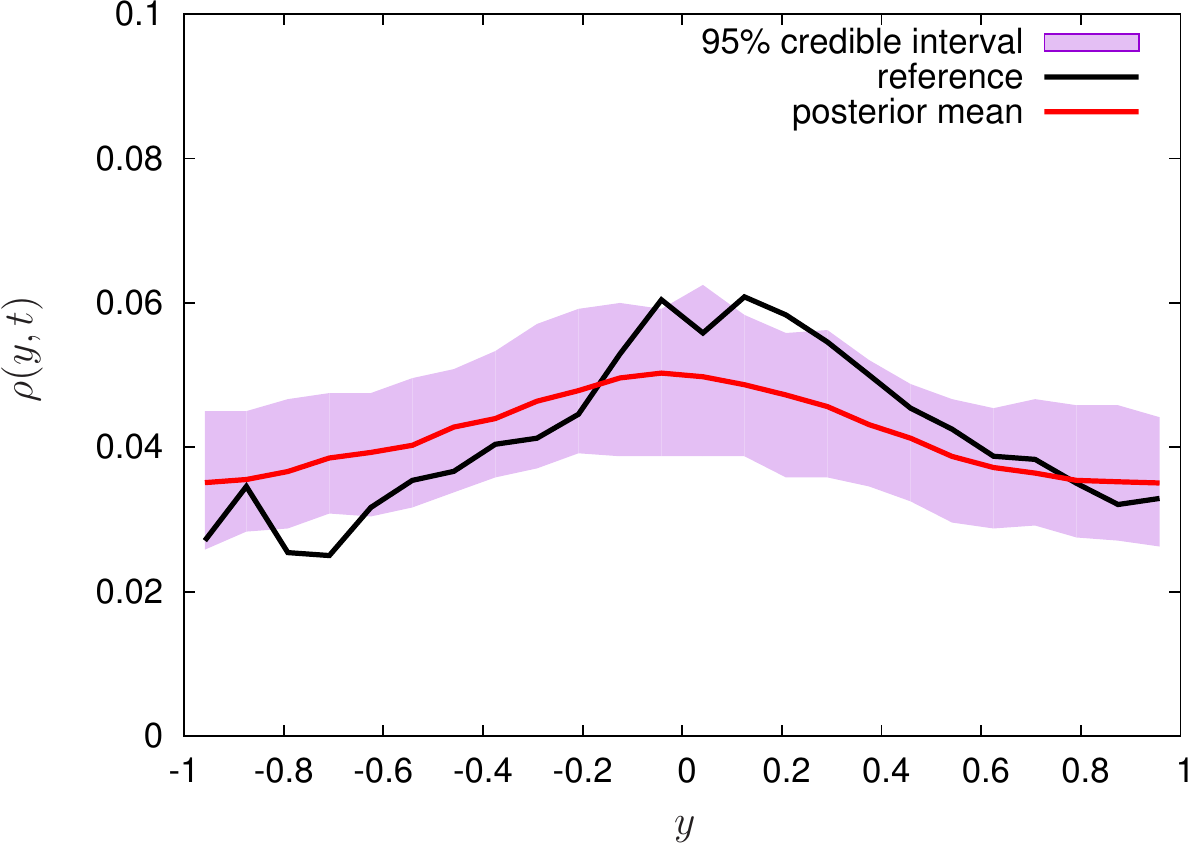}
 \caption{$t=50\Delta t$}
  \end{subfigure}
 \end{minipage}
 \caption{Predictive estimates of walker density  for $24$ bins (same as in CG evolution) for various future  times. The reference density profile was computed by simulating the FG model of walkers using the FG time step $\delta t$. ($N=64$) }
 \label{fig:adpredrho}
\end{figure}

\begin{figure}[!h]
\begin{minipage}[c]{0.5\textwidth}
\centering
\begin{subfigure}[b]{.9\textwidth}
\includegraphics[width=\textwidth]{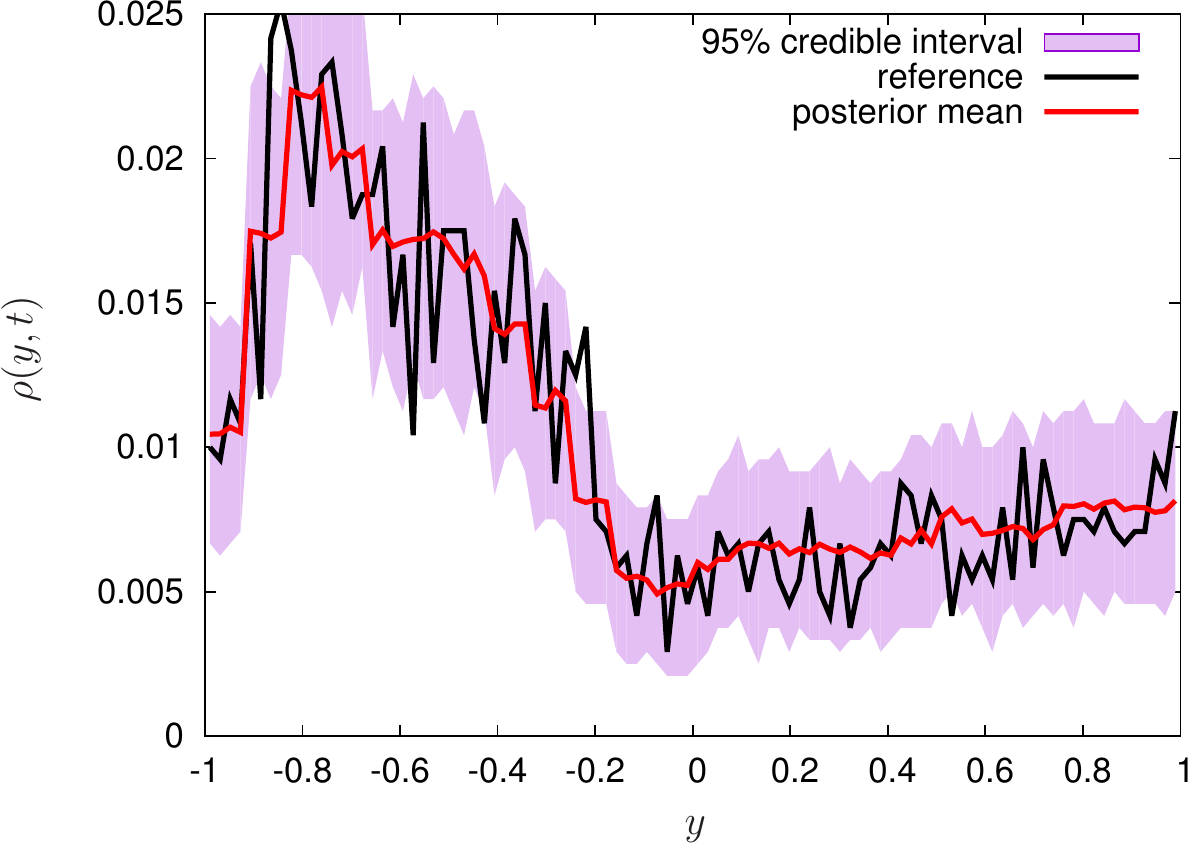}
 \caption{$t=2\Delta t$}
 \end{subfigure}
 \begin{subfigure}[b]{.9\textwidth}
\includegraphics[width=\textwidth]{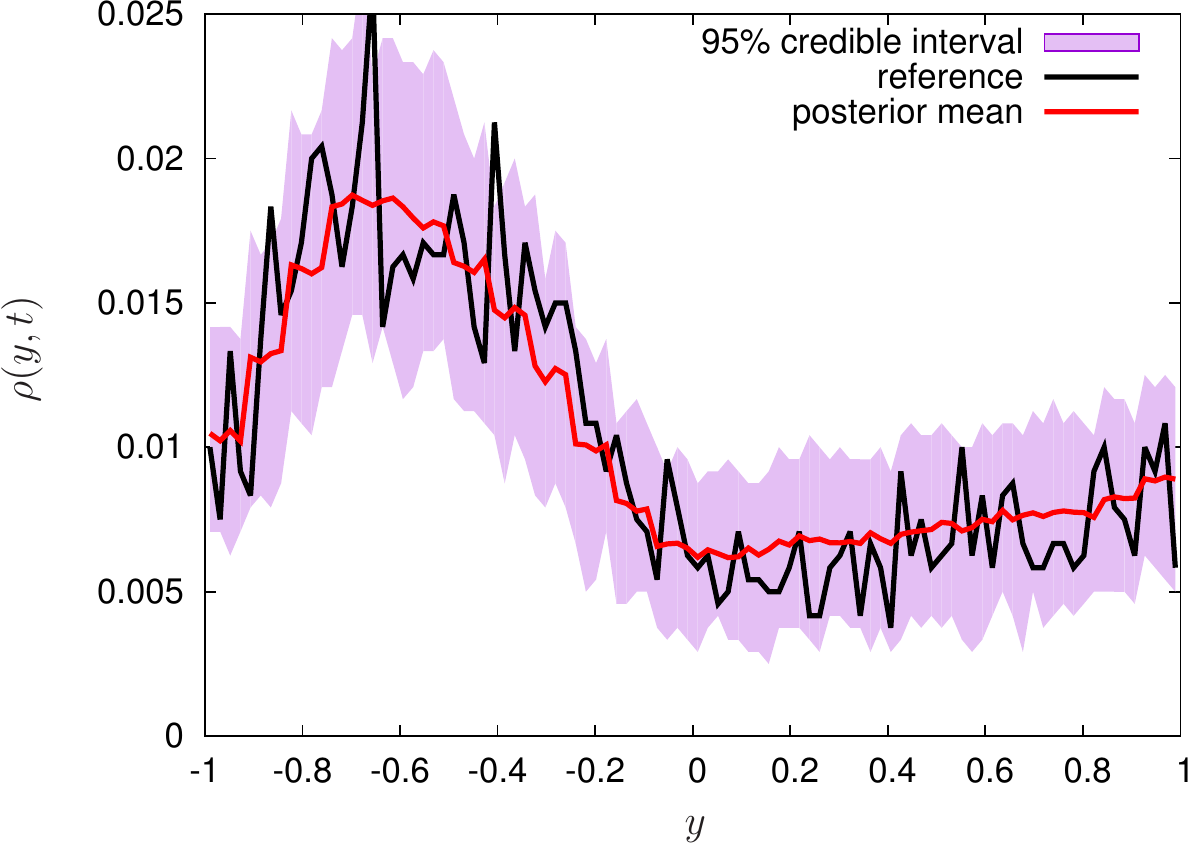}
 \caption{$t=6\Delta t$}
 \end{subfigure}
 \begin{subfigure}[b]{.9\textwidth}
\includegraphics[width=\textwidth]{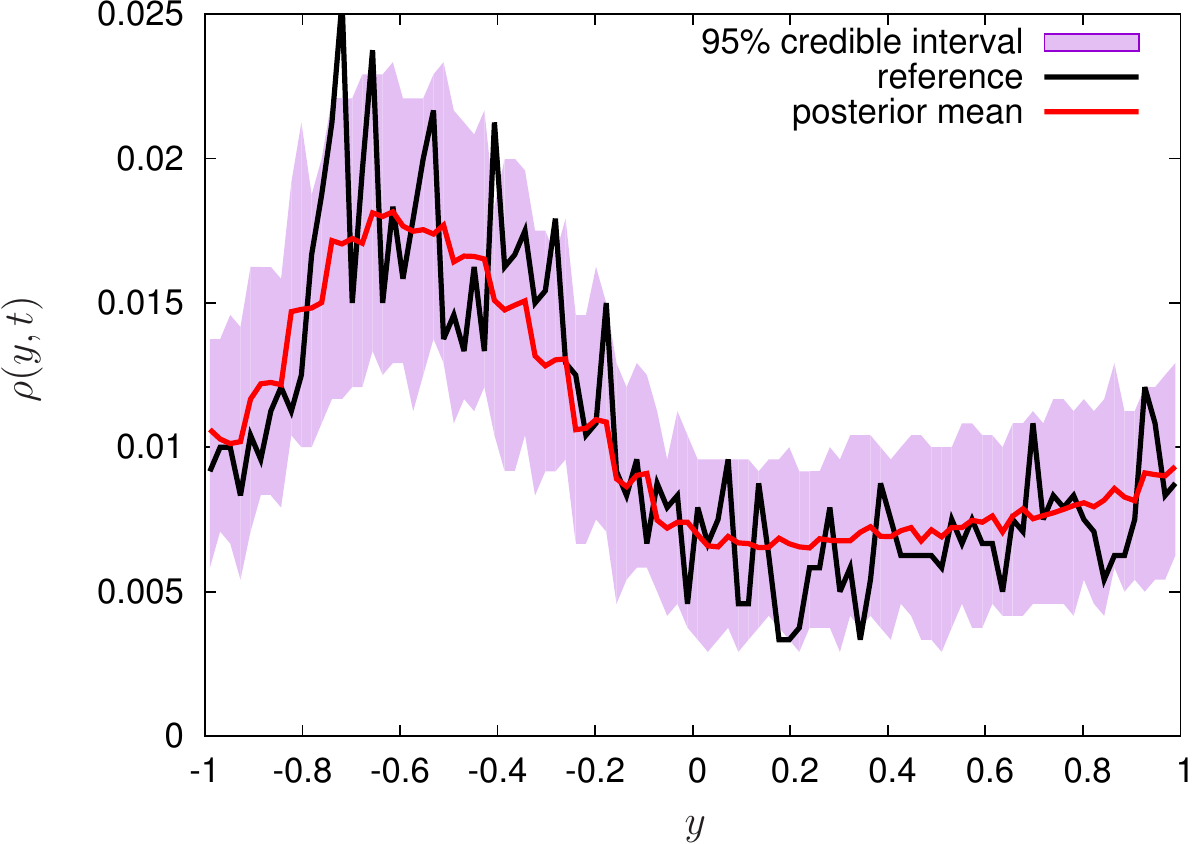}
 \caption{$t=8\Delta t$}
 \end{subfigure}
 \begin{subfigure}[b]{.9\textwidth}
\includegraphics[width=\textwidth]{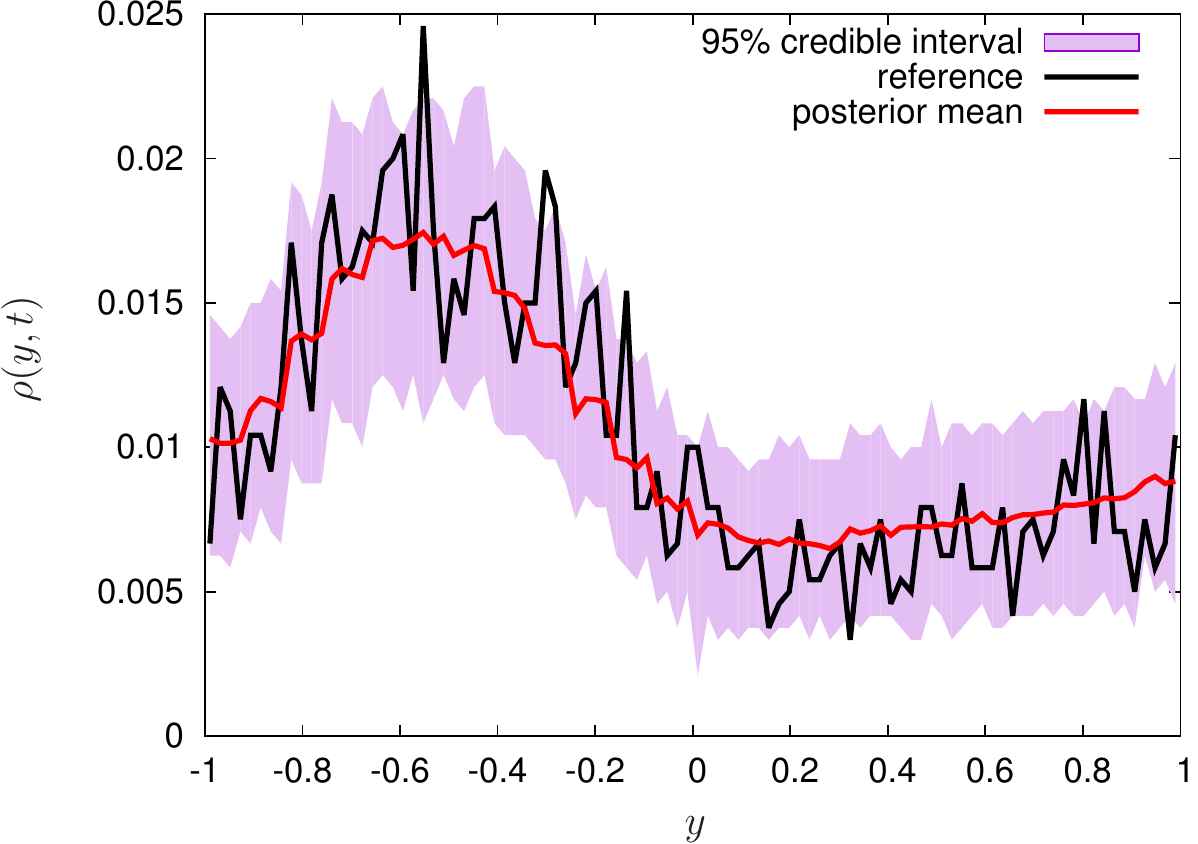}
 \caption{$t=10\Delta t$}
 \end{subfigure}
 \end{minipage}
 \begin{minipage}[c]{0.5\textwidth}
\centering
\begin{subfigure}[b]{.9\textwidth}
\includegraphics[width=\textwidth]{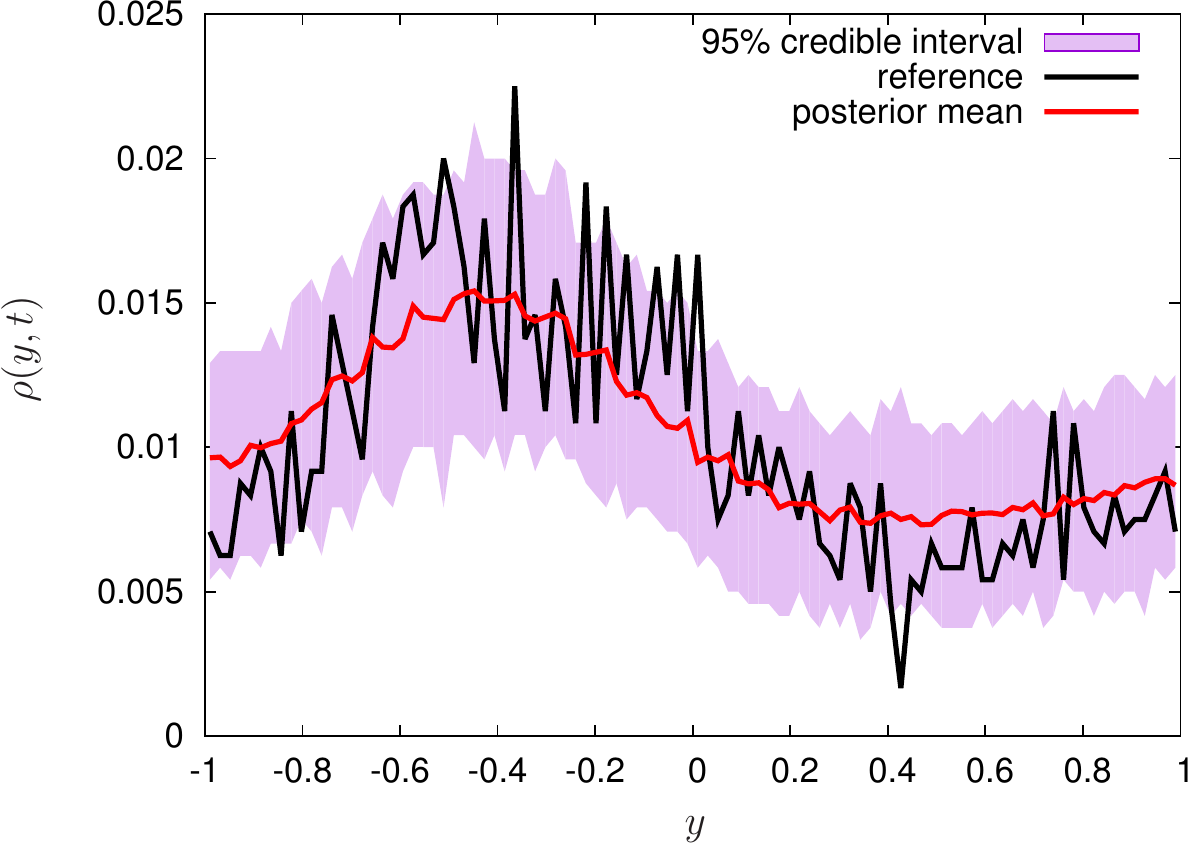}
 \caption{$t=20\Delta t$}
 \end{subfigure}
 \begin{subfigure}[b]{.9\textwidth}
\includegraphics[width=\textwidth]{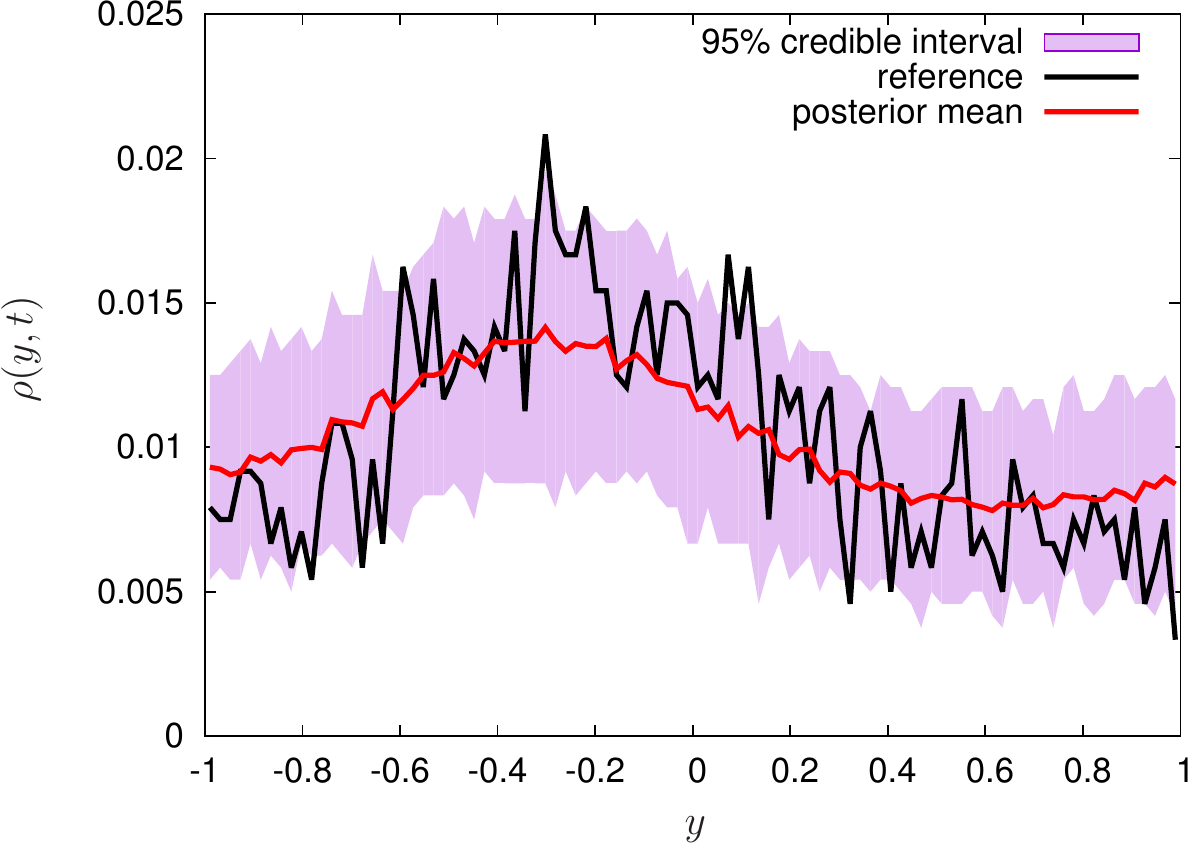}
 \caption{$t=30\Delta t$}
 \end{subfigure}
 \begin{subfigure}[b]{.9\textwidth}
\includegraphics[width=\textwidth]{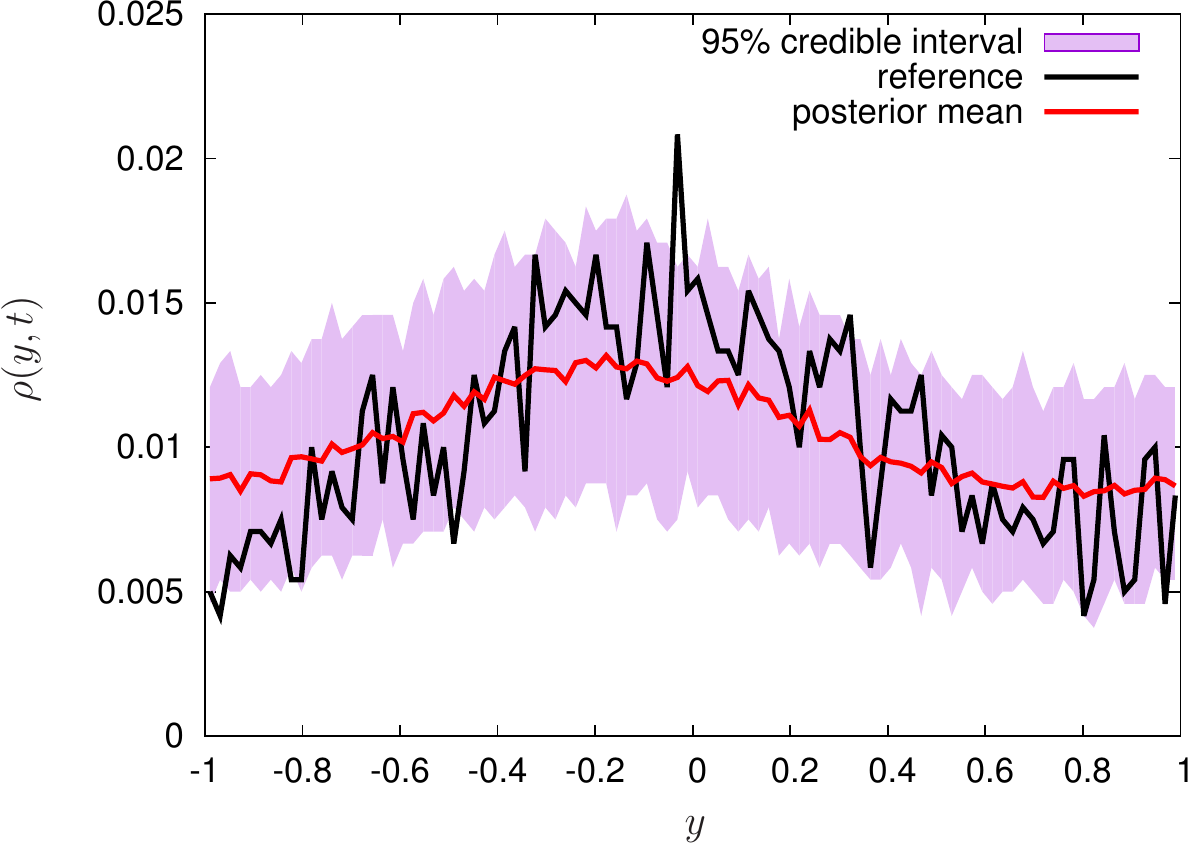}
 \caption{$t=40\Delta t$}
 \end{subfigure}
 \begin{subfigure}[b]{.9\textwidth}
\includegraphics[width=\textwidth]{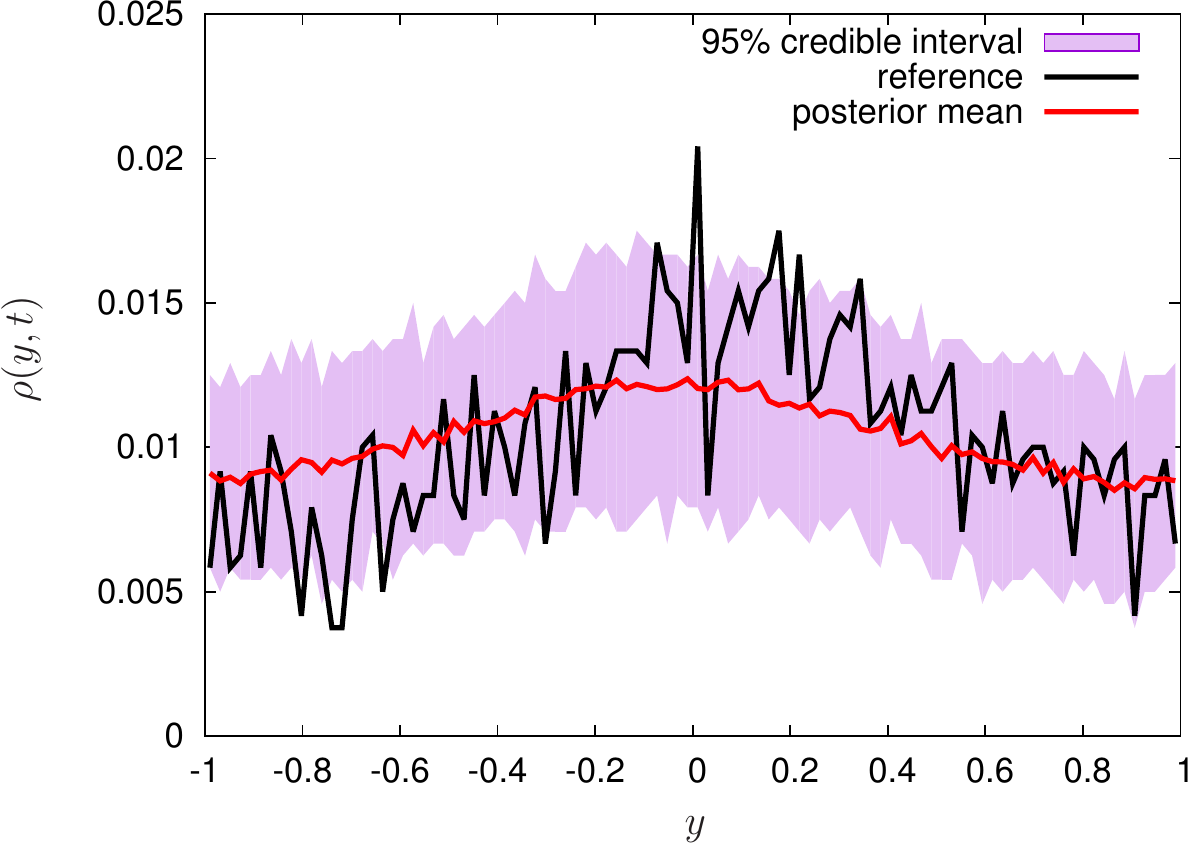}
 \caption{$t=50\Delta t$}
  \end{subfigure}
 \end{minipage}
 \caption{Predictive estimates of walker density {\em for $96$ bins} ($4$ times greater than the $24$ used in CG evolution), for various times in the future. The reference density profile was computed by simulating the FG model of walkers using the FG time step $\delta t$.  ($N=64$)}
 \label{fig:adpredrhofine4}
\end{figure}

In Figure \ref{fig:adpredrho}, we compare the evolution of the walker density at various future times with the predictive posterior estimates (section \ref{sec:probpred}) of our model. One observes that despite the fluctuations at the fine-scale, the CG model learned from $N=64$ training data is able to capture the basic trend of the density profile and that  the 95\% credible interval almost always envelops the ground truth. This is true as shown  even for future times up to $50 \Delta t$ (i.e. $20000 \delta t$) even though the training data consisted of a single $\Delta t$.  In Figure \ref{fig:adpredrhofine4} we perform the same comparison but for a 4-times finer binning. As in the synthetic example previously, very good predictive accuracy is attained despite the wider credible intervals which can be attributed to the  more detailed character of the observable of interest.

\begin{figure}[!h]
\begin{minipage}[c]{0.5\textwidth}
\centering
\begin{subfigure}[b]{.9\textwidth}
\includegraphics[width=\textwidth]{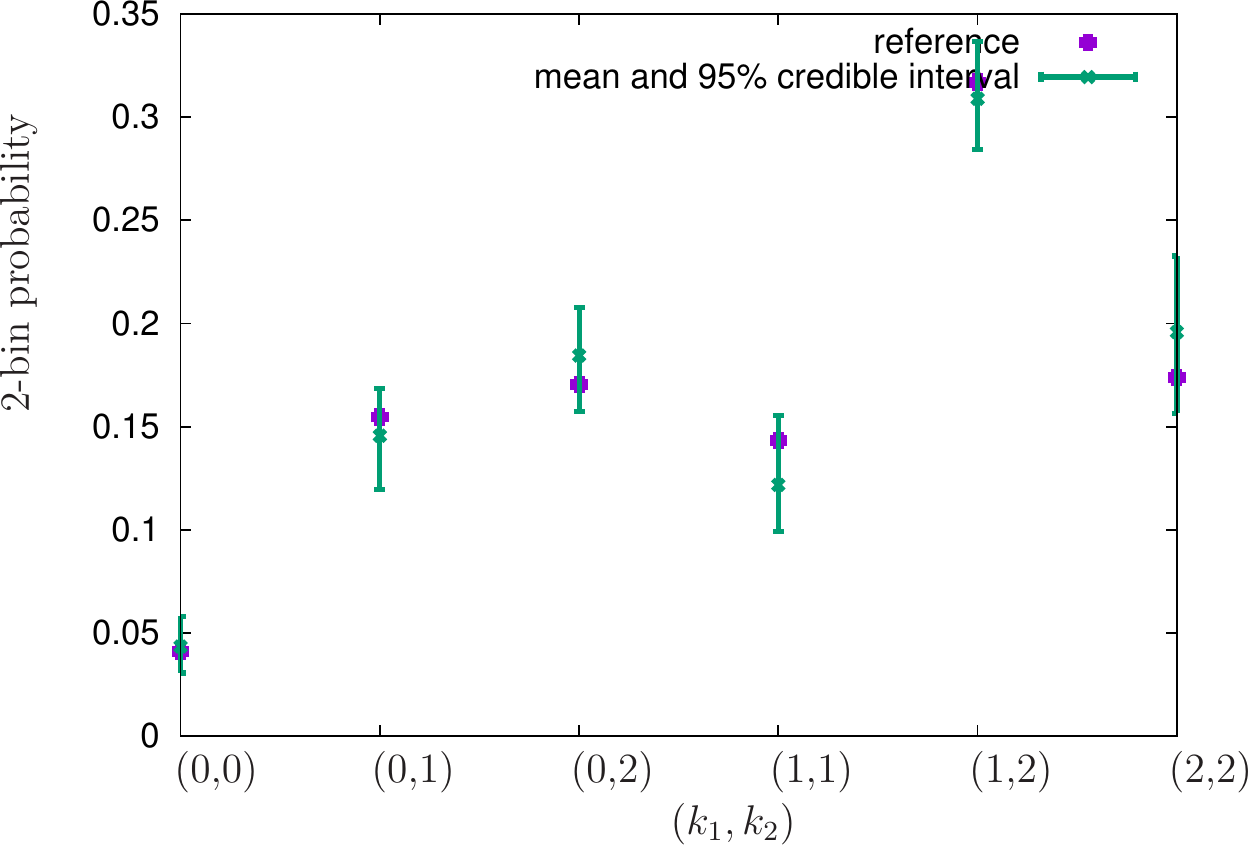}
\caption{$t=2\Delta t$}
 \end{subfigure}
 \begin{subfigure}[b]{.9\textwidth}
\includegraphics[width=\textwidth]{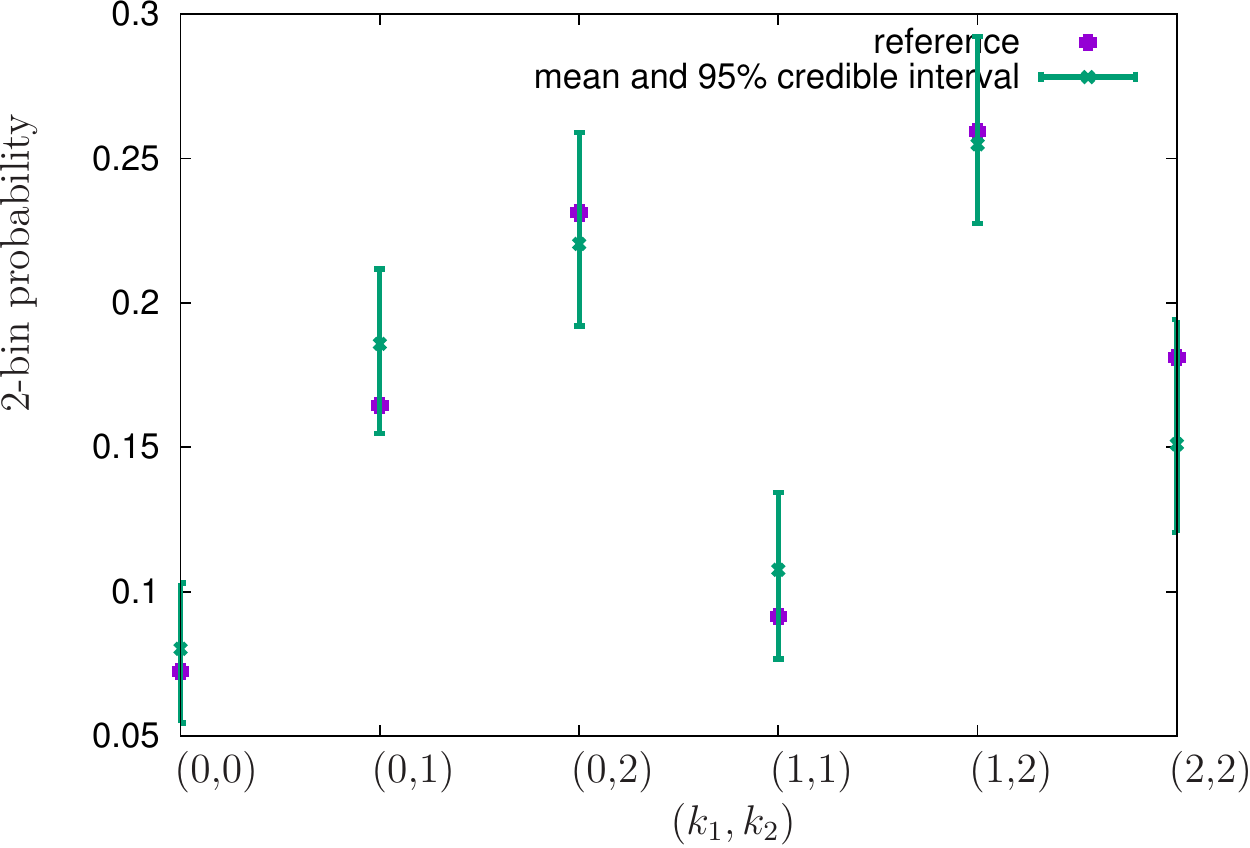}
 \caption{$t=10\Delta t$}
 \end{subfigure}
 \begin{subfigure}[b]{.9\textwidth}
\includegraphics[width=\textwidth]{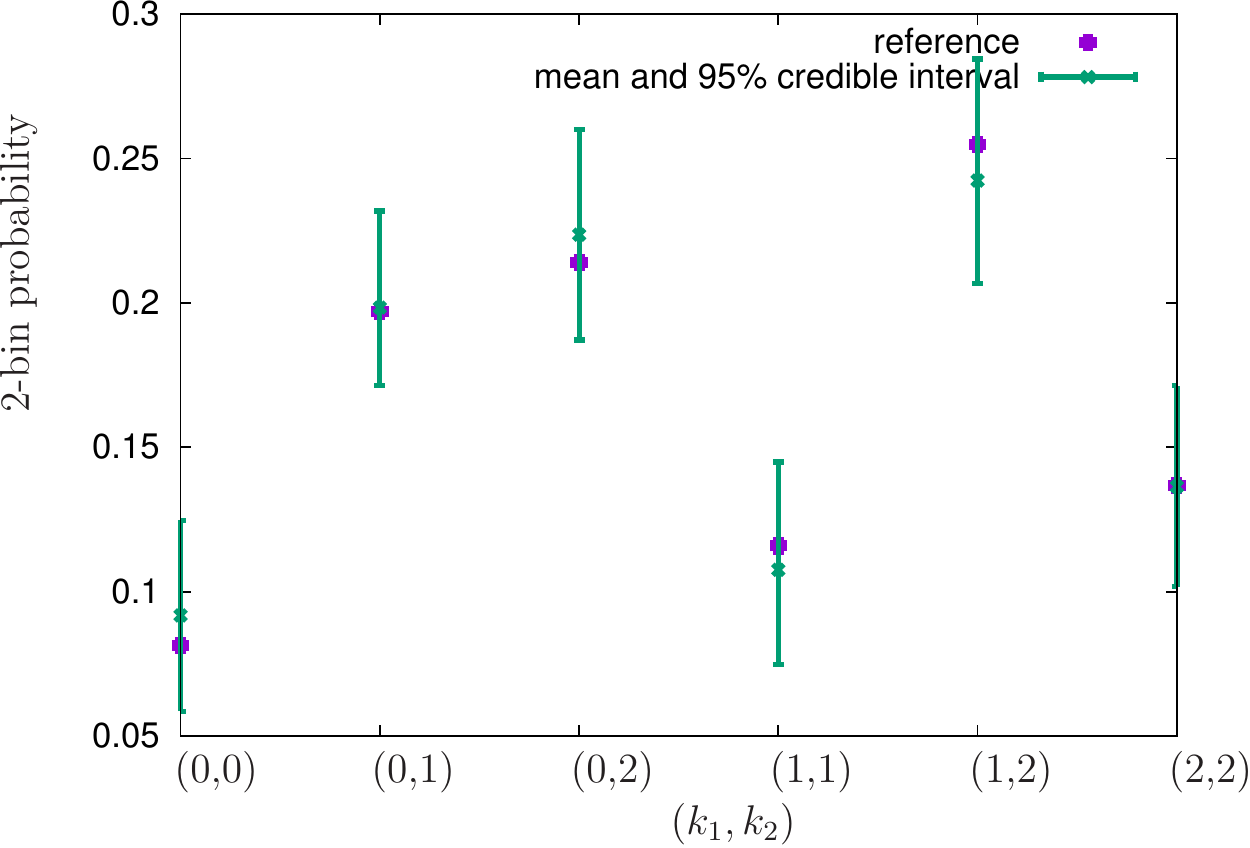}
 \caption{$t=20\Delta t$}
 \end{subfigure}
 \begin{subfigure}[b]{.9\textwidth}
\includegraphics[width=\textwidth]{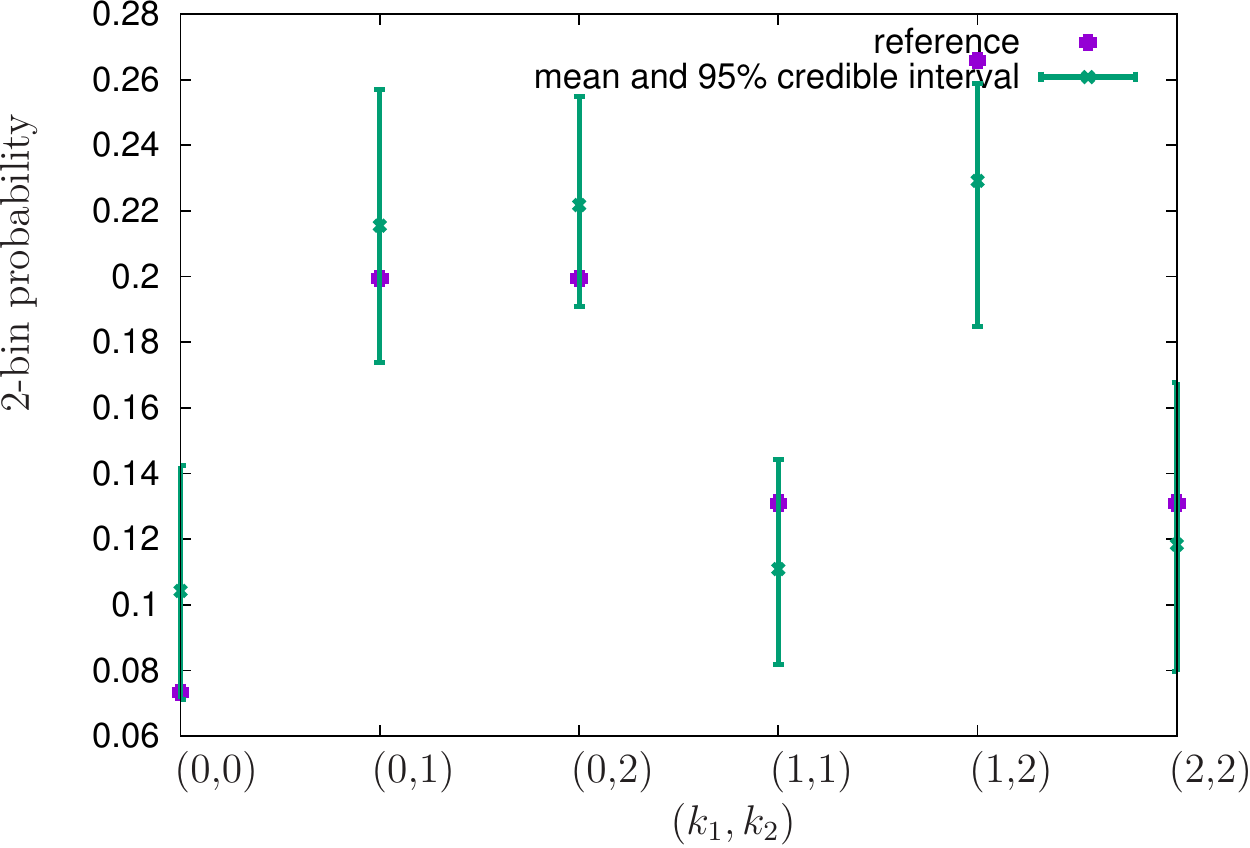}
 \caption{$t=40\Delta t$}
 \end{subfigure}
 \end{minipage}
 \begin{minipage}[c]{0.5\textwidth}
\centering
\begin{subfigure}[b]{.9\textwidth}
\includegraphics[width=\textwidth]{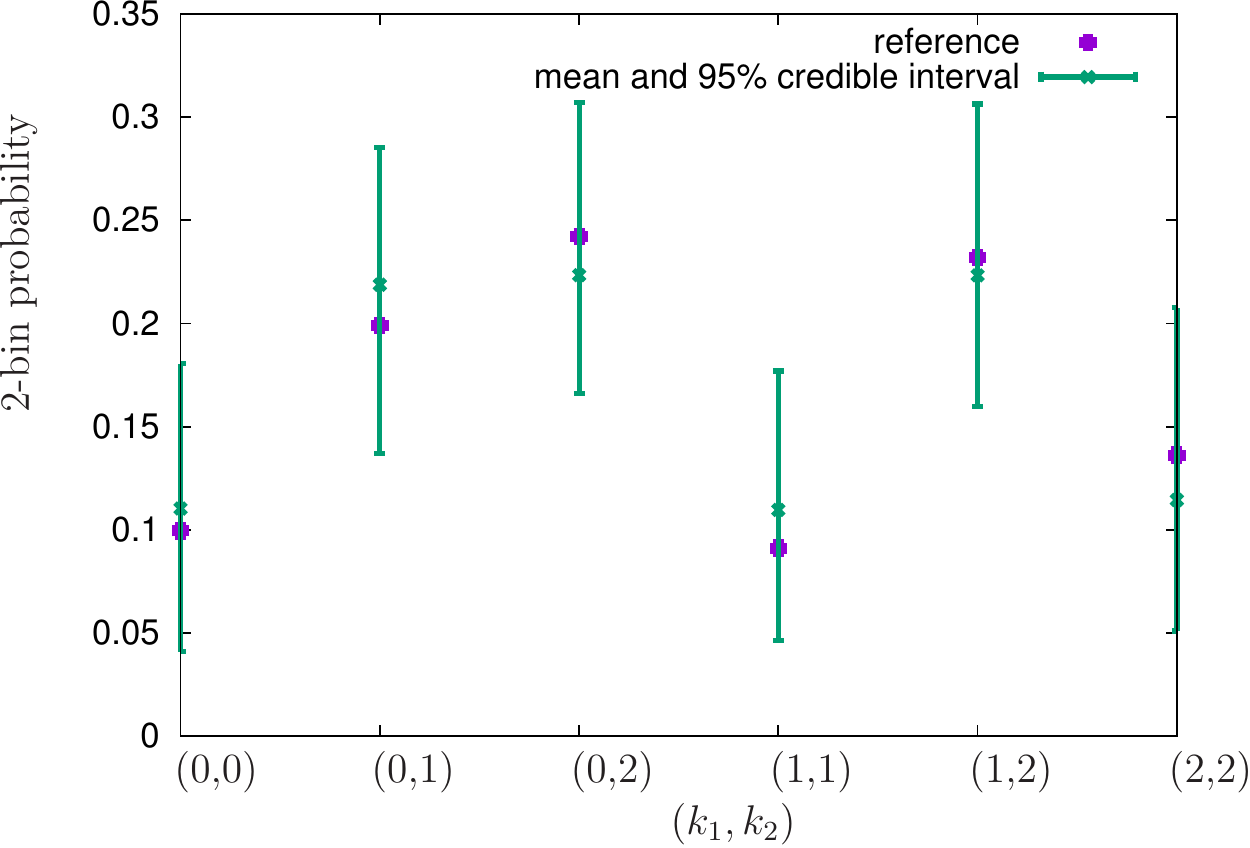}
\caption{$t=2\Delta t$}
 \end{subfigure}
 \begin{subfigure}[b]{.9\textwidth}
\includegraphics[width=\textwidth]{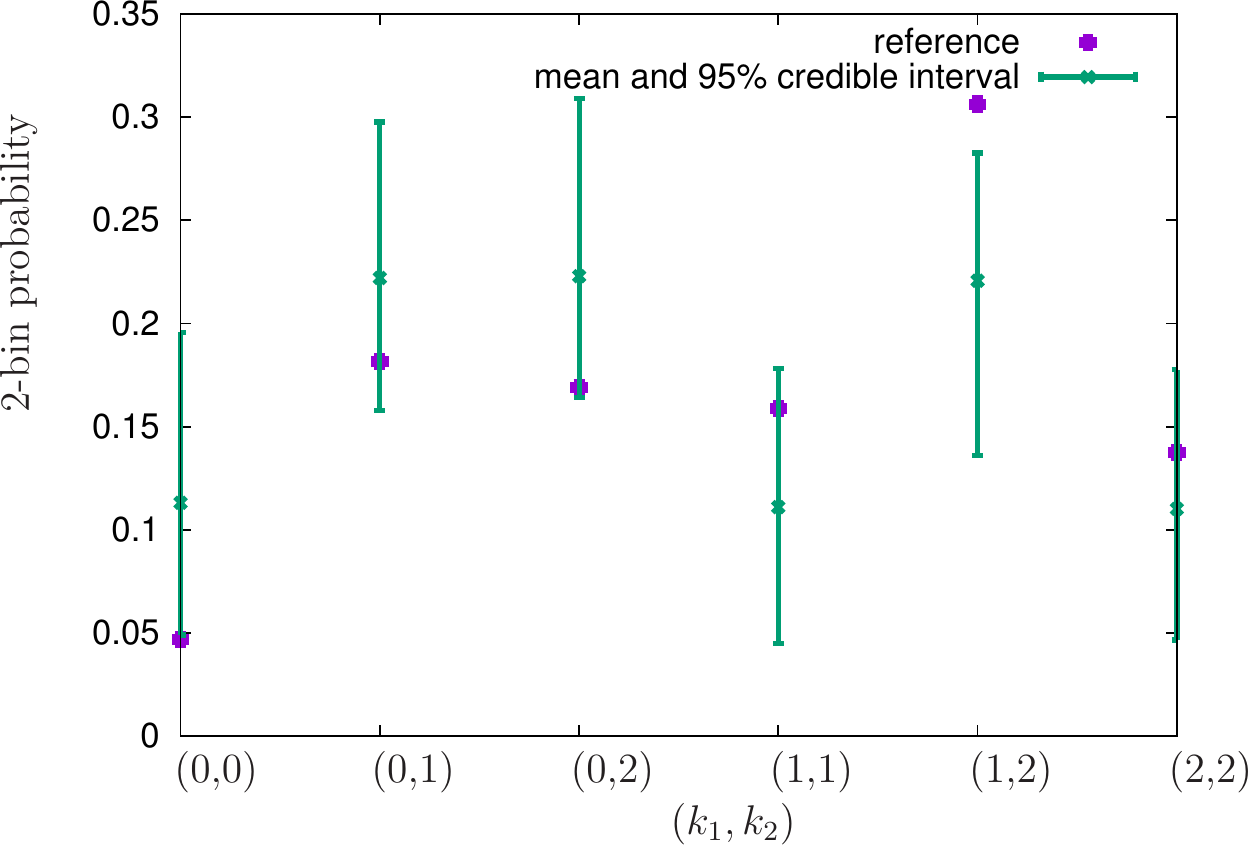}
 \caption{$t=10\Delta t$}
 \end{subfigure}
 \begin{subfigure}[b]{.9\textwidth}
\includegraphics[width=\textwidth]{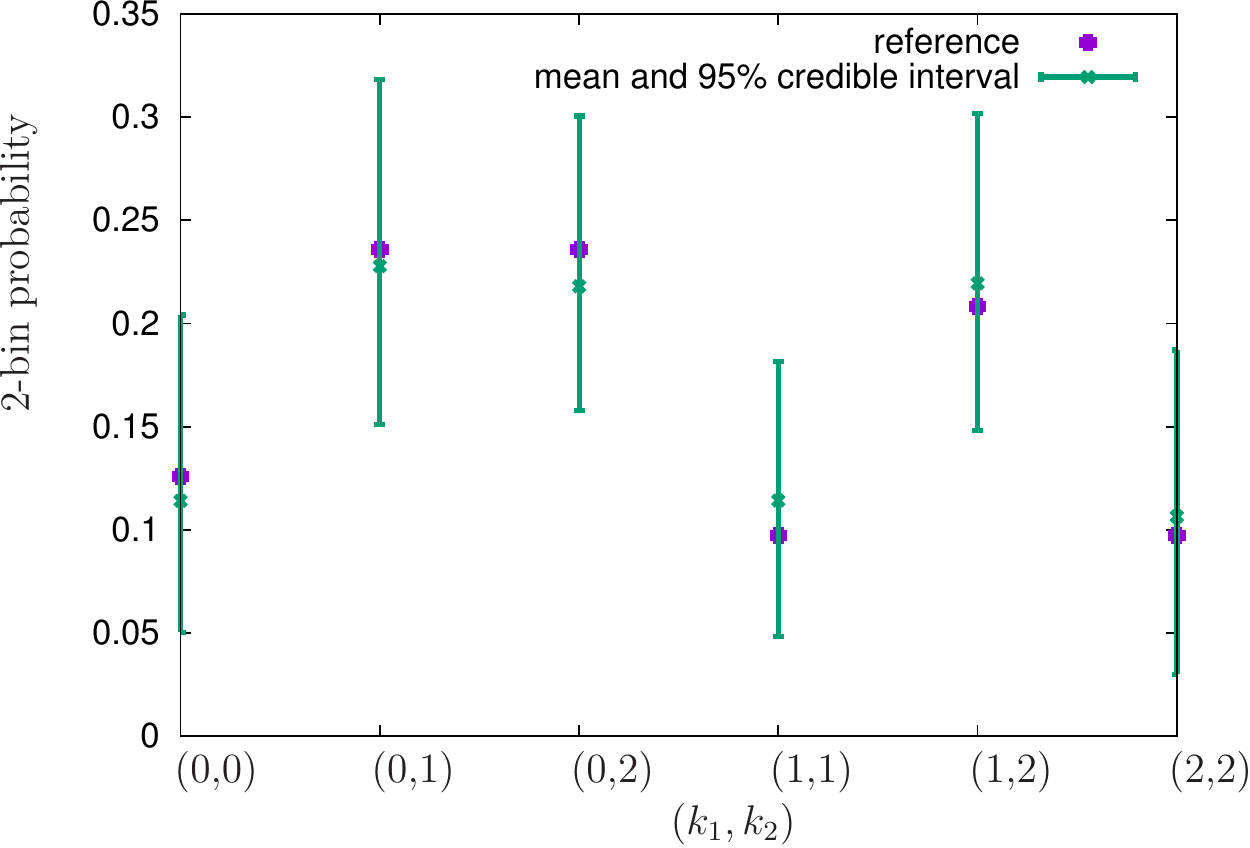}
 \caption{$t=20\Delta t$}
 \end{subfigure}
 \begin{subfigure}[b]{.9\textwidth}
\includegraphics[width=\textwidth]{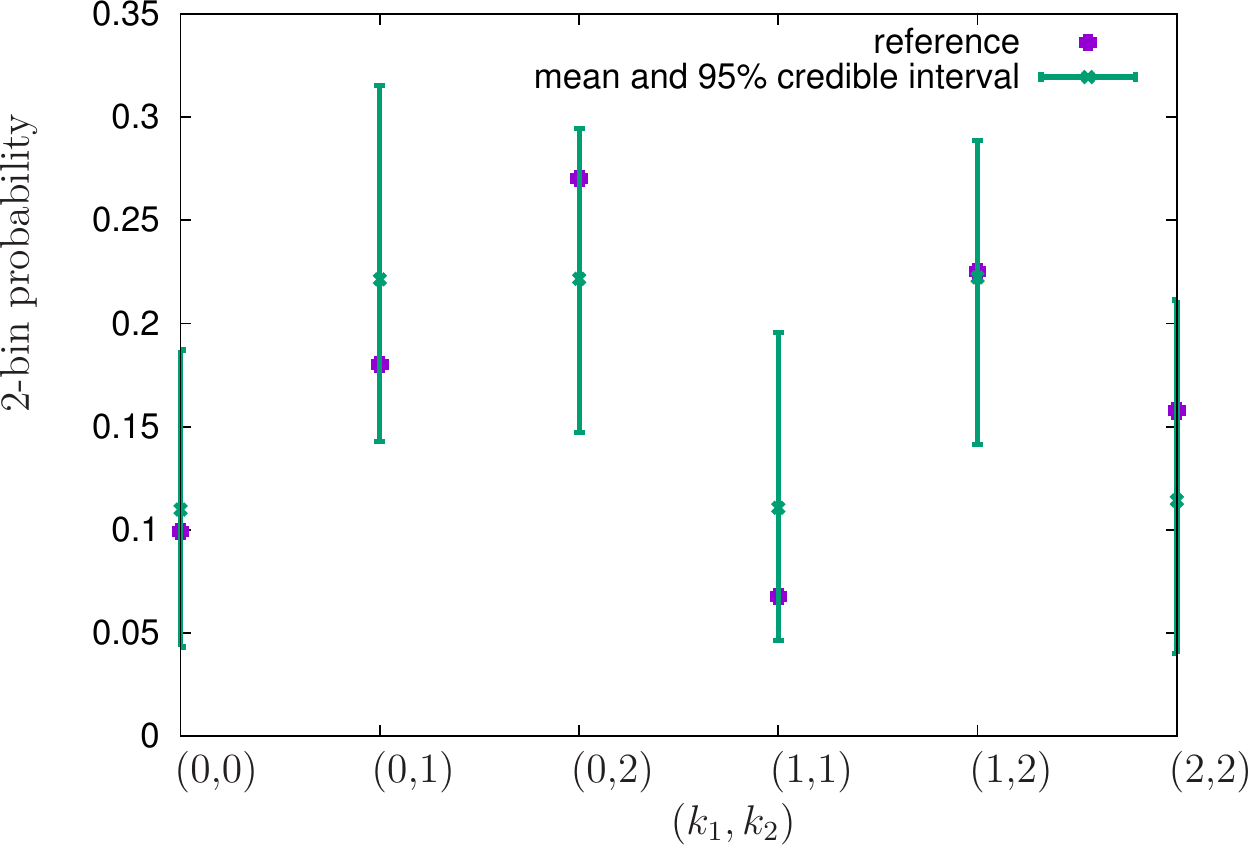}
 \caption{$t=40\Delta t$}
 \end{subfigure}
 \end{minipage}
 \caption{Predictive estimates of 2-bin probability for various times in the future i.e. $t=2\Delta t, 10 \Delta t, 20\Delta t, 40\Delta t$ . The left column corresponds to  the  original binning, whereas the right column to  4 times finer binning.  The reference density profile was computed by simulating the FG model of walkers using the FG time step $\delta t$. ($N=64$) }
 \label{fig:adpred2drho}
\end{figure}

Finally, in Figure \ref{fig:adpred2drho} we depict predictive estimates pertaining to second-order statistics, namely  the probability of finding two walkers, simultaneously, at two bins  $(k_1,k_2)$ (2-bin probability as in the synthetic example). The comparison of the reference values (i.e. those obtained by simulating the FG model) with the probabilistic predictive estimates of the proposed model is carried out for some indicative  values of $(k_1,k_2)$, for different time instants in the future (up to $40 \Delta t = 16000 \delta t$), and for two different numbers of bins i.e. $24$ (left column) and $96$ (right column). Accurate predictive estimates are observed and, as expected, the credible intervals  are  wider  when a finer binning is employed.

\subsection{Inviscid Burgers' equation}
The third and last example investigated consists at  the fine-scale of $n_f=2400$ walkers\footnote{As in the previous example, we employed  $\delta y=3.875 \times 10^{-3}$, $\delta t=2.5 \times 10^{-3}$.  The latter two numbers should be compared with the size of the problem domain which is $[-1,1]$ and the CG time-step which was set to $\Delta t =1$ i.e. $400$ times larger than the FG one.}
 which perform {\em interactive} random walks i.e. the jump performed at each fine-scale time-step $\delta t=2.5 \times 10^{-3}$ depends on the positions of the other walkers. In particular we adopted interactions as described in   \cite{roberts_convergence_1989,chertock_particle_2001,li_deciding_2007}
 so as, in the limit, the walker density $\rho(t,y)$ follows the inviscid Burgers' equation:
 \be
  \cfrac{\pa \rho }{\pa t} +\rho \cfrac{\pa \rho}{\pa y}=0.
  \label{eq:inviscidbu}
  \ee

At the CG level we employed $n_c=24$ variables (i.e. $n_f/n_c=100$) and as described in section \ref{sec:train} initialized the walkers using $N$ randomly selected initial densities and propagated them for $400$ FG time-steps $\delta t$ (i.e. one CG time-step $\Delta t$) in order to generate $N$ training data.
In this investigation we employed the same first- and second-order  feature function (Equations (\ref{eq:phi1order}), (\ref{eq:phi2order})) for $M=5$. This gave rise to $132$ features in total and the same number of coefficients $\bt_c$.

\begin{figure}
\includegraphics[width=.5\textwidth]{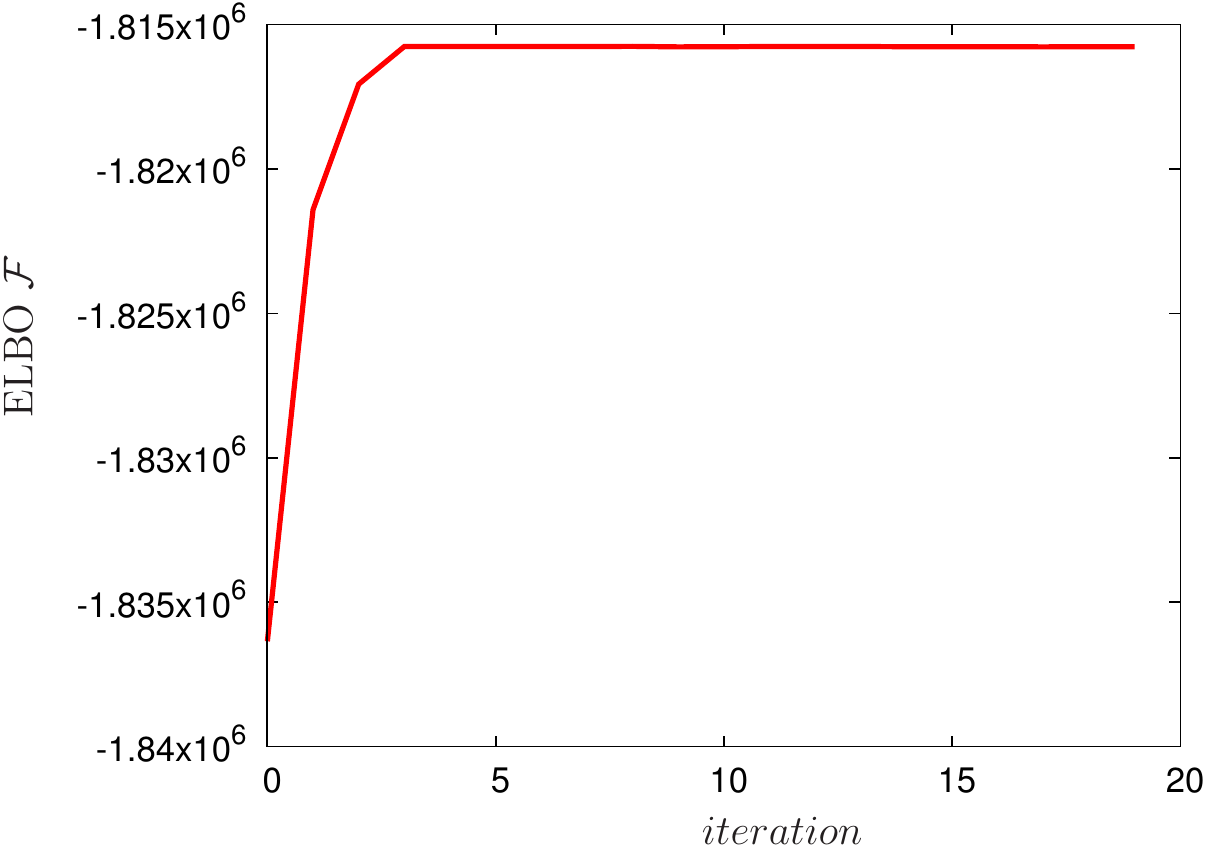}
 \caption{Evolution  of the ELBO  $\mathcal{F}$ as estimated at each iteration based on \refeq{eq:elbofinal} for $N=128$.}
 \label{fig:buelbo}
\end{figure}

\begin{figure}
\begin{subfigure}[b]{.5\textwidth}
\includegraphics[width=\textwidth]{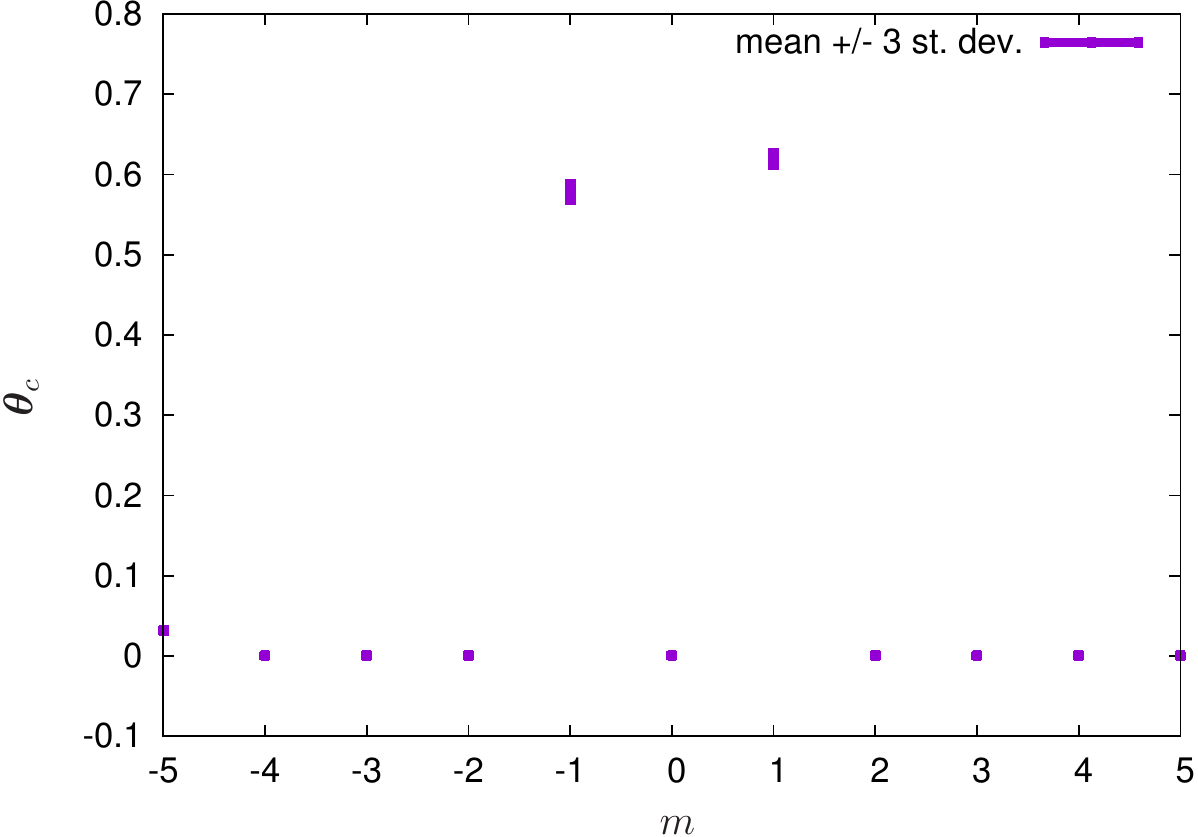}
 \end{subfigure}
\begin{subfigure}[b]{.5\textwidth}
\includegraphics[width=\textwidth]{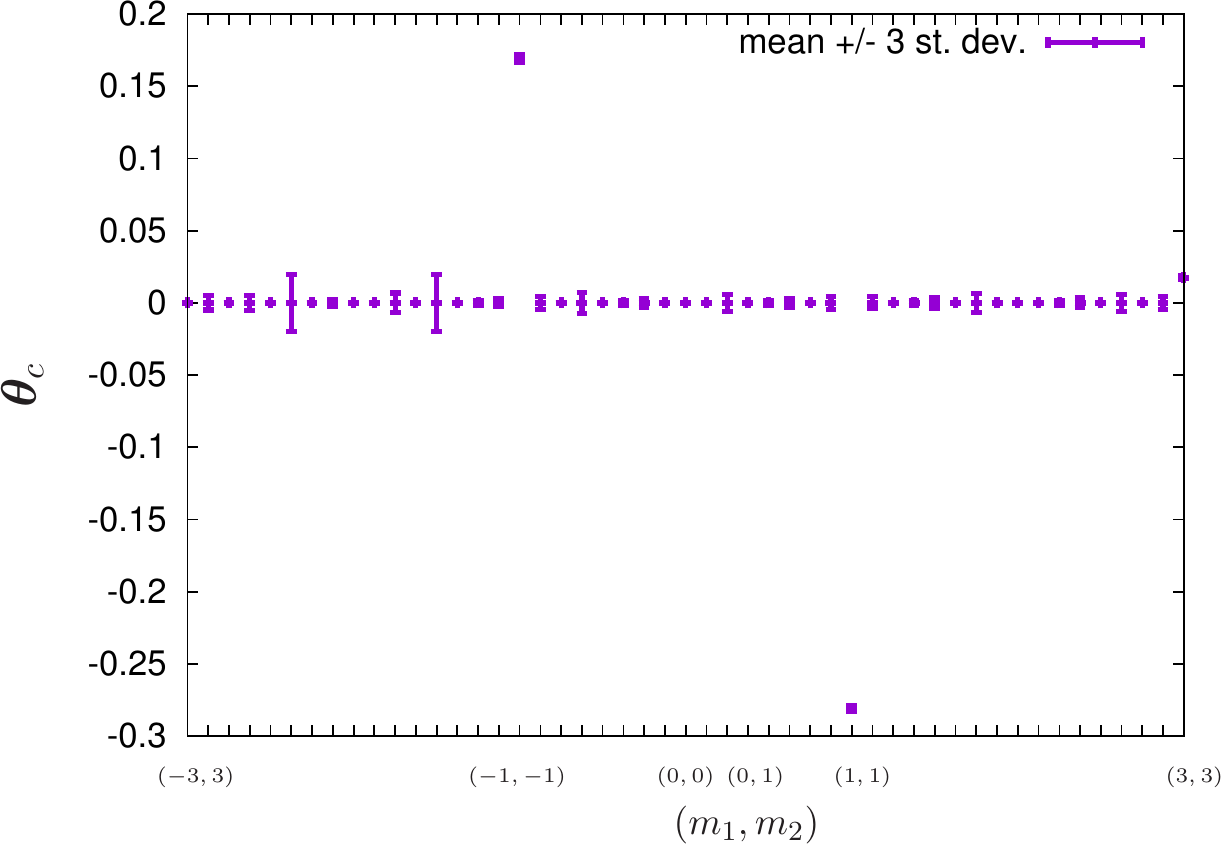}
\end{subfigure}
 \caption{Posterior estimates for $N=128$ of $\bt_c$ associated with first- (\refeq{eq:phi1order}) and second-order (\refeq{eq:phi2order}) feature functions. We plot only a subset of the coefficients of the second-order features for clarity. The omitted coefficients were practically zero (deactivated).}
 \label{fig:buposttheta}
\end{figure}

Figure \ref{fig:buelbo} depicts the evolution   of the ELBO  $\mathcal{F}$ as estimated at each iteration based on \refeq{eq:elbofinal} for $N=64$ training data sequences.
Figure \ref{fig:buposttheta} depicts the posterior mean and standard deviation of the $\bt_c$ associated with the aforementioned feature functions for various values of $m$ (\refeq{eq:phi1order}) and for $N=128$. We note that the notably  active ones are for    $m=\pm 1$ (first-order) and for $(-1,-1)$ and $(1,1)$ (second-order). This could be compared with the Lax-Friedrichs discretization scheme (with a spatial step  $\delta y$  and time-step $\delta t$) of \refeq{eq:inviscidbu} which is:
\be
\rho_j^{t+1}=\frac{1}{2}\left( \rho_{j-1}^t + \rho_{j+1}^t   \right) -\frac{\delta y}{4 \delta t} 
\left( (\rho_{j+1}^t)^2 - (\rho_{j-1}^t)^2 \right)
\label{eq:lax}
\ee
While, as previously noted, an immediate comparison is not possible (due to the fact the CG evolution is written with respect to $\bxx$ which are transformed with the softmax function in \refeq{eq:softmax}), we cannot help but note that the same type of interactions are identified i.e. the same general structure for the (discretized) CG evolution law. The values of the corresponding $\bt_c$ are not as \refeq{eq:lax} would suggest (e.g. the coefficients for the first order features at $m=\pm 1$ are not equal) which  could be justified on the basis of the previous argument.  
 
\begin{figure}[!h]
\begin{minipage}[c]{0.5\textwidth}
\centering
\begin{subfigure}[b]{.9\textwidth}
\includegraphics[width=\textwidth]{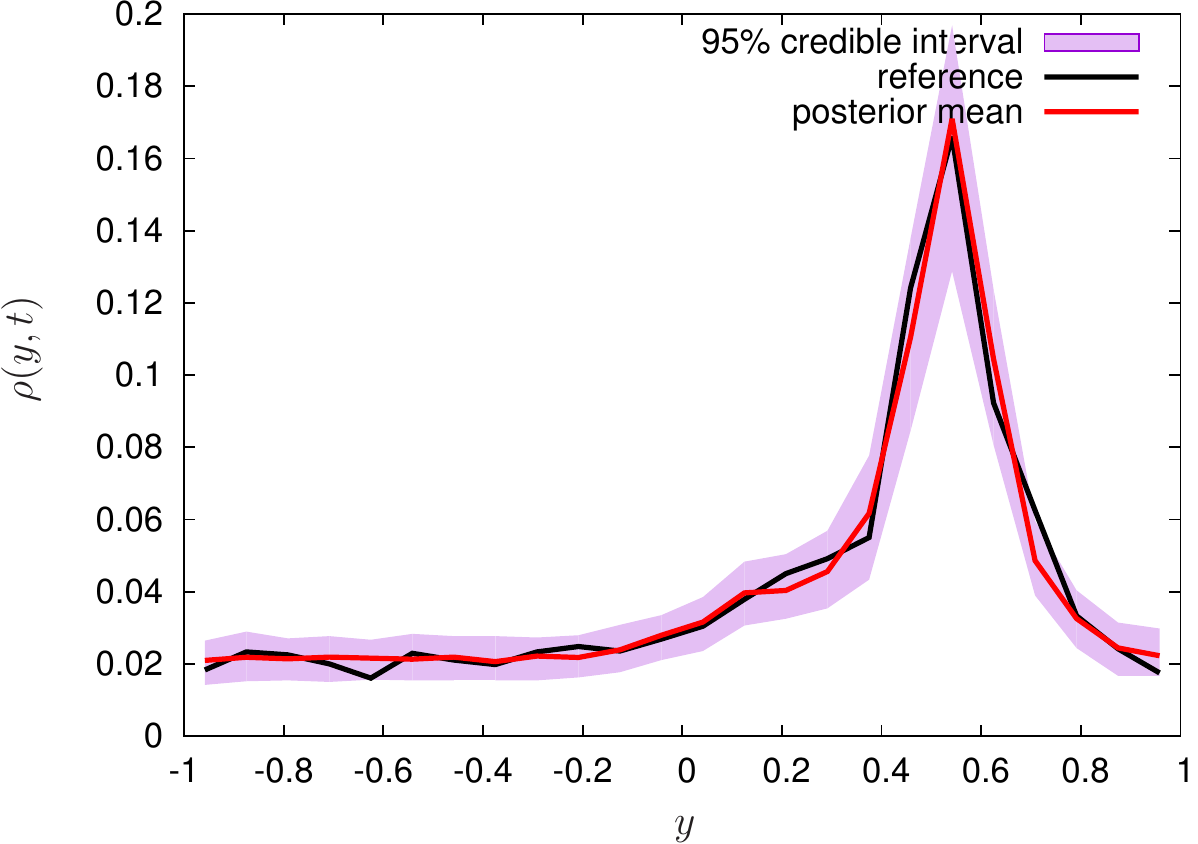}
 \caption{$t=2\Delta t$}
 \end{subfigure}
 \begin{subfigure}[b]{.9\textwidth}
\includegraphics[width=\textwidth]{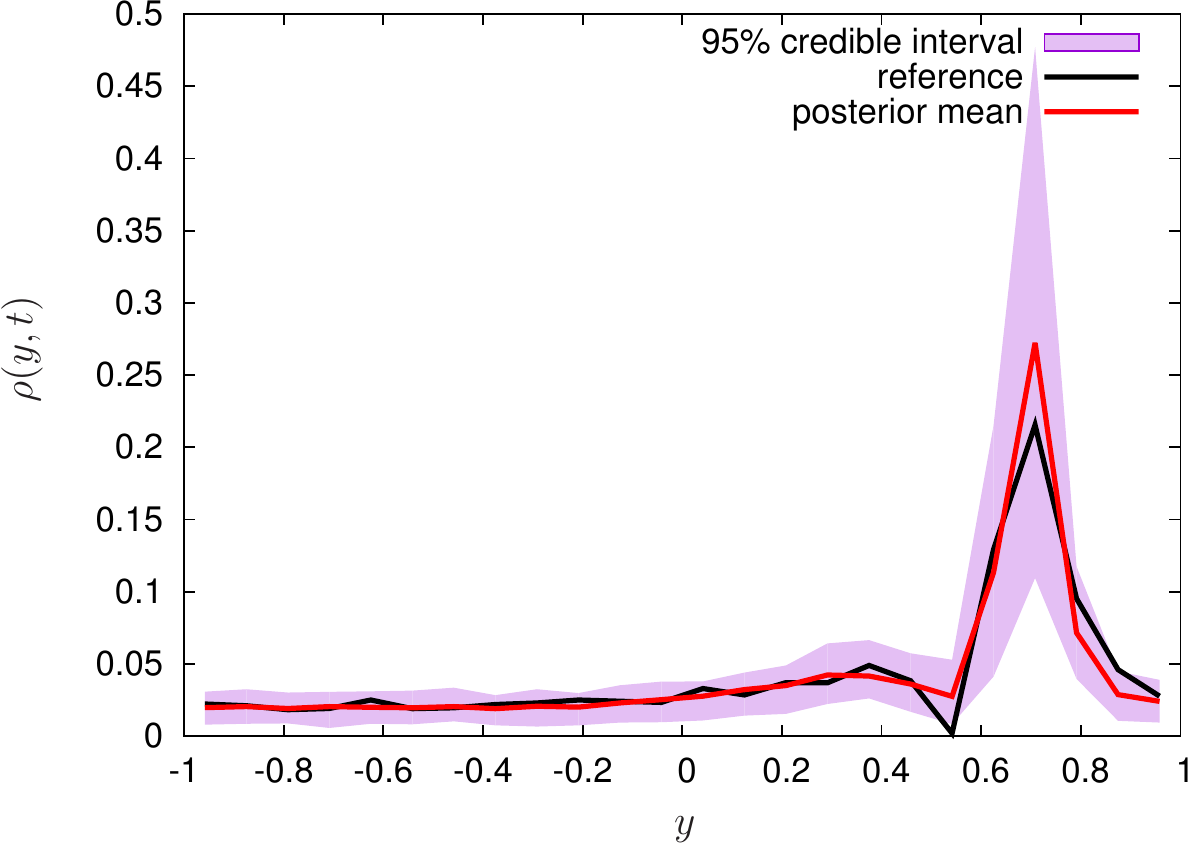}
 \caption{$t=4\Delta t$}
 \end{subfigure}
 \begin{subfigure}[b]{.9\textwidth}
\includegraphics[width=\textwidth]{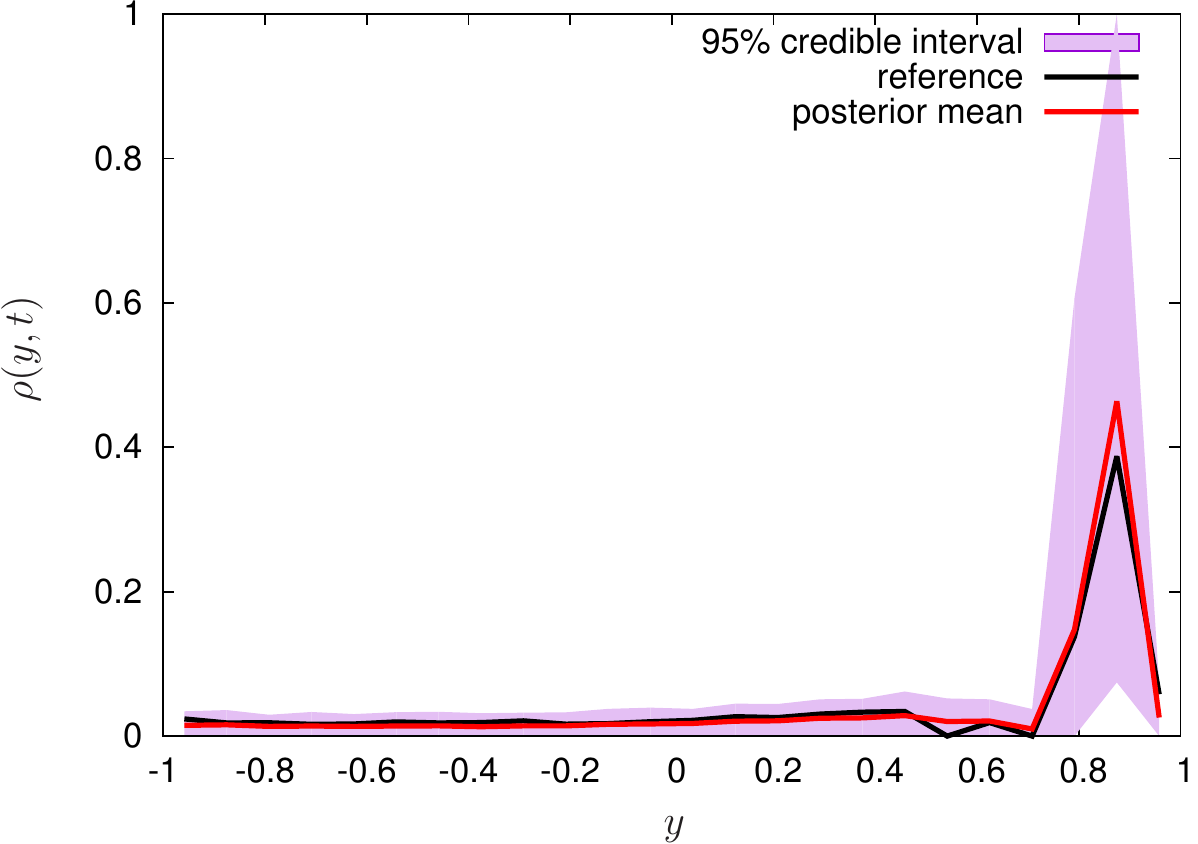}
 \caption{$t=6\Delta t$}
 \end{subfigure}
 \begin{subfigure}[b]{.9\textwidth}
\includegraphics[width=\textwidth]{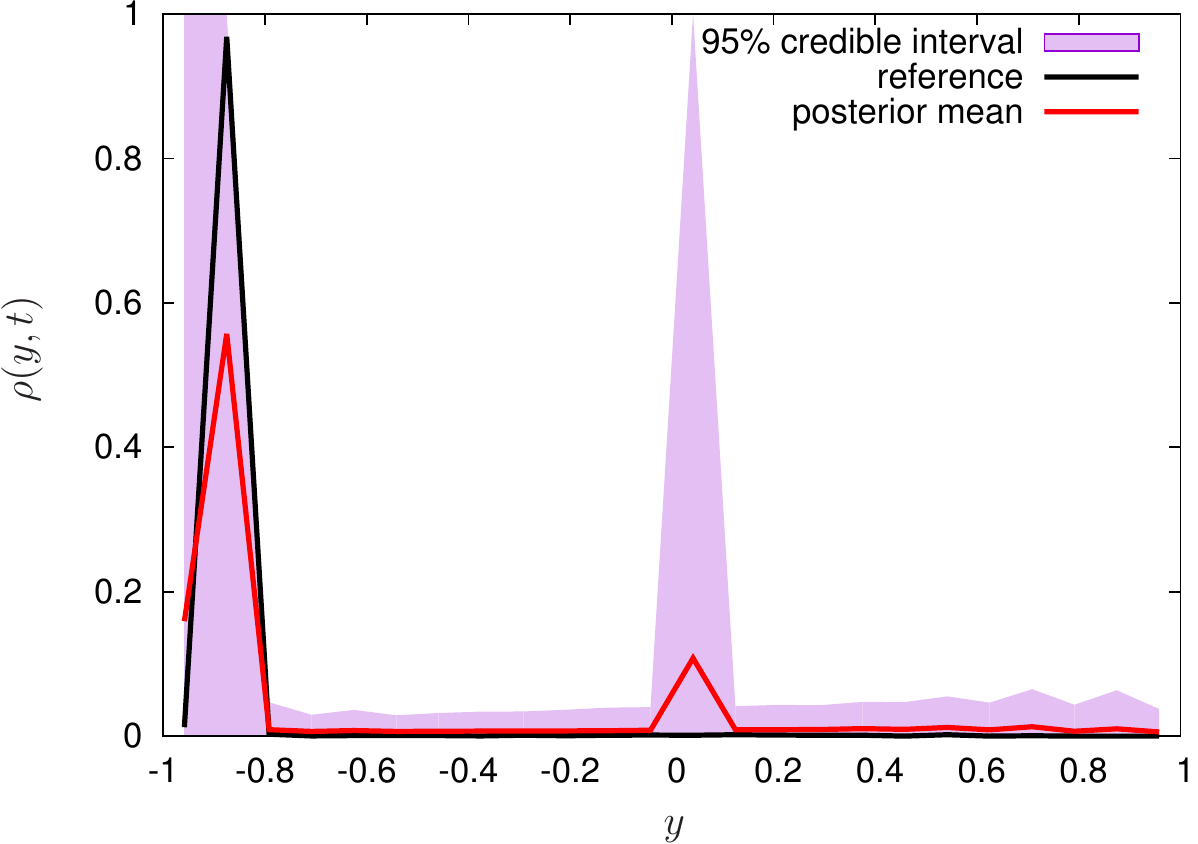}
 \caption{$t=9\Delta t$}
 \end{subfigure}
 \end{minipage}
 \begin{minipage}[c]{0.5\textwidth}
\centering
\begin{subfigure}[b]{.9\textwidth}
\includegraphics[width=\textwidth]{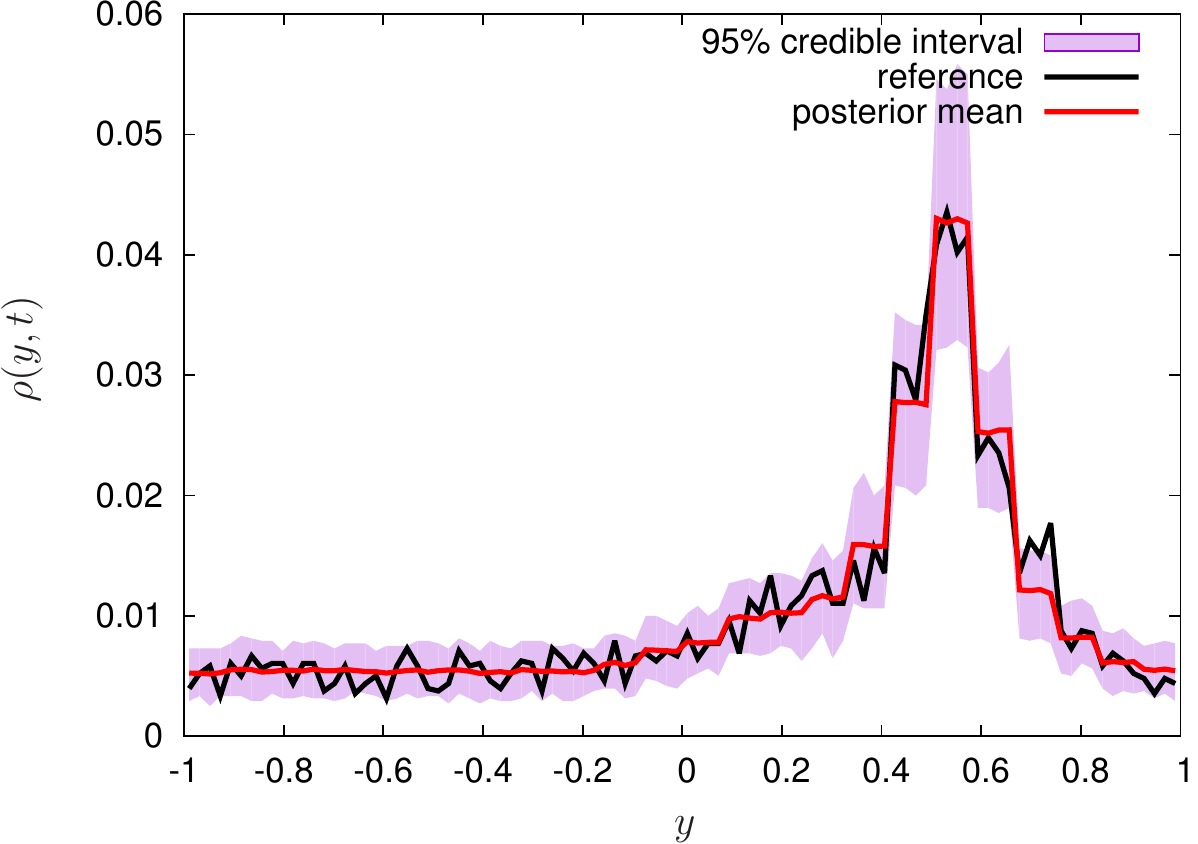}
 \caption{$t=2\Delta t$}
 \end{subfigure}
 \begin{subfigure}[b]{.9\textwidth}
\includegraphics[width=\textwidth]{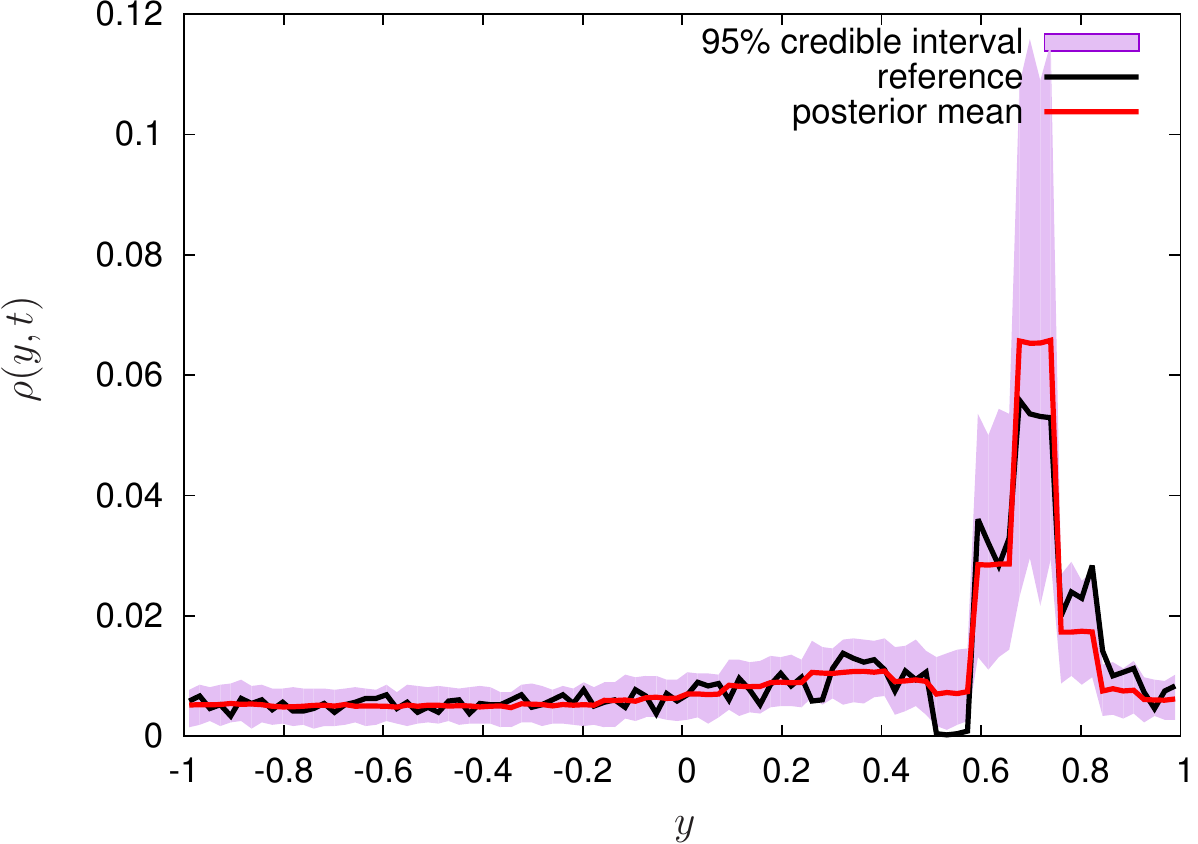}
 \caption{$t=4\Delta t$}
 \end{subfigure}
 \begin{subfigure}[b]{.9\textwidth}
\includegraphics[width=\textwidth]{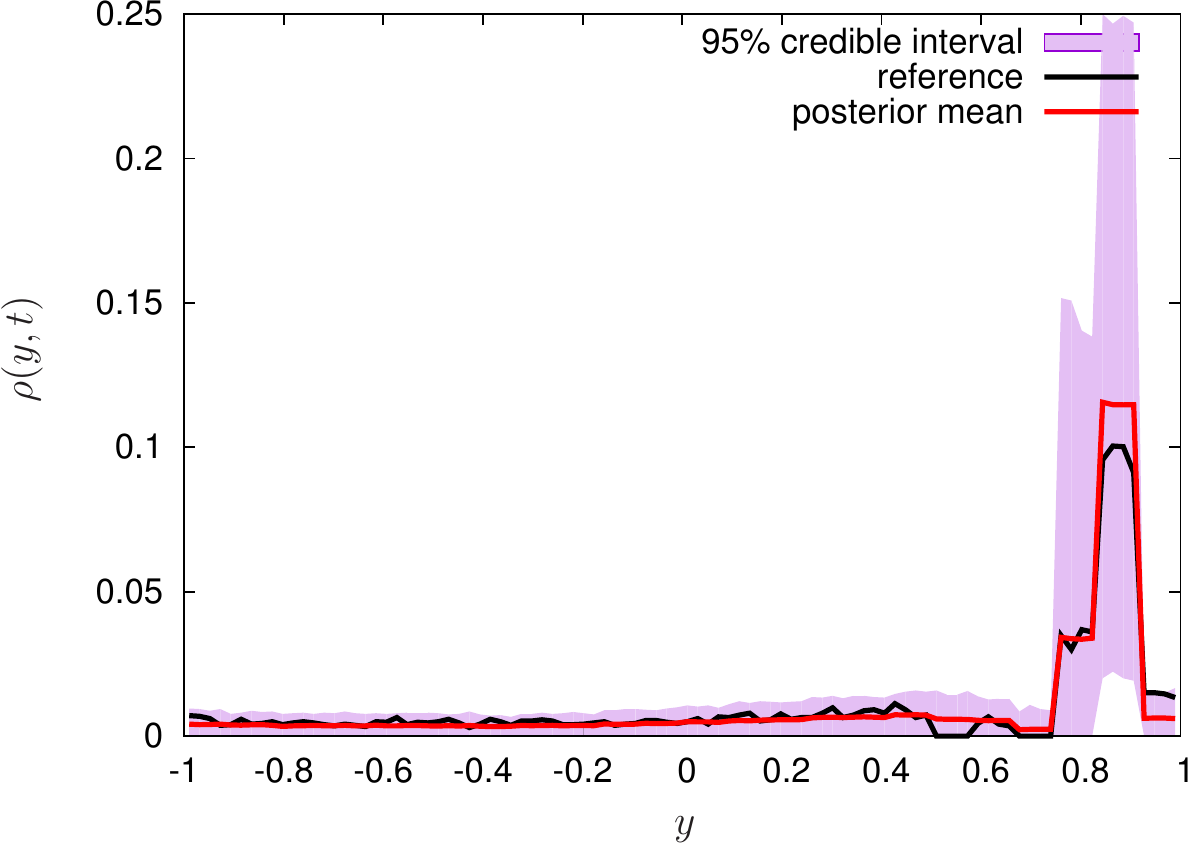}
 \caption{$t=6\Delta t$}
 \end{subfigure}
 \begin{subfigure}[b]{.9\textwidth}
\includegraphics[width=\textwidth]{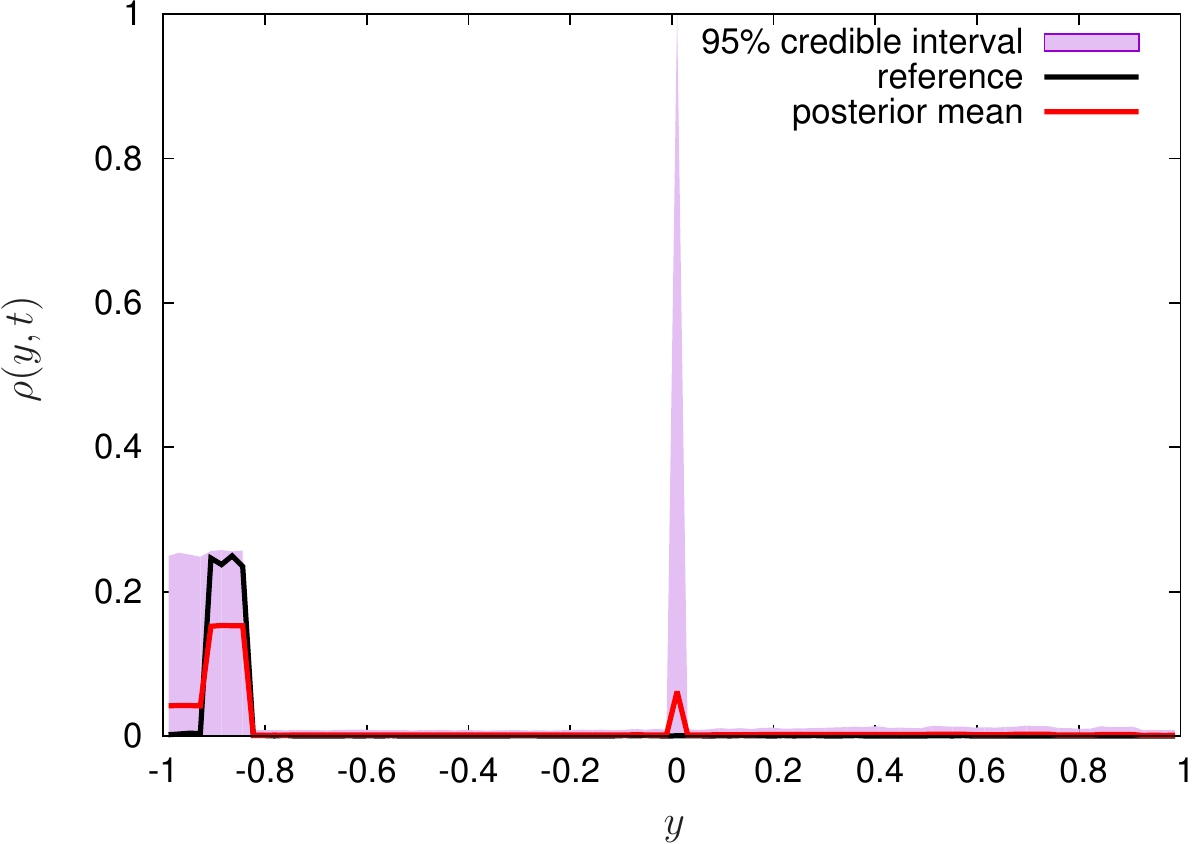}
 \caption{$t=9\Delta t$}
 \end{subfigure}
 \end{minipage}
 \caption{ Predictive estimates of walker density for various future  times. The left column corresponds to  the  original binning, whereas the right column to  4 times finer binning.  The reference density profile was computed by simulating the FG model of walkers using the FG time step $\delta t$. ($N=128$) }
 \label{fig:bupredrho64}
\end{figure}

\begin{figure}[!h]
\begin{minipage}[c]{0.5\textwidth}
\centering
\begin{subfigure}[b]{.9\textwidth}
\includegraphics[width=\textwidth]{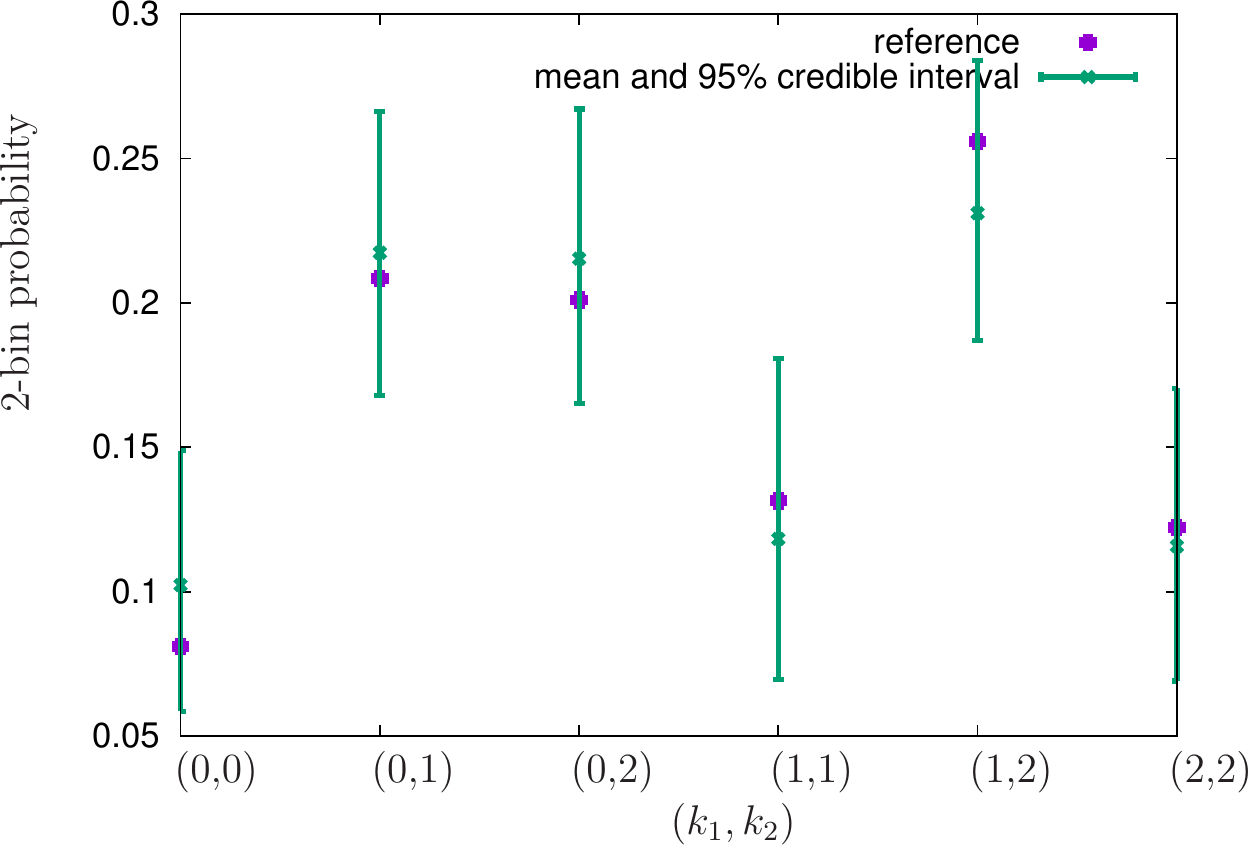}
\caption{$t=2\Delta t$}
 \end{subfigure}
 \begin{subfigure}[b]{.9\textwidth}
\includegraphics[width=\textwidth]{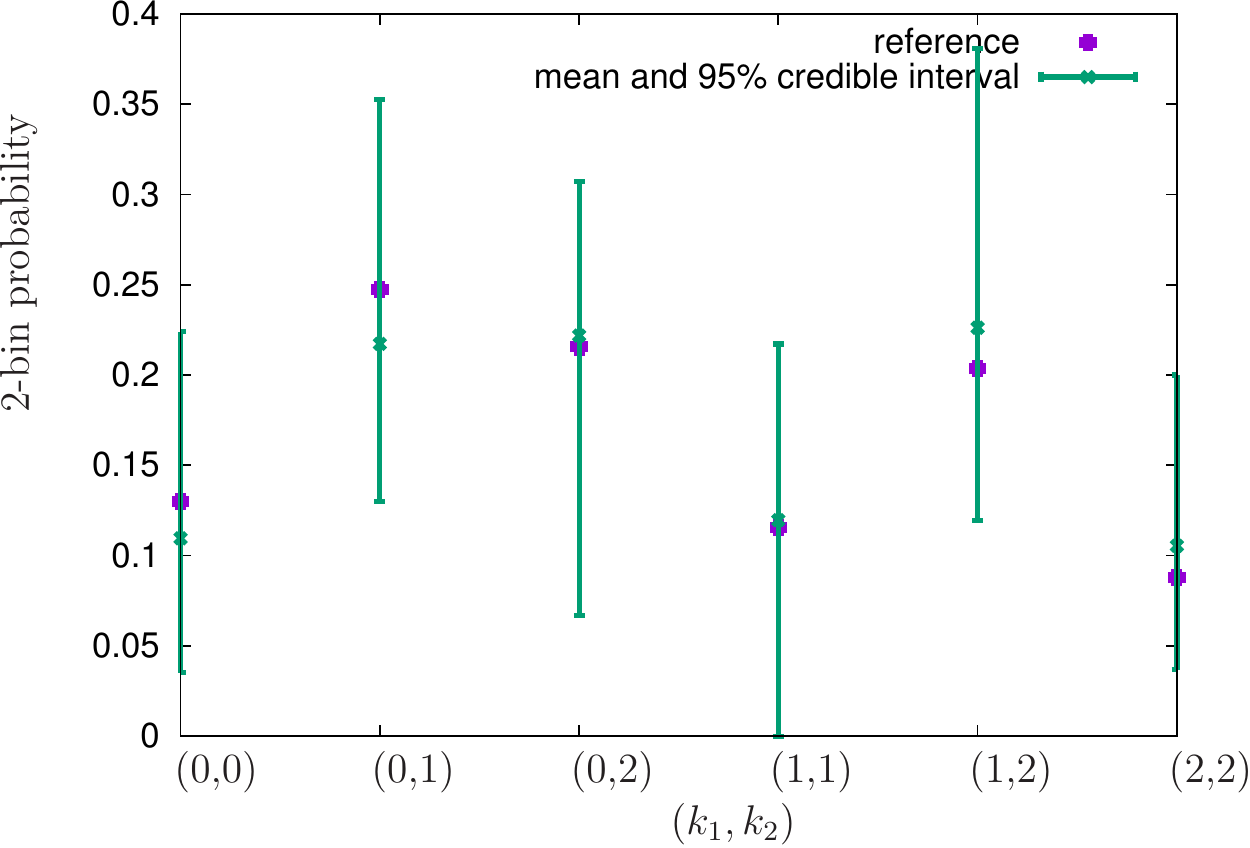}
 \caption{$t=4\Delta t$}
 \end{subfigure}
 \begin{subfigure}[b]{.9\textwidth}
\includegraphics[width=\textwidth]{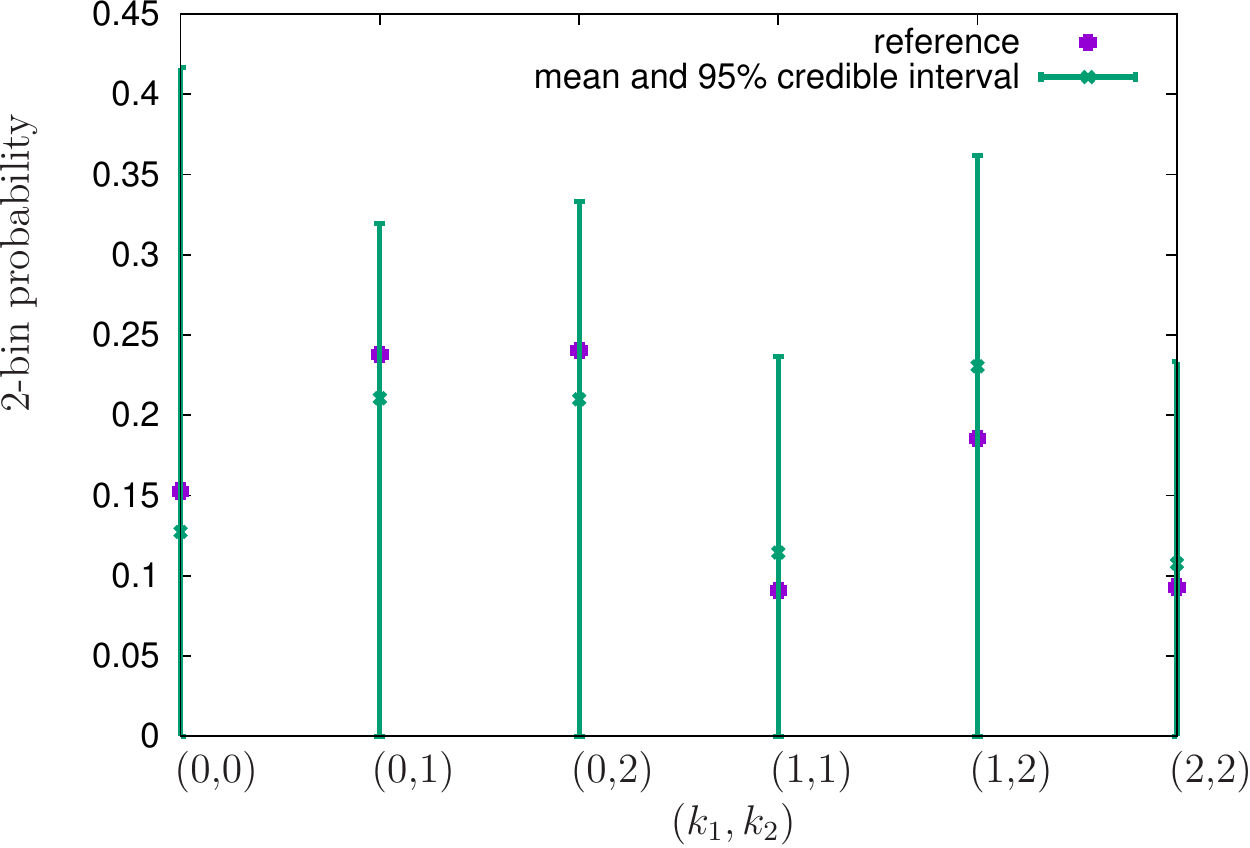}
 \caption{$t=6\Delta t$}
 \end{subfigure}
 \begin{subfigure}[b]{.9\textwidth}
\includegraphics[width=\textwidth]{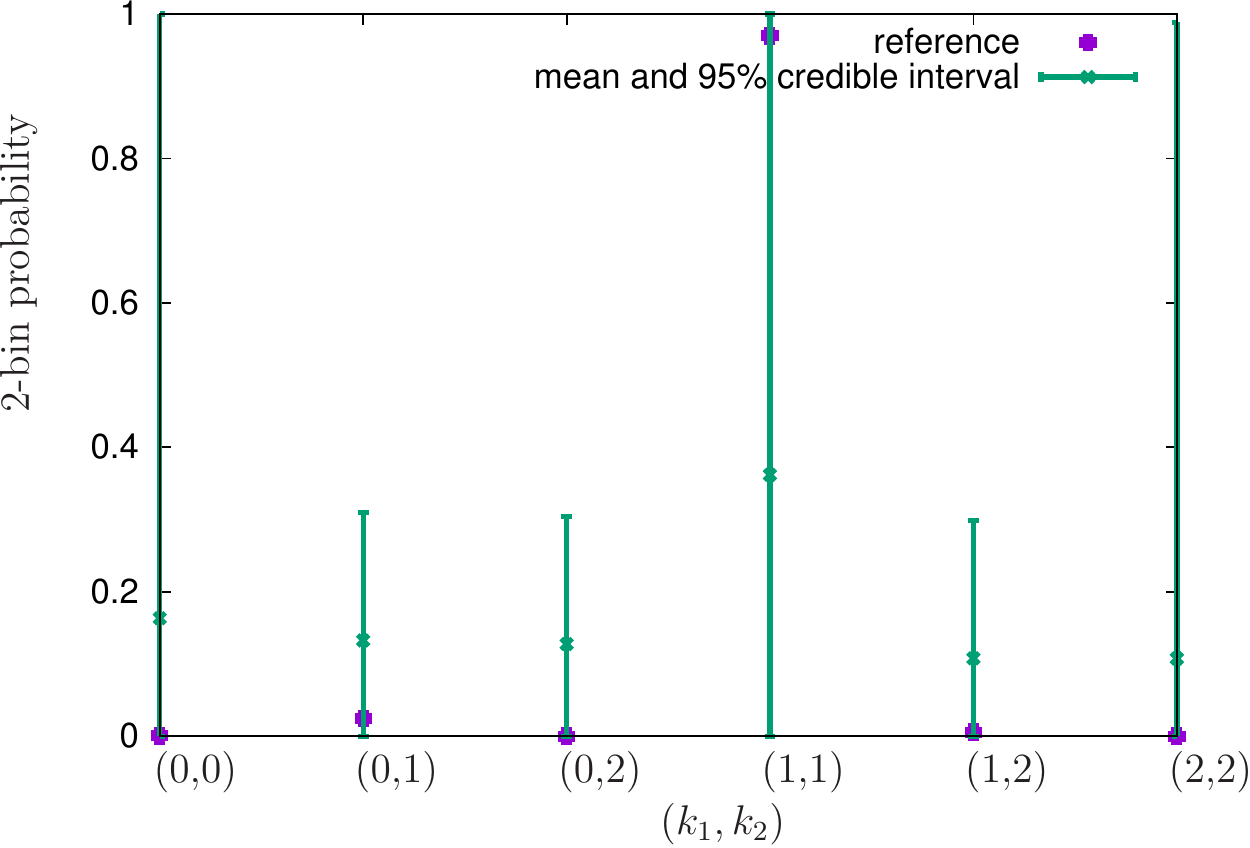}
 \caption{$t=9\Delta t$}
 \end{subfigure}
 \end{minipage}
 \begin{minipage}[c]{0.5\textwidth}
\centering
\begin{subfigure}[b]{.9\textwidth}
\includegraphics[width=\textwidth]{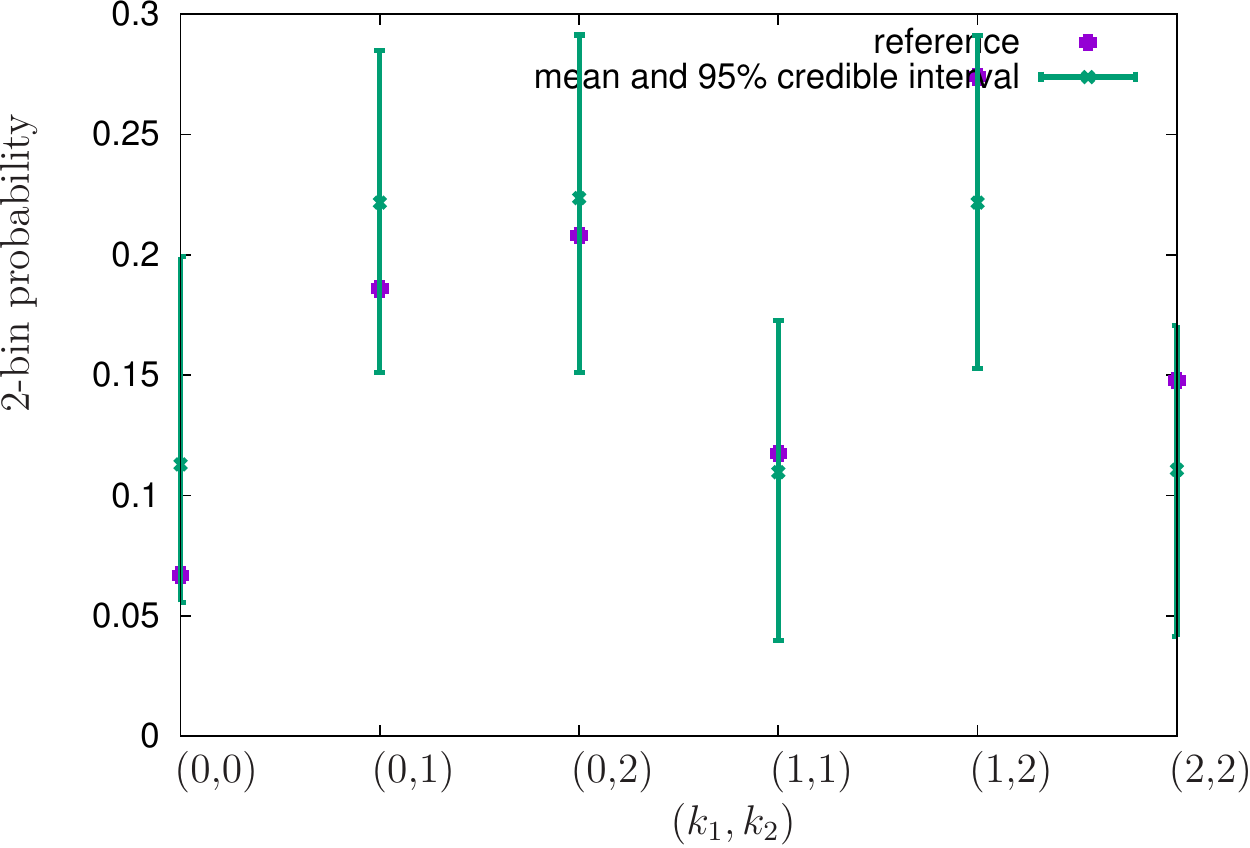}
\caption{$t=2\Delta t$}
 \end{subfigure}
 \begin{subfigure}[b]{.9\textwidth}
\includegraphics[width=\textwidth]{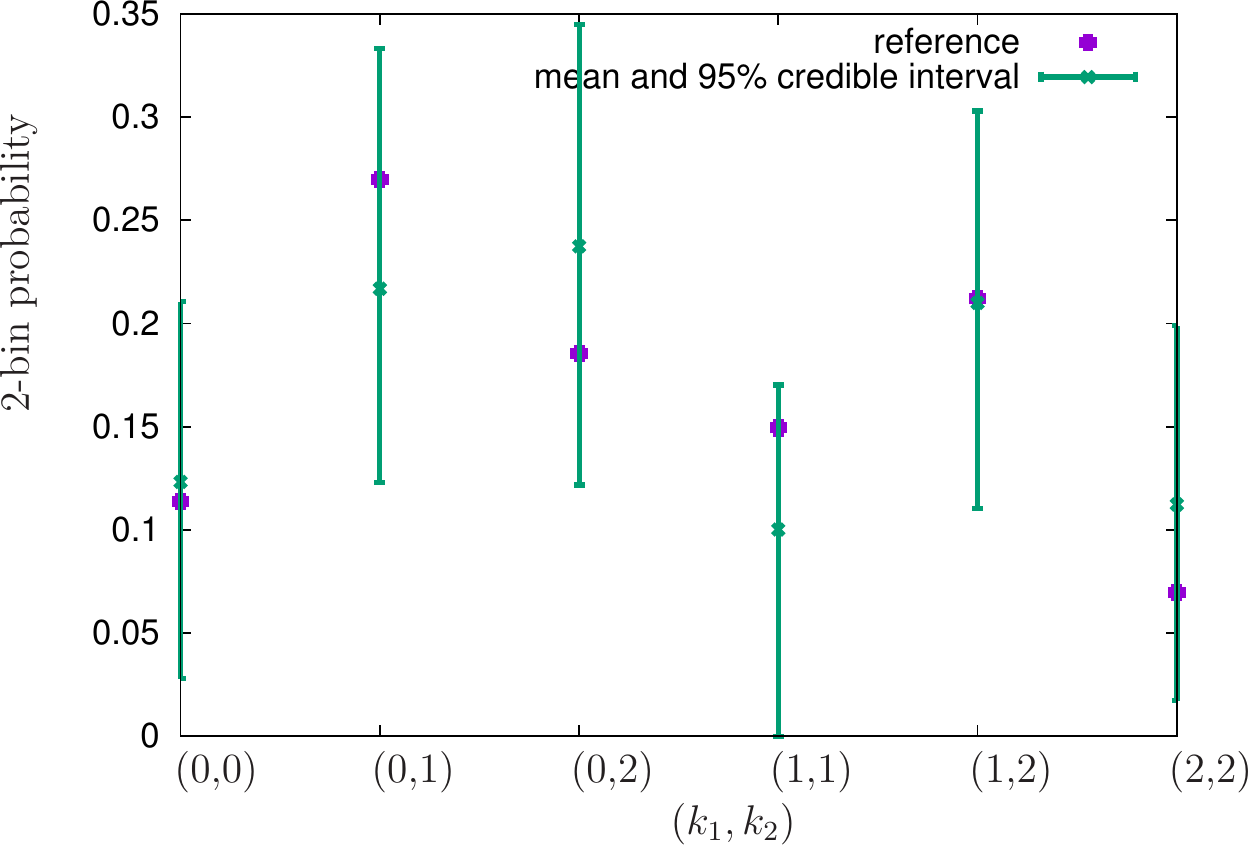}
 \caption{$t=4\Delta t$}
 \end{subfigure}
 \begin{subfigure}[b]{.9\textwidth}
\includegraphics[width=\textwidth]{FIGURES/Burgers/rho_pred_fine_4/rho_pred2d_04_29.pdf}
 \caption{$t=6\Delta t$}
 \end{subfigure}
 \begin{subfigure}[b]{.9\textwidth}
\includegraphics[width=\textwidth]{FIGURES/Burgers/rho_pred_fine_4/rho_pred2d_04_29.pdf}
 \caption{$t=9\Delta t$}
 \end{subfigure}
 \end{minipage}
 \caption{ Predictive estimates of 2-bin probability for various times in the future and for the same initial conditions used in Figure \ref{fig:bupredrho64}. The left column corresponds to  the  original binning, whereas the right column to  4 times finer binning.  The reference 2-bin  probability values were computed by simulating the FG model of walkers using the FG time step $\delta t$. ($N=64$) }
 \label{fig:bupred2drho}
\end{figure}

The most critical perhaps aspect pertains to the predictive ability of the CG model learned. In Figure \ref{fig:buposttheta} we compare the reference walker density (as computed from the FG model) with the predictive estimates (posterior mean and credible intervals) for an indicative initial condition. The comparison (as in the previous example) was also performed for 4 times finer binning than that used in the CG law which indicates the ability of the model to reconstruct, albeit with uncertainty, the full fine-scale picture. As before, the predictive uncertainty  is bigger when finer details are sought. In Figure \ref{fig:bupredrho64} we perform the same comparison, but for an initial condition that  leads to a shock formation (in terms of the walker density) i.e. the walkers tend to concentrate in a (very) small region of the problem domain. As it can be seen in Figure \ref{fig:bupredrho64} and particularly at time $9 \Delta t$ when the shock is more clearly formed, the CG model is able to follow its evolution and envelop it in the  predictive estimates. Interestingly the results obtained suggest the possibility of a shock at an alternative position ($\approx y=0$) which warrants further investigation.
Finally and for the same initial conditions as in Figure \ref{fig:bupredrho64}, we depict in Figure \ref{fig:bupred2drho} predictive estimates of the 2-bin probability at various time instants. As in the previous examples, good agreement is observed.

\begin{figure}[!h]
\begin{minipage}[c]{0.5\textwidth}
\centering
\begin{subfigure}[b]{.9\textwidth}
\includegraphics[width=\textwidth]{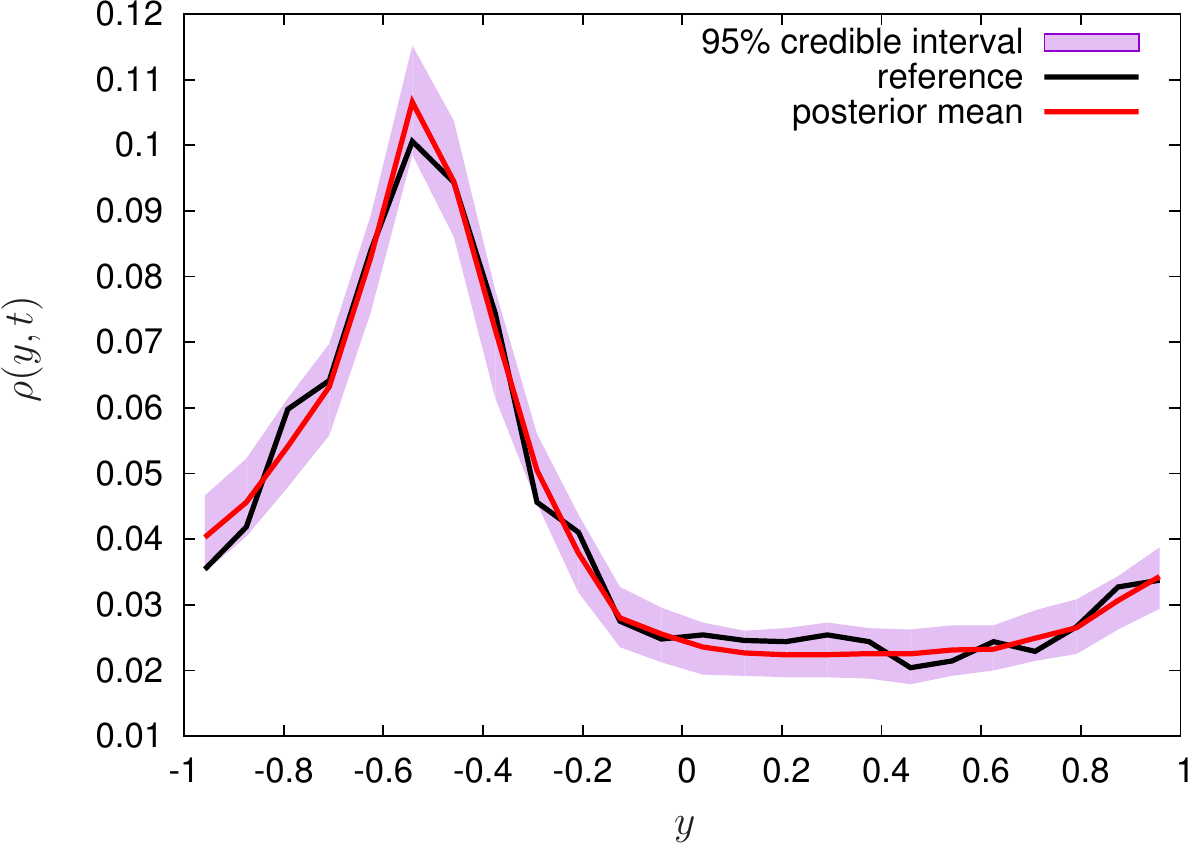}
 \caption{$t=2\Delta t$}
 \end{subfigure}
 \begin{subfigure}[b]{.9\textwidth}
\includegraphics[width=\textwidth]{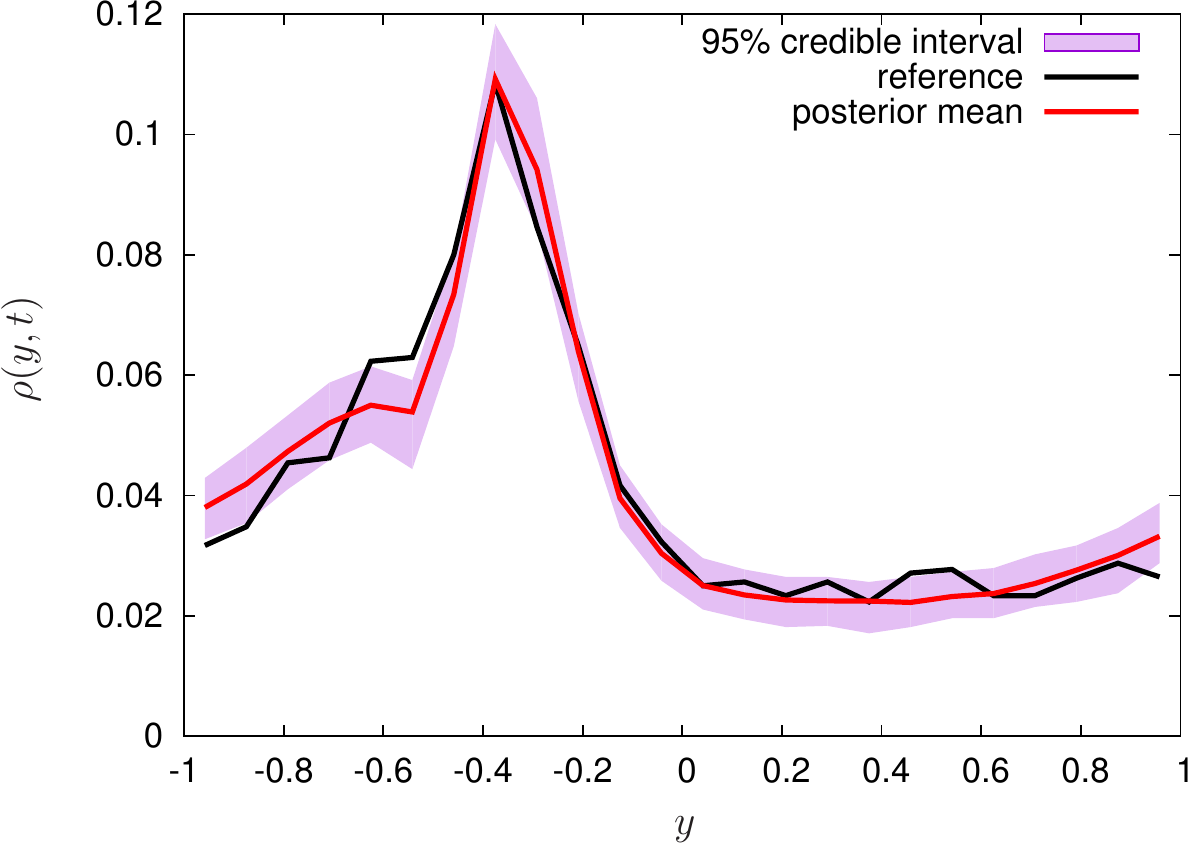}
 \caption{$t=4\Delta t$}
 \end{subfigure}
 \begin{subfigure}[b]{.9\textwidth}
\includegraphics[width=\textwidth]{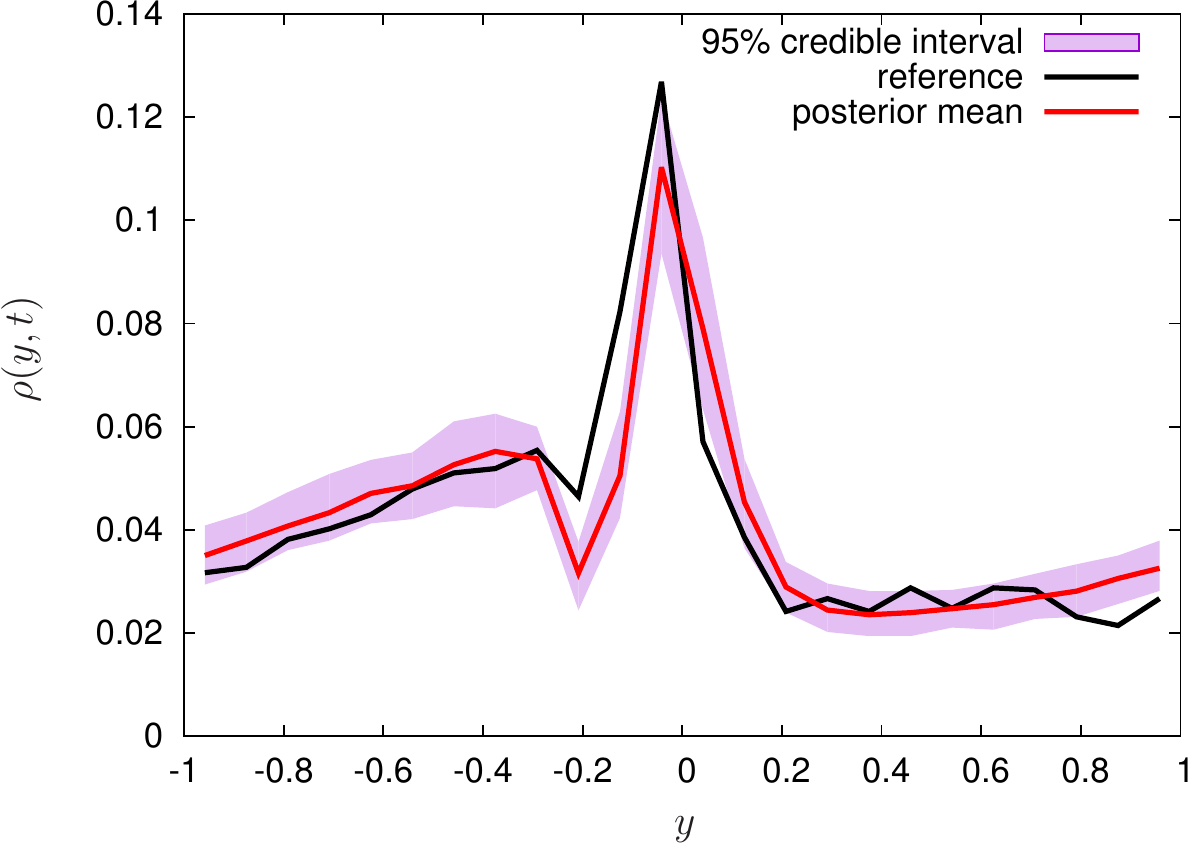}
 \caption{$t=8\Delta t$}
 \end{subfigure}
 \begin{subfigure}[b]{.9\textwidth}
\includegraphics[width=\textwidth]{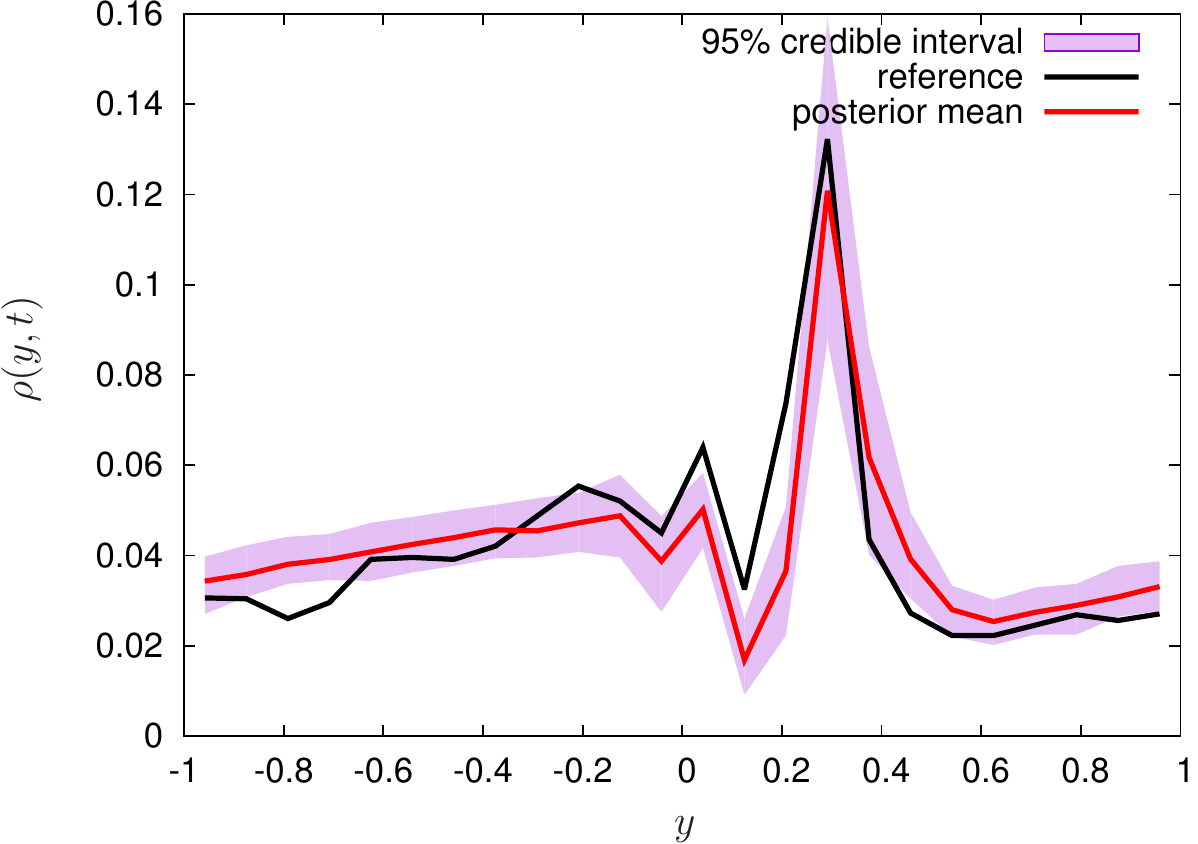}
 \caption{$t=12\Delta t$}
 \end{subfigure}
 \end{minipage}
 \begin{minipage}[c]{0.5\textwidth}
\centering
\begin{subfigure}[b]{.9\textwidth}
\includegraphics[width=\textwidth]{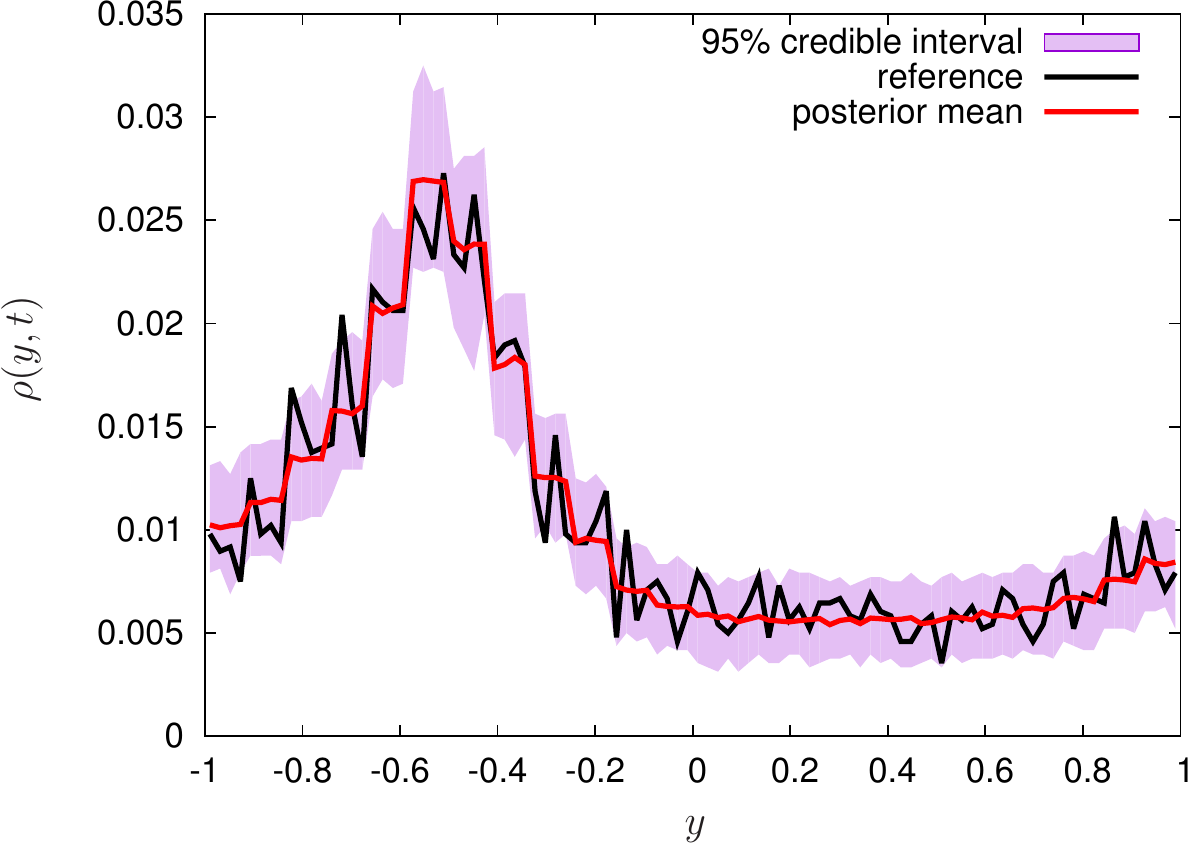}
 \caption{$t=2\Delta t$}
 \end{subfigure}
 \begin{subfigure}[b]{.9\textwidth}
\includegraphics[width=\textwidth]{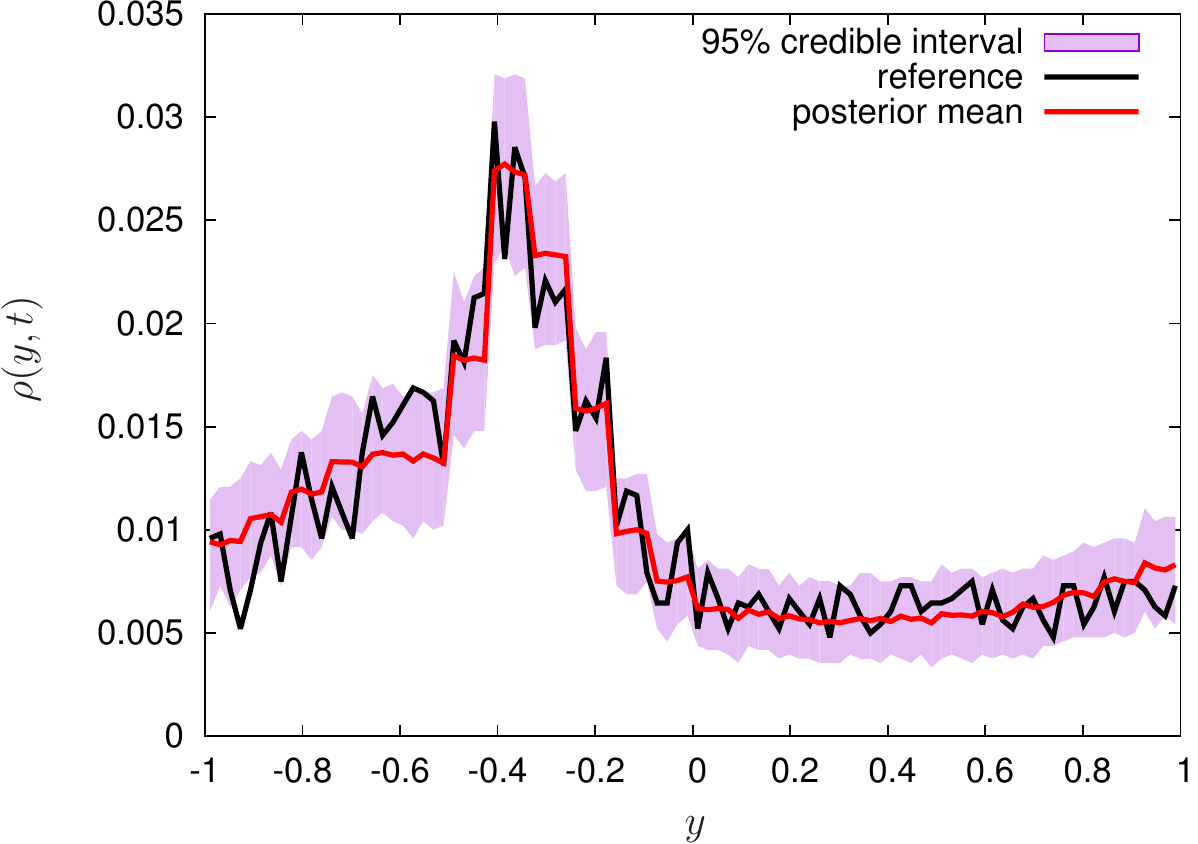}
 \caption{$t=4\Delta t$}
 \end{subfigure}
 \begin{subfigure}[b]{.9\textwidth}
\includegraphics[width=\textwidth]{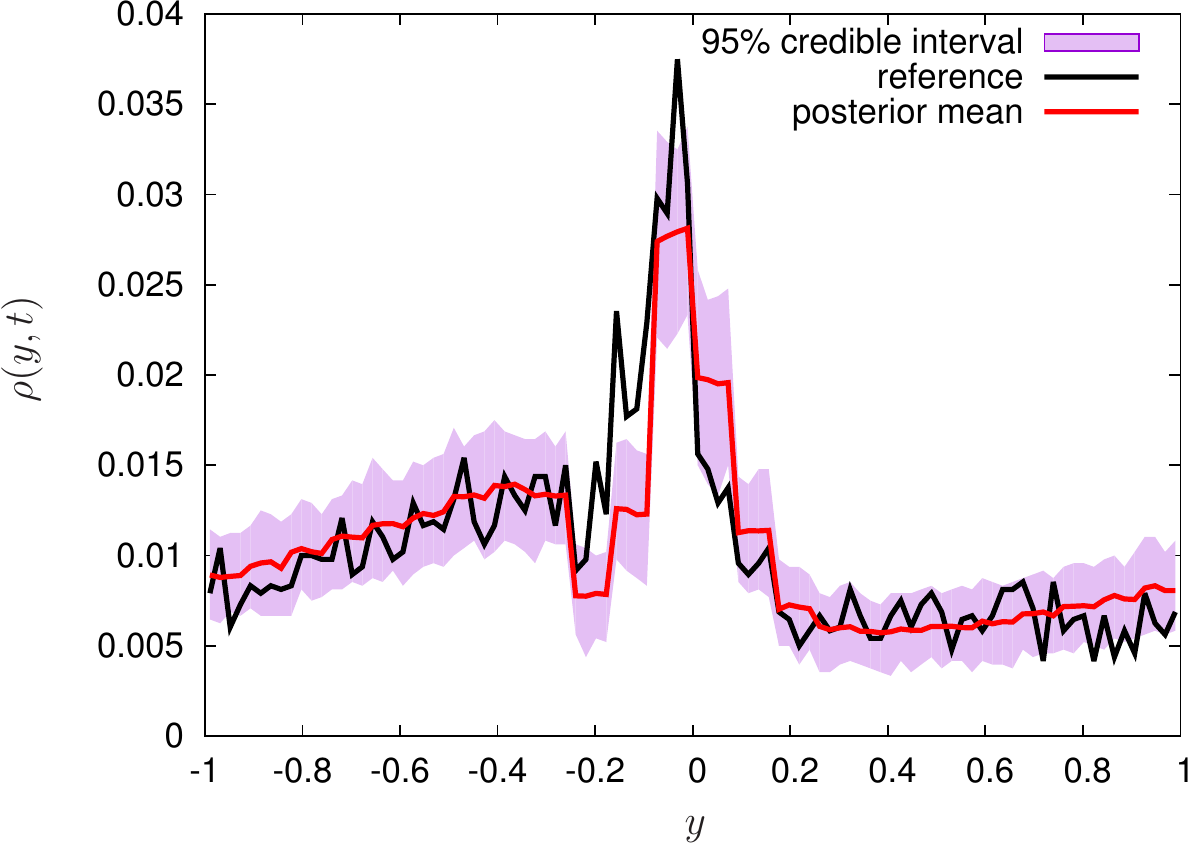}
 \caption{$t=8\Delta t$}
 \end{subfigure}
 \begin{subfigure}[b]{.9\textwidth}
\includegraphics[width=\textwidth]{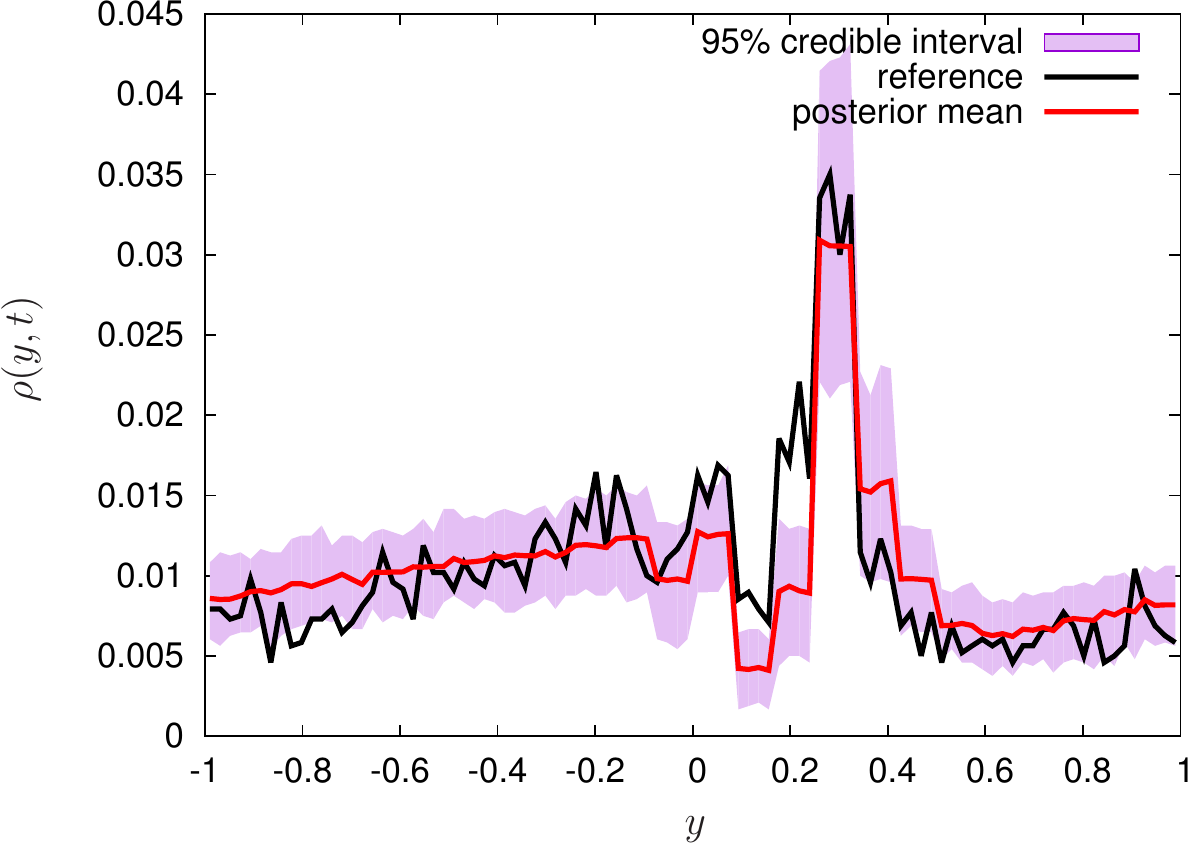}
 \caption{$t=12\Delta t$}
 \end{subfigure}
 \end{minipage}
 \caption{Predictive estimates of walker density for various future  times. The left column corresponds to  the  original binning, whereas the right column to  4 times finer binning.  The reference density profile was computed by simulating the FG model of walkers using the FG time step $\delta t$. ($N=128$) }
 \label{fig:bupredrho128}
\end{figure}

%

\section{Conclusions}
\label{sec:con}

We presented a probabilistic, generative probabilistic model for the automated construction of predictive, coarse-grained models from data generated by controlled, computational simulations of fine-grained models. Rather than trying to project the fine-scale state variables and governing equations to a reduced description/model, we follow the opposite route according to which the coarse states give rise to the fine-scale ones. In particular we employ a state-space model that consists of two  probability densities, the first  describing the evolution of the coarse state variables and the second the map from the coarse description to the the fine. We adopt a Bayesian perspective in order to infer the hidden, coarse states corresponding to the data and calibrate the parameters of the model. As a result the uncertainty arising from information loss as well as the finiteness of the data available, can be quantified. More importantly  this uncertainty can be propagated in order to produce probabilistic predictive estimates of the {\em full} fine-scale picture at any future time. Hence observables not contained in the coarse description can be predicted and the certainty in these predictions can be quantified. 

We demonstrated the efficacy of the proposed model for high-dimensional systems of random walkers which, by adjusting the type of interactions, can give rise to a wide range of collective behaviors. In all cases, with small to modest data, very good predictive estimates have been obtained even in cases where the mass density of the walkers forms shocks. 
An important component of the overall formulation is the discovery of salient features in the coarse model's evolution law. In the absence of prior information, this poses a formidable model selection problem which we address by employing sparsity-inducing, hierarchical priors. As demonstrated in the numerical illustrations, despite the availability of hundreds  of feature functions which encode a multitude of interaction types and orders at the coarse scale, the proposed model is able to prune most of them and reveal a small subset that encodes the appropriate evolution characteristics.

Perhaps the most critical question in coarse-graining endeavors in physical modeling, pertains to the enforcement of physical constraints. In the context of the models examined, these reduce to the conservation of mass which we were able to incorporate a priori using the softmax transformation. When the coarse-grained description adopted,  corresponds to continuum scales, it is well-known that additional such constraints are available from (continuum) thermodynamics in the form of conservation laws or the entropy inequality. Incorporating such constraints/invariances is of primary importance not only because it ensures physically-meaningful predictions but also because it has the potential of significantly reducing the amount of training data needed. Particularly since the data are generated from expensive, fine-scale simulations which unavoidably explore a (small) portion of the configuration space. An automated, general framework for enforcing such constraints is currently missing from the vocabulary of probabilistic modeling, and in our opinion, represents the most important challenge along this modeling direction.

\section*{Appendix A}
\label{sec:appF}
In this section we derive the expression of the ELBO $\mathcal{F}$ based on the original formula in \refeq{eq:elbo} and the optimal approximating densities $q$ for the latent variables of the model as presented in section \ref{sec:inference}. We note that (up to constants):
\be
\begin{array}{ll}
 \mathcal{F} & =   \int q( \bxx_{\Delta t}^{(1:N)}, \bt_c, \bs{\tau}, \bs{v})  \log p(\bx_{\Delta t}^{(1:N)} | \bxx_{\Delta t}^{(1:N)} ) ~d\bxx_{\Delta t}^{(1:N)}~d\bt_c~d\bs{\tau}~d\bs{v} \\
 & +\int q( \bxx_{\Delta t}^{(1:N)}, \bt_c, \bs{\tau}, \bs{v})  \log \frac{p( \bxx_{\Delta t}^{(1:N)}, \bt_c, \bs{\tau}, \bs{v}) }{q( \bxx_{\Delta t}^{(1:N)}, \bt_c, \bs{\tau}, \bs{v})} ~d\bxx_{\Delta t}^{(1:N)}~d\bt_c~d\bs{\tau}~d\bs{v} \\
 & = \sum_{i=1}^N < \log p_{cf}(\bx_{\Delta t}^{(i)} | \bxx_{\Delta t}^{(i)} )>_{q(\bxx_{\Delta t}^{(1:N)})} \hfill \textrm{(from \refeq{eq:like})} \\
 & + \sum_{i=1}^N <\log p_{cf}( \bxx_{\Delta t}^{(i)} | \bxx_{0}^{(i)},\bt_c, \bs{v})>_{q(\bxx_{\Delta t}^{(i)}) q(\bt_c) q(\bs{v})} \hfill \textrm{(from Equations (\ref{eq:prior1}), (\ref{eq:prior2})  )} \\
 & + < \log p(\bt_c | \bs{\tau})>_{q(\bt_c) q(\bs{\tau})} + <\log p(\bs{\tau})>_{q(\bs{\tau}} +<\log p(\bs{v})>_{q(\bs{v})}\hfill \textrm{(from  \refeq{eq:prior1})} \\
 & -\sum_{i=1}^N <q(\bxx_{\Delta t}^{(i)})>_{q(\bxx_{\Delta t}^{(i)})} -<\log q(\bt_c)>_{q(\bt_c)} \\
 & -<\log q(\bs{\tau})>_{q(\bs{\tau})}-<\log q(\bs{v})>_{q(\bs{v})} \hfill \textrm{(from  \refeq{eq:mf})}  \\
 & =   \sum_{i=1}^N < \log p_{cf}(\bx_{\Delta t}^{(i)} | \bxx_{\Delta t}^{(i)} )>_{q(\bxx_{\Delta t}^{(1:N)})} \\
 & +\frac{N}{2} \sum_{j=1}^{n_c} <log v_j>_{q(\bs{v})} \\
 & -\frac{1}{2} \sum_{i=1}^N \sum_{j=1}^{n_c} <v_j>_{q(\bs{v})} <(X_{(k+1)\Delta t, ~j}^{(i)}-\bt_c^T \bs{\phi}^{(j)}(\bxx_0^{(i)}) )^2>_{q(\bxx_{\Delta t}^{(i)}) q(\bt_c)}  \hfill \textrm{(from  \refeq{eq:pc})}  \\
 & +\frac{1}{2}\sum_{l=1}^L <\log \tau_l>_{q(\bs{\tau})} -\frac{1}{2} <\tau_l>_{q(\bs{\tau})} <\theta_{c,l}^2>_{q(\bt_c)} \hfill \textrm{(from  \refeq{eq:ard1})} \\
 & + (\alpha_0-1) \sum_{l=1}^L <\log \tau_l>_{q(\bs{\tau})}-\beta_0  \sum_{l=1}^L <\tau_l>_{q(\bs{\tau})} \hfill \textrm{(from  \refeq{eq:ard2})}  \\
 & + (\gamma_0-1)\sum_{j=1}^{n_c} <\log v_j>_{q(\bs{v})} -\zeta_0 \sum_{j=1}^{n_c} <v_j>_{q(\bs{v})} \hfill \textrm{(from  \refeq{eq:vprior})}  \\
 & +\frac{1}{2} \sum_{i=1}^N \log |\bs{S}_i| \hfill \textrm{(from  \refeq{eq:qxi})} \\
 & + \frac{1}{2} \log | \bs{S}_{\bt}|   \hfill \textrm{(from  \refeq{eq:qopttheta1})} \\
 & -\sum_l^L (\alpha_l-1)<\log \tau_l>_{q(\bs{\tau})}+ \sum_{l=1}^L \beta_l <\tau_l>_{q(\bs{\tau})} +\sum_{l=1}^L \log Z(\alpha_l, \beta_l) \hfill \textrm{(from  \refeq{eq:qopttau1})}  \\
 & - \sum_{j=1}^{n_c} (\gamma_j-1)<\log v_j>_{q(\bs{v})} +\zeta_0 \sum_{j=1}^{n_c} <v_j>_{q(\bs{v})}+\sum_{j=1}^{n_c} \log Z(\gamma_j,\zeta_j) \hfill \textrm{(from  \refeq{eq:qoptv1})}  \\
\end{array}
\ee
where $Z(\alpha, \beta)=\frac{\Gamma(\alpha)}{\beta^{\alpha}}$ is the normalization constant of the Gamma distribution with parameters $\alpha,\beta$.
Using the forms of optimal $q$'s as presented in Equations (\ref{eq:qopttheta1}), (\ref{eq:qopttau1}) and (\ref{eq:qoptv1}), we can establish that:
\be
\begin{array}{ll}
 \mathcal{F} & = \sum_{i=1}^N <\log p_{cf}(\bx_{\Delta t}^{(i)} | \bxx_{\Delta t}^{(i)} )>_{q(\bxx_{\Delta t}^{(1:N)})} \\
 &  +\frac{1}{2} \sum_{i=1}^N \log |\bs{S}_i| + \frac{1}{2} \log | \bs{S}_{\bt}|    \\
 & +\sum_{l=1}^L \log Z(\alpha_l, \beta_l)+\sum_{j=1}^{n_c} \log Z(\gamma_j,\zeta_j) 
\end{array}
\ee
which based on the fact that $\alpha_j$ and $\gamma_j$ are constant (given $N$), can be further simplified to:
\be
\begin{array}{ll}
 \mathcal{F} & =  \sum_{i=1}^N <\log p_{cf}(\bx_{\Delta t}^{(i)} | \bxx_{\Delta t}^{(i)} )>_{q(\bxx_{\Delta t}^{(1:N)})}> +\frac{1}{2} \sum_{i=1}^N \log | \bs{S}_i|      +\frac{1}{2} \log | \bs{S}_{\bt}| \\
  & - \sum_l^{L} \alpha_l \log \beta_l -\sum_{j=1}^{n_c} \gamma_j \log \zeta_j
\end{array}
\label{eq:elbofinal1}
\ee
The first expectation is analytically intractable given the form of the $p_{cf}$ in \refeq{eq:pcf}. For that purpose it is estimated with Monte Carlo. We finally note that lower bounds have been proposed for the log of the softmax function \cite{bouchard_efficient_2007,titsias_one-vs-each_2016} appearing in $p_{cf}$ which could lead to closed-form updates for $q(\bxx_{\Delta t}^{(1:N)})$ at the expense of a poorer approximation to the log-evidence.

\section*{Appendix B}
\label{sec:appadam}
The present section contains further details  for the computation of the gradient of $\mathcal{F}_i$ (\refeq{eq:fvari}) with respect to the variational parameters $\bs{\mu}_i, \bs{S}_i$ in \refeq{eq:qxi}. These derivatives are  used in the context of the ADAM algorithm for updating the latter parameters.
We recall from \refeq{eq:fvari} that:
\be
\begin{array}{ll}
\mathcal{F}_i & = <\log p_{cf}(\bx_{\Delta t}^{(i)} | \bxx_{\Delta t}^{(i)}) >_{q(\bxx_{\Delta t}^{(i)})}
 +  <\log p_c(\bxx_{\Delta t}^{(i)} | \bxx_{0}^{(i)},\bt_c,\bs{v}) >_{q(\bxx_{\Delta t}^{(i)})q(\bt_c) q(\bs{v})} - <\log q(\bxx_{\Delta t}^{(i)})>_{q(\bxx_{\Delta t}^{(i)})}
\end{array}
\label{eq:fvari2}
\ee

In the following we {\em drop the super/sub-script $i$ referring to the data point considered} in order to simplify the notations. Furthermore we employ a diagonal covariance for $\bs{S} =diag(\bs{s^2_j})$  and denote with $\bs{s} \in \RR^{n_c, +}$ the vector of the  diagonal standard deviations.  
The latter two terms in \refeq{eq:fvari2} can be analytically computed. In particular (up to a constant):
\be
<\log q(\bxx_{\Delta t})>_{q(\bxx_{\Delta t})}=-\frac{1}{2}\log |\bs{S}|
\ee
 
Furthermore,  from Eq. (\ref{eq:pc}) and (\ref{eq:pcmean}) (up to a constant)::
\be
\begin{array}{ll}
 <\log p_c(\bxx_{\Delta t} | \bxx_{0},\bt_c,\bs{v}) >_{q(\bxx_{\Delta t})q(\bt_c) q(\bs{v})} & =  -\frac{1}{2}\sum_{j=1}^{n_c} <v_j>_{q_j({v}_j)} ( \mu_{j}^2+s_j^2-2\mu_j < \bt_c^T>_{q(\bt_c)} \bs{\phi}^{(j)}(\bxx_0))
 \\
\end{array}
\ee
Finally for the first term in \refeq{eq:fvari2} we employ the reparametrization trick \cite{kingma_auto-encoding_2014}. This consists of the observation that for any function $f(\bxx_{\Delta t})$, its expectation with respect to $q(\bxx_{\Delta t})=\mathcal{N}(\bxx_{\Delta t}; \bs{\mu}, \bs{S})$ can be expressed as:
\be
<f(\bxx_{\Delta t})>_{q(\bxx_t)}= <f(\bs{\mu}+\sqrt{\bs{S}} \bs{\epsilon})>_{q(\bs{\epsilon})}
\ee
where $q(\bs{\epsilon})=\mathcal{N}(\bxx_t; \bs{0}, \bs{I})$ is the standard normal. Hence:
\be
\cfrac{\pa <f(\bxx_{\Delta t})>_{q(\bxx_{\Delta t})}}{\pa \bs{\mu}} = < \nabla_{\bxx_{\Delta t}} f(\bs{\mu}+\sqrt{\bs{S}}\bs{\epsilon})>_{q(\bs{\epsilon})}
\ee
and:
\be
\cfrac{\pa <f(\bxx_{\Delta t})>_{q(\bxx_{\Delta t})}}{\pa \bs{s}} = <  \nabla_{\bxx_{\Delta t}} f(\bs{\mu}+\sqrt{\bs{S}}\bs{\epsilon}) \otimes \bs{\epsilon}>_{q(\bs{\epsilon})}
\ee
where $\otimes$ is the Hadamard (element-wise) product.
The derivatives of the first term of \refeq{eq:fvari2} are computed by substituting $\log p_{cf}(\bx_{\Delta t} | \bxx_{\Delta t})$ in the place of $f$ above and using the functional form of $p_{cf}$ from \refeq{eq:pcf}.

\newpage
\bibliographystyle{h-physrev3}
\bibliography{paper}

\end{document}